\PassOptionsToPackage{pagebackref=false}{hyperref}
\documentclass[aip,jap,reprint,nofootinbib,10pt,longbibliography]{revtex4-2}
\usepackage[T1]{fontenc}
\usepackage[utf8]{inputenc}
\usepackage{booktabs}
\usepackage{bm}
\usepackage{amsmath}
\usepackage{amssymb}
\usepackage{graphicx}
\usepackage[dvipsnames]{xcolor}
\usepackage{subcaption}  
\usepackage[unicode=true,
 bookmarks=false, breaklinks=false,pdfborder={0 0 1},backref=false,colorlinks=false]
 {hyperref}

\hypersetup{colorlinks=true,linkcolor=blue!60!black,
            citecolor=blue!60!black,urlcolor=blue!50!black}

\usepackage{amsfonts}
\usepackage{textcomp}

\usepackage[bottom]{footmisc}
\usepackage{perpage}
\MakePerPage{footnote} 

\usepackage{float}

\newcommand{\taus}{\tau_{\mathrm s}}
\newcommand{\zetaR}{\zeta_{\mathrm R}}

\newcommand{\acg}{a_{\mathrm{cg}}}
\newcommand{\bcg}{b_{\mathrm{cg}}}

\begin{document}

\title{Local-to-global heating crossover in chains of nanomagnets: A two-scale
analytical framework}
\author{H. Kachkachi}
\affiliation{Université de Perpignan Via Domitia, Laboratoire PROMES-CNRS (UPR
8521), 66100 Perpignan, France}
\date{\today}

\begin{abstract}
We develop a two-scale analytical formalism to study heat generation and thermal transport in one-dimensional systems of nanomagnets subjected to a uniform alternating magnetic field.
At the nanoscale, each nanomagnet acts as a localized, temperature-dependent heat source governed by its magnetic response,
dipolar interactions, and interfacial coupling to the matrix, characterized by
a nanoscale volumetric loss coefficient $L_m$. After spatial and temporal averaging, we obtain a coarse-grained assembly-scale equation with effective heating terms and a macroscopic loss coefficient $L_N$.

Using modal decomposition, we solve both equations exactly under Dirichlet and Neumann boundary conditions and establish explicit conditions for a local-to-global heating crossover; this is governed by the competition between heat generation, diffusion, dipolar coupling, and hierarchical losses. The crossover is quantified through the spatial correlation length and temperature variance, with stability criteria
incorporating both diffusion and nanoscale losses. The coarse-graining procedure is derived rigorously, and its systematic approximation errors are quantified.

For prototypical magnetic hyperthermia systems, such as magnetite nanomagnets in water, our formalism reveals that realistic parameters place these systems firmly in the collective heating regime, with local temperature variations at the
$\sim\mu$K level, which is currently unresolvable experimentally.
The continuum Fourier description used here is validated by a Knudsen-number analysis ($\mathrm{Kn} \ll 1$ for amorphous polymer and aqueous
matrices).
\end{abstract}
\maketitle

\section{Introduction}
\label{sec:Introduction}
Thermal transport in nanostructures, driven by nanoscale heat sources, is central to a wide range of physical, chemical, and biological processes\cite{Pop2010EnergyDissipation, Chen2005, benenti23rivi} and applications, including magnetic hyperthermia\cite{rosensweig02j3m, OrtegaPankhurst_nsc13, Hergt_etal_JPMC2006}, nanoscale catalysis\cite{Bell2003, Haruta1997, Norskov2009}, spin-caloritronics\cite{
Bauer2012, Uchida2008, Adachi2013}, and thermally assisted switching\cite{ElBarajietal2009, Herzogetal2010, Granitzkaetal2017} in nanostructured materials. In assemblies of nanomagnets subjected to alternating magnetic fields, each nanomagnet dissipates energy through magnetic relaxation mechanisms, converting the electromagnetic energy it absorbs into heat. This dissipation is commonly quantified by the Specific Absorption Rate (SAR), or Specific Loss Power (SLP), which depends on intrinsic magnetic parameters, such as particle size and anisotropy, interparticle dipolar interactions (DI), as well as on the amplitude and frequency of the applied field.

A fundamental question that arises in such systems concerns the spatial organization of heating. At sufficiently short times and small length scales, heat generation should \textit{a priori} be inherently local. This implies that each nanomagnet acts as an individual source, giving
rise to localized temperature elevations in its immediate vicinity
\cite{riedinger2013subnanometer, DongZink2014, serantes2020local}, thus constituting a \textit{hotspot}.
On the other hand, at longer times and larger scales, thermal diffusion, interparticle interactions, and heat exchange with the surrounding medium couple these sources, leading to collective heating and spatial homogenization of the temperature field
\cite{rosensweig02j3m, Weaveretal2009, tan14prb, AstAle_jap19, Guetal2023}.
Understanding the conditions under which a system could transition from localized to global heating, and whether such conditions are realizable in practice, is essential for controlling thermal effects in applications ranging from targeted hyperthermia to nanoscale chemical activation
\cite{OrtegaPankhurst_nsc13, mai19acs, riedinger2013subnanometer}.
In dense nanomagnet assemblies, magnetic dipolar coupling between the nanomagnets can significantly alter their collective magnetic response, thereby modifying the power dissipation under an AC field \cite{Mehdaoui_prb2013, dejardin17jap, boucheretal11apl}.
This anisotropic and long-range interaction not only shifts the effective anisotropy and relaxation times \cite{ilg20phys, figueiredo07jpcm, sanchez22small, vernayetal14prb} but also introduces a temperature-dependent feedback that competes with purely thermal diffusion and interfacial coupling.

In this work we aim to establish a unified theoretical framework that accounts for both electromagnetic (dipolar) and thermal interplay in nanomagnet arrays, clarifying how interparticle spacing, which modulates dipolar coupling strength, affects the transition between local and global heating regimes.
In this respect, a two-scale description is necessary. Indeed, estimating the orders of magnitude for an isolated nanomagnet reveals a number of stringent physical constraints.
For example, the energy required to raise its temperature by 1~K is only $\sim 10^{-17}$~J, and the associated thermal relaxation time is on the order of 0.1--1~ns, which is far shorter than typical AC-field periods ($\sim \mu$s).
This means that any temperature spike (or hotspot) generated during one field cycle decays completely before the next, making heating effectively impulsive at the nanomagnet level \cite{davis20biop, Guetal2023, DongZink2014}.
Furthermore, maintaining a steady elevation of even 1~K a few nanometers from an isolated nanomagnet would require an unrealistically large dissipated power (SLP $\sim 10^9$~W~g$^{-1}$), far beyond experimentally attainable values ($\sim 10^2$--$10^3$~W~g$^{-1}$ for
state-of-the-art magnetite) \cite{Hergt_etal_JPMC2006, OrtegaPankhurst_nsc13,
hergt04jmmm}.

A conclusion of this short analysis is that observable heating at realistic power levels necessarily involves the collective contribution of many nanomagnets across multiple length and time scales.
However, most of the existing theoretical descriptions implicitly adopt a coarse-grained approach in which sources are pre-averaged over space and time and dissipation is modeled through effective volumetric loss terms. While adequate for macroscopic or long-time behavior \cite{Weaveretal2009, riedinger2013subnanometer, DongZink2014, DejKac2022, Guetal2023}, this approach washes out local temperature variations by construction \cite{davis20biop, tan14prb, AstAle_jap19} and may require unphysically large loss coefficients to match observations.
To address this issue, we develop a physically consistent description that explicitly separates, and then rigorously reconnects, two relevant scales. At the nanomagnet scale, individual nanomagnets are resolved as localized, temperature-dependent heat sources embedded in a thermal matrix. Heat exchange with the surrounding medium is characterized by both an interfacial (Newton-type) heat transfer coefficient $h_s$
\cite{lyeo06prb, Cahill_et_al_APR2014, davis20biop} and a nanoscale volumetric loss coefficient $L_m$. Then, the nanomagnet-scale heat equation, which incorporates these effects through renormalized coefficients $\tilde{a}$ and $\tilde{b}$ [see Eq.~\eqref{eq:HE_nm_final}], is solved exactly under Dirichlet and Neumann boundary conditions.
At the assembly scale, we derive a coarse-grained heat equation by explicit spatial and temporal averaging (Section~\ref{subsec:scale_link}). At this scale, microscopic sources are replaced by effective heating terms consistent with the time-averaged SLP, and the environmental coupling is described by an assembly-scale loss coefficient $L_N$.
Then, the relationship $L_N = L_m + L_{\rm emergent}$ implies that assembly-scale losses incorporate both direct nanoscale coupling and emergent large-scale contributions \cite{dejkac_as24, RytovEtAl_sr22, AstAle_jap19}.
This approach allows us to
(i)~establish the mathematical conditions under which localized thermal hotspots can exist and persist, and
(ii)~quantify the crossover toward collective heating as diffusion, dipolar coupling, feedback ($\tilde{b}$), and hierarchical losses ($L_m, L_N$) compete.

Although developed explicitly for one-dimensional nanomagnet chains, a geometry that is analytically tractable and experimentally relevant
\cite{talapatra20nanoscale, anand19prb, huizarfelix16apl}, the formalism is quite general.
For typical hyperthermia ferrofluids, the present analysis places the system firmly in the collective regime, with local variations suppressed to the $\sim\mu$K level \cite{davis20biop, Guetal2023, riedinger2013subnanometer}. Our formalism thus identifies what parameter combinations, such as ultra-high SLP materials, sub-nanometer spacings, or MHz-range fields that outpace diffusion, would be required to access genuinely localized heating regimes.

\textit{Organization of the paper}:
The paper is organized as follows. Section~\ref{sec:validity} validates the continuum Fourier diffusion approximation via Knudsen number analysis. Section~\ref{sec:framework} introduces the two-scale theoretical framework: Section~\ref{sec:nanoscale} formulates the nanomagnet-resolved heat equation with nanoscale losses and interfacial coupling, presenting its exact analytical solution in terms of renormalized coefficients; Section~\ref{sec:HE_assembly} introduces the coarse-grained, assembly-scale heat equation; and Section~\ref{subsec:scale_link} explicitly derives the mathematical connection between the two descriptions. Section~\ref{sec:parameters} provides explicit scaling relations between physical parameters and dimensionless control variables, with reference values for magnetite-PMMA and magnetite-water systems.
Section~\ref{sec:Results} analyzes the resulting thermal dynamics, identifies the indicators of the local-to-global heating crossover, quantifies the two-scale consistency and its systematic limitations, and provides physical interpretation of boundary condition effects.
The expressions of the heating terms and their relation to SLP calculations are addressed in Appendix~\ref{app:sar_coeffs}, a comparison between Dirichlet boundary conditions (DBC) and Neumann boundary conditions (NBC) is presented in Appendix~\ref{app:BC_comparison}, and for completeness, Appendix~\ref{app:GF_nanoscale} presents an alternative Green's function formulation of the nanomagnet-scale problem.

\section{Validity of the continuum approximation at the nanoscale}
\label{sec:validity}

A critical requirement for modeling thermal transport at the nanoscale is to determine the applicability of classical Fourier diffusion \cite{Grmela2005Ballisticdiffusive, Yang2010, Pop2010EnergyDissipation, Jou2011, Cahill_et_al_JAP2003, Cahill_et_al_APR2014}. In low-dimensional crystalline systems, such as silicon nanowires, the phonon mean free path ($\Lambda_{\text{mfp}}$) can exceed hundreds of nanometers, leading to ballistic transport and a breakdown of Fourier's law when the system dimensions are comparable to $\Lambda_{\text{mfp}}$. However, the validity of the continuum approximation depends intrinsically on the material's crystalline order and the resulting phonon scattering rates. Here, we explicitly define the physical regime in which our two-scale framework is valid.

\subsection{Length scale hierarchy and Knudsen number}
The transition from diffusive to ballistic transport is governed by the Knudsen number, $Kn = \Lambda_{\text{mfp}} / d$, where $d$ is the characteristic system length scale (here, the interparticle spacing). Fourier's law is valid only in the limit $Kn \ll 1$ \cite{Ziman1960, Rezgui2023, Ziambarasetal2005}.

We consider nanomagnets embedded in two distinct environments: an amorphous polymer matrix (PMMA) and an aqueous fluid.
\begin{enumerate}
    \item \textbf{Amorphous matrices (PMMA):} In amorphous solids, the lack of long-range structural order leads to strong phonon scattering. The phonon mean free path is typically on the order of the inter-atomic spacing or short-range correlation lengths. For PMMA at room temperature, $\Lambda_{\text{PMMA}} \simeq 0.5 - 2$ nm.
    \item \textbf{Liquid matrices (water):} In aqueous ferrofluids, heat transport is dominated by molecular collisions with extremely short mean free paths, $\Lambda_{\text{water}} < 0.5$ nm.
\end{enumerate}

For the typical interparticle spacings considered in this work ($d \simeq 30 - 50$ nm, i.e., at least three times the NM diameter), we calculate the Knudsen number:
\begin{equation}
    Kn_{\text{PMMA}} \simeq \frac{2 \text{ nm}}{30 \text{ nm}} \simeq 0.07, \quad
    Kn_{\text{water}} \simeq \frac{0.3 \text{ nm}}{30 \text{ nm}} \simeq 0.01.
\end{equation}
In both cases, $Kn \ll 1$ (or at most $Kn \sim 0.1$ for very dense PMMA arrays). Consequently, the phonon transport is dominated by scattering events within the matrix rather than boundary scattering at the nanomagnets. This places the system firmly in the diffusive regime, justifying the use of the parabolic heat equation \eqref{eq:HE_nm_dim} and the definition of an effective thermal conductivity $\kappa_m$.

It is important to note that our framework would not apply without modification to nanomagnets embedded in crystalline matrices (e.g., Silicon or Graphene)\cite{Rezgui2023, Ziambarasetal2005}, where $\Lambda_{\text{mfp}} \sim 100$ nm would yield $Kn > 1$, necessitating a ballistic-diffusive approach such as the Cattaneo-Vernotte equation \cite{Vernotte1958, Cattaneo1958} or the Boltzmann Transport Equation~\cite{Majumdar1993, Chen2000}.

\subsection{Timescale separation and thermal inertia}
The validity of the parabolic diffusion model also requires that the observation timescale dominates the thermal relaxation time of the heat carriers (phonons), $\tau_{\text{ph}}$. The Cattaneo-Vernotte equation~\cite{Cattaneo1958, Vernotte1958}, which accounts for the finite velocity of heat propagation (second sound), introduces a relaxation term $\tau_{\text{ph}} \partial \mathbf{q}/\partial t$.
For amorphous polymers, $\tau_{\text{ph}}$ is on the order of $1-10$ ps. In contrast, the magnetic heating occurs on the timescale of the AC driving field period, $\mathcal{T}_{\text{AC}} = 1/f$. For a typical frequency $f = 200$ kHz, the excitation period is $\mathcal{T}_{\text{AC}} \simeq 5 \mu\text{s}$. The diffusive time scale across the interparticle spacing is $t_d \simeq 0.1 - 1$ ns. The hierarchy of time scales thus is:
\begin{equation}
    \tau_{\text{ph}} (\sim 10 \text{ ps}) \ll t_d (\sim 1 \text{ ns}) \ll \mathcal{T}_{\text{AC}} (\sim 5000 \text{ ns}).
\end{equation}
Since the thermal inertia timescale $\tau_{\text{ph}}$ is orders of magnitude shorter than the diffusion time $t_d$, the hyperbolic terms in the heat equation can be safely neglected. The system operates in a quasi-static thermal regime relative to the phonon dynamics, ensuring thermodynamic consistency without the need for generalized hydrodynamic terms.


\section{Theoretical framework}
\label{sec:framework}

\subsection{Nanomagnet-scale heat equation}
\label{sec:nanoscale}

Now, we formulate the thermal problem at the scale of individual nanomagnets, where heat is generated from magnetic dissipation under an alternating magnetic field and is transferred locally to the surrounding matrix. This description resolves spatial variations on nanometer length scales and applies to time scales larger than the AC-field period but short compared to macroscopic thermal relaxation times. At this scale, each nanomagnet acts as a localized heat source embedded in a continuous host medium, and exchanges heat with its immediate surroundings through interfacial thermal coupling.


We consider a one-dimensional domain $0\le x\le L$ containing a chain of $\mathcal{N}$ nanomagnets located at positions
\begin{equation}
x_{n}=nd,\qquad n=0,\dots,\mathcal{N}-1,
\end{equation}
where $d$ is the interparticle spacing and $L=(\mathcal{N}-1)d$. The nanomagnets are embedded in a homogeneous matrix characterized by mass density $\rho_{m}$, specific heat $c_{v,m}$, and thermal conductivity $\kappa_{m}$.

The temperature field $T(x,t)$ denotes the continuous temperature in the embedding medium. It obeys the heat equation
\begin{align}
\rho_{m}c_{v,m}\,\frac{\partial T}{\partial t}-\kappa_{m}\,\frac{\partial^{2}T}{\partial x^{2}}= & \sum_{n=0}^{\mathcal{N}-1}P_{n}(t)\,\delta(x-x_{n})\nonumber \\
 & -L_{m}\left[T(x,t)-T_{0}\right],\label{eq:HE_nm_dim}
\end{align}
where $P_{n}(t)$ is the \emph{power density per unit transverse area}
(W/m$^{2}$) dissipated by the nanomagnet $n$ and transferred to the surrounding
matrix, and $L_{m}$ is a volumetric Newton cooling coefficient representing
heat leakage from the matrix to the external environment at the nanoscale.
More precisely, it characterizes direct environmental coupling of the immediate matrix surrounding each nanomagnet. At the nanoscale and short times ($t\sim t_{d}$, where $t_{d}$ is the diffusion time across one interparticle spacing), heat remains largely confined near the NM, and environmental losses are relatively weak.


For each nanomagnet $n$, we distinguish: (i) its internal (assumed
uniform) temperature $T_{i}^{(n)}(t)$, (ii) its surface temperature
$T_{s}^{(n)}(t)$, and (iii) the matrix temperature $T(x,t)$ evaluated
at the nanomagnet position, $T(x_{n},t)$.

At the nanomagnet scale, we model heat exchange between each nanomagnet and the matrix
via an interfacial (Newtonian) thermal conductance $h_{s}$ (W m$^{-2}$K$^{-1}$).
Because of the small nanomagnet size ($R\sim10$ nm) and high internal thermal
conductivity $\kappa_{p}$, the Biot number\footnote{The Biot number (Bi) is defined as the ratio of internal conduction resistance to external convection resistance [See Ref. \onlinecite{Pop2010EnergyDissipation}]: $\mathrm{Bi}=h L /\kappa$, where $h$ is the heat transfer coefficient (W/m$^2$/K), $L$ the characteristic length (m), and $\kappa$ the thermal conductivity of the solid (W/m/K).} $\mathrm{Bi}=R/(\kappa_{p}/h_s)\ll 1$,
justifying the approximation,
\begin{equation}
T_{i}^{(n)}(t)=T_{s}^{(n)}(t),\label{eq:lumped_nm}
\end{equation}
so that the nanomagnet is characterized by a single internal temperature.

Therefore, heat transfer between a nanomagnet $n$ and the surrounding matrix
is governed by (Newtonian) interfacial exchange,
\begin{equation}
P_{n}(t)=h_{s}\left[T_{s}^{(n)}(t)-T(x_{n},t)\right],\label{eq:Newton_NM}
\end{equation}

This equation defines how the nanomagnet injects heat into the matrix and ensures that thermal coupling between different nanomagnets is mediated through the temperature field $T(x,t)$.


Each nanomagnet is subjected to an AC magnetic field of angular frequency
$\omega$ and amplitude $h_{0}$. Magnetic dissipation occurs on the
time scale of the AC period $\mathcal{T}_{\text{AC}}=2\pi/\omega$. Since thermal
diffusion and interfacial heat exchange are much slower than the field
oscillations, we replace the instantaneous dissipated power by its
cycle-averaged value,
\begin{equation}
P_{n}(t)\;\longrightarrow\;\overline{P}_{n}\!\left(T_{i}^{(n)}\right).
\end{equation}

The cycle-averaged magnetic power dissipated inside a nanomagnet $n$ is related to the specific loss power (SLP), in W/kg, as
\begin{equation}
\overline{P}_{n}\!\left(T_{i}^{(n)}\right)=\frac{V_{p}\rho_{p}}{A_{p}}\,\mathrm{SLP}\!\left(T_{i}^{(n)}\right),\label{eq:Pn_SAR}
\end{equation}
where $A_{p}$ is the nanomagnet surface area, $V_{p}$ its volume and $\rho_{p}$
its mass density. For a spherical nanomagnet of radius $R$, the coefficient in Eq. (\ref{eq:Pn_SAR})
becomes $\text{R}\rho_{p}/3$.

$\mathrm{SLP}\!\left(T_{i}^{(n)}\right)$ depends on the nanomagnet's internal temperature through its magnetic relaxation parameters, as derived in Appendix~\ref{app:sar_coeffs} from the linear response theory, taking into account (weak) dipolar interactions \cite{rosensweig02j3m, Hergt_etal_JPMC2006, dejardin17jap, DejKac2022, DejKac2024},
\begin{equation}
\rho_{p}\,\mathrm{SLP}\!\left(T\right)=\frac{\mu_{0}h_{0}^{2}\omega}{2}\chi_{\mathrm{eq}}\left(T\right)\frac{\eta\left(T\right)}{1+\eta^{2}},\quad\eta\left(T\right)=\frac{\omega}{\Gamma\left(T\right)},\label{eq:SAR_LRT}
\end{equation}
where $\chi_{\mathrm{eq}}$ is the equilibrium susceptibility and $\Gamma\left(T\right)$ the relaxation rate.

Combining Eqs.~\eqref{eq:Newton_NM} and \eqref{eq:Pn_SAR} gives the self-consistent relation for the nanomagnet's internal temperature:
\begin{equation}
\frac{V_{p}\rho_{p}}{A_{p}}\,\mathrm{SLP}\!\left(T_{i}^{(n)}\right)=h_{s}\left[T_{i}^{(n)}-T(x_{n},t)\right],\label{eq:self_consistent_T}
\end{equation}
which implicitly determines $T_{i}^{(n)}$ in terms of the local matrix temperature $T(x_{n},t)$ [see below]. This coupling indirectly links all nanomagnets through the matrix temperature field $T(x,t)$.


In a prototypical situation, we have moderate (relative) temperature elevations
($|T_{i}^{(n)}-T_{0}|\ll T_{0}$)~\cite{DejKac2022}, so that the SLP can be linearized
about the ambient temperature $T_{0}$ and written in terms of the relative temperature elevation
\begin{equation}
\theta_{i}^{(n)}(t)=\frac{T_{i}^{(n)}(t)-T_{0}}{T_{0}},
\end{equation}
as
\begin{equation}
\rho_{p}\,\mathrm{SLP}\!\left(T_{i}^{(n)}\right)\simeq a_{p}\left(T_{0}\right)+b_{p}\left(T_{0}\right)\,\theta_{i}^{(n)},\label{eq:SLP_linear}
\end{equation}
where the coefficients $a_{p}\left(T_{0}\right)$ and $b_{p}\left(T_{0}\right)$ which encode, respectively, the baseline heating strength and the thermo-magnetic feedback; their explicit expresions, their plots, and other details are given in Appendix~\ref{app:sar_coeffs}.

Therefore, using Eqs. (\ref{eq:Pn_SAR}, \ref{eq:SLP_linear}), the cycle-averaged power injected by nanomagnet $n$ can be written as
\begin{align}
\overline{P}_{n}\!\left(T_{i}^{(n)}\right) & =\frac{V_{p}}{A_{p}}\left[a_{p}+b_{p}\,\theta_{i}^{(n)}\right].\label{eq:Pn_linear}
\end{align}

The relationship between the nanomagnet's internal temperature $\theta_{i}^{(n)}$ and the matrix temperature at the nanomagnet position, $\theta(x_{n},t)$, is obtained from the self-consistent condition equating the magnetic power dissipated inside the nanomagnet to the interfacial heat flux into the matrix, Eqs.~(\ref{eq:Pn_SAR}, \ref{eq:self_consistent_T}). Substituting the linearized power expression Eq.~\eqref{eq:Pn_linear} and the Newtonian exchange law Eq.~\eqref{eq:Newton_NM} yields:
\begin{align}
\frac{V_{p}}{A_{p}}\left[a_{p}+b_{p}\,\theta_{i}^{(n)}(t)\right]=T_{0}h_{s}\left[\theta_{i}^{(n)}-\theta(x_{n},t)\right],\label{eq:Temp_BC}
\end{align}
where we have also introduced the relative temperature elevation

\[
\theta(x,t)=\frac{T(x,t)-T_{0}}{T_{0}}.
\]

Using the (renormalized) interfacial coupling strength (in W/m$^{3}$), $\gamma_{s}=A_{p}h_{s}T_{0}/V_{p}$, and introducing the following
dimensionless coefficients
\begin{equation}
 \gamma\equiv\frac{\gamma_{s}}{a_{p}},\quad\varepsilon\equiv\frac{b_{p}}{a_{p}},\label{gameps}
\end{equation}
Eq.~\eqref{eq:Temp_BC} can be solved for $\theta_{i}^{(n)}$ to obtain the explicit relation

\begin{equation}
\theta_{i}^{(n)}(t)=\frac{1+\gamma\,\theta(x_{n},t)}{\gamma-\varepsilon},\label{eq:theta_relation_full}
\end{equation}
which is valid for arbitrary $\gamma$.


Let us now introduce the following dimensionless space and time variables
\begin{equation}
\xi=\frac{x}{d},\qquad\tau=\frac{t}{t_{d}},\qquad t_{d}=\frac{\rho_{m}c_{v,m}d^{2}}{\kappa_{m}},\label{eq:xi-tau-td}
\end{equation}
where $t_{d}$ is interpreted as the \emph{thermal diffusion time} over one interparticle spacing $d$.

Next, substituting Eq.~\eqref{eq:theta_relation_full} into Eq. \eqref{eq:Pn_linear} eliminates the explicit dependence
on the nanomagnet's internal temperature $\theta_{i}^{(n)}$ and leads to the following closed equation (dimensionless) for the matrix temperature
$\theta(\xi,\tau)$
\begin{equation}
\frac{\partial\theta}{\partial\tau}-\frac{\partial^{2}\theta}{\partial\xi^{2}}=\sum_{n=0}^{\mathcal{N}-1}\left[\tilde{a}+\tilde{b}\,\theta(\xi_{n},\tau)\right]\delta(\xi-\xi_{n})-\beta\,\theta(\xi,\tau),\label{eq:HE_nm_final}
\end{equation}
where the renormalized heating coefficients
\begin{equation}
\tilde{a}=\frac{\gamma}{\gamma-\varepsilon}a,\qquad\tilde{b}=\frac{\gamma}{\gamma-\varepsilon}b,\label{eq:renorm_coeffs}
\end{equation}
incorporate the effect of finite interfacial thermal resistance through the dimensionless coupling strength $\gamma$. For later convenience,
we have introduced, respectively, the renormalized baseline heating strength, the thermo-magnetic feedback and the nanoscale loss coefficient
\begin{equation}
a=\Upsilon_{0}\,a_{p},\quad b=\Upsilon_{0}\,b_{p},\quad\beta\equiv\frac{L_{m}t_{d}}{\rho_{m}c_{v,m}}\label{eq:nano-loss-coeffs},
\end{equation}
with
\begin{equation}
 \Upsilon_{0}\equiv\frac{V_{p}\,t_{d}}{A_{p}\rho_{m}c_{v,m}T_{0}\,d}\label{eq:Upsilon}.
\end{equation}

The parameter $\beta$ can also be written as $\beta = t_d/t_l$, where $t_l=\rho_{m}c_{v,m}/L_{m}$ represents the leakage time.

In the strong interfacial-coupling limit ($\gamma\gg|\varepsilon|$), Eq.~\eqref{eq:theta_relation_full} reduces to $\theta_{i}^{(n)}\simeq\theta(x_{n},t)$, implying that the nanomagnet and matrix temperatures are essentially equal, and the renormalized coefficients recover their bare values, $\tilde{a}\to a$, $\tilde{b}\to b$. For finite $\gamma$, however, the nanomagnet's internal temperature differs from the matrix temperature: the parameter $\varepsilon$ is renormalized by the interfacial thermal resistance through the denominator $\gamma-\varepsilon$, and the resulting coefficients $\tilde{a}$, $\tilde{b}$ modify both the baseline heating and the thermo-magnetic feedback, thereby capturing the influence of interfacial resistance on the effective heat injection into the matrix\footnote{Furthermore, Kapitza-interface effects at nanomagnet/matrix contact can be included by introducing the Kapitza resistance $R_{\mathrm{K}}$ (or conductance $G=1/R_{\mathrm{K}}$, and replacing $b\to b_{\mathrm{eff}}=\lambda b$, $a\to a_{\mathrm{eff}}=\lambda a$, with $\lambda=\Lambda_{3}/(1+\Lambda_{3})$, $\Lambda_{3}=GR/\kappa$; $R$ is the radius of the nanomagnet.}.

Note that, by using the delta function on the right-hand side of Eqs. (\ref{eq:HE_nm_dim}, \ref{eq:HE_nm_final}), we are considering the nanomagnets as point heat sources. However, to reflect finiteness of the nanomagnet size, we may replace the point sources at $\xi_{n}$ by narrow Gaussian functions
\begin{align}
 g_{\sigma_{g}}\left(\xi-\xi_{n}\right)\;&=\;\frac{1}{\sqrt{2\pi}\,\sigma_{g}}\,\exp\!\Big(-\frac{\left(\xi-\xi_{n}\right)^{2}}{2\sigma_{g}^{2}}\Big),\\
 \int_{\mathbb{R}}&g_{\sigma_{g}}(\xi)\,d\xi=1,
\end{align}
so that the first term on right-hand side of~\eqref{eq:HE_nm_final} becomes
\[
\sum_{n=0}^{\mathcal{N}-1}\Big[\tilde{a}+\tilde{b}\,\theta(\xi_{n},\tau)\Big]\;g_{\sigma_{g}}(\xi-\xi_{n}).
\]
As such, the feedback is still local in temperature (evaluated at the particle center $\xi_{n}$), but spatially distributed over a small neighborhood by $g_{\sigma_{g}}$.

We now impose Dirichlet boundary conditions,
\begin{equation}
\theta(0,\tau)=\theta(\Lambda,\tau)=0,\qquad\Lambda=\frac{L}{d},\label{eq:Dirichlet_nm}
\end{equation}
corresponding to a nanomagnet chain in contact with an ideal thermal bath.

The choice of boundary conditions---Dirichlet (DBC) versus Neumann (NBC)---has non-trivial consequences for the thermal response of the assembly, and a detailed comparison is deferred to Sections~(\ref{sec:DBC_NBC_summary}, \ref{sec:Results}) and Appendix~\ref{app:BC_comparison}. Here we proceed with DBC by way of illustration and build the exact modal solution, the structure of which carries over directly to NBC with the substitution of the corresponding eigenfunctions.


The solution of Eq.~\eqref{eq:HE_nm_final} is obtained by expanding the temperature field in the orthonormal Dirichlet eigenfunctions
\begin{equation}
\phi_{r}(\xi)=\sqrt{\frac{2}{\Lambda}}\sin\!\left(\frac{r\pi\xi}{\Lambda}\right),\qquad r=1,2,\dots,\label{eq:eigenfunctions}
\end{equation}
as
\begin{equation}
\theta(\xi,\tau)=\sum_{r=1}^{\infty}c_{r}(\tau)\,\phi_{r}(\xi).\label{eq:modal_expansion}
\end{equation}

The modal amplitudes satisfy the coupled system
\begin{equation}
\frac{d\mathbf{c}}{d\tau}=\mathbf{B}\mathbf{c}+\mathbf{d},\label{eq:modal_nm_full}
\end{equation}
where $\mathbf{c}=(c_{1},\dots,c_{R})^{\mathrm{T}}$,
\begin{align}
B_{rs} & =-(\lambda_{r}+\beta)\delta_{rs}+\tilde{b}\sum_{n=0}^{\mathcal{N}-1}\phi_{r}(\xi_{n})\phi_{s}(\xi_{n}),\label{eq:B_matrix}\\
d_{r} & =\tilde{a}\sum_{n=0}^{\mathcal{N}-1}\phi_{r}(\xi_{n}),\label{eq:d_vector}
\end{align}
and $\lambda_{r}=(r\pi/\Lambda)^{2}$.

For vanishing initial (relative) temperature, $\mathbf{c}(0)=\mathbf{0}$, the exact solution is
\begin{equation}
\mathbf{c}(\tau)=\mathbf{B}^{-1}\left(e^{\mathbf{B}\tau}-\mathbf{I}\right)\mathbf{d}.\label{eq:exact_modal_solution_full}
\end{equation}

Substitution into the eigenfunction expansion \eqref{eq:modal_expansion} yields the full space--time temperature field $\theta(\xi,\tau)$. This exact solution provides complete access to transient dynamics, steady states, and stability properties of nanomagnet-scale heating, and forms the basis for the analysis of localized hotspots and thermal overlap presented in the following sections.

%
\subsection{Assembly-scale heat equation and coarse-grained description}
\label{sec:HE_assembly}

We now turn to the description of heat transport at the scale of the entire nanomagnet assembly and its embedding matrix. This second level
of description is obtained by coarse-graining the nanoscale temperature field over space and time scales large compared to the nanomagnet
size and the AC-field period. It is intended to capture the collective thermal response of the array rather than the detailed structure of
local hotspots.


At the assembly scale, we assume that: (i) the temperature field varies
smoothly on length scales larger than the interparticle distance $d$;
(ii) fast nanoscale temperature oscillations over individual AC cycles
have been averaged out; (iii) the surrounding matrix and environment
act as effective thermal reservoirs characterized by macroscopic transport
and leakage parameters.

Under these conditions, the detailed nanomagnet-resolved source term
$P_{n}(t)\,\delta(\bm{r}-\bm{r}_{n})$ introduced in Sec.~\ref{sec:nanoscale}
is replaced by an effective volumetric heat source $Q_{\mathrm{eff}}(\bm{r},t)$,
representing the spatial and temporal average of the power dissipated
by many nanomagnets within a coarse-graining volume. The resulting
temperature field $T_{\mathrm{cg}}(\bm{r},t)$ describes the collective
thermal response of the assembly; see Eq.~\eqref{eq:Tcg}.


The assembly-scale temperature field obeys the standard heat equation with effective material parameters,
\begin{align}
\rho_{\mathrm{eff}}c_{v,\mathrm{eff}}\frac{\partial T_{\mathrm{cg}}}{\partial t}(\bm{r},t)= & \kappa_{\mathrm{eff}}\nabla^{2}T_{\mathrm{cg}}(\bm{r},t)+Q_{\mathrm{eff}}(\bm{r},t)\nonumber \\
 & -L_{N}\big[T_{\mathrm{cg}}(\bm{r},t)-T_{0}\big],\label{eq:HE_coarse}
\end{align}
where $\rho_{\mathrm{eff}}c_{v,\mathrm{eff}}$ and $\kappa_{\mathrm{eff}}$ are the effective volumetric heat capacity and thermal conductivity of the nanocomposite, and $L_{N}$ is an effective volumetric Newtonian cooling coefficient describing heat leakage from the assembly to the external environment.

The solution of Eq. \eqref{eq:HE_coarse}, with and without the diffusion term, was investigated in Refs. \onlinecite{DejKac2022, dejkac_as24} for the prototypical ferrofluids of magnetite and maghemite nanomagnet in water, where the parameter $L_{N}$ was estimated by adjusting the solution to the experimental data.


The coefficient $L_{N}$ represents the total effective environmental coupling after coarse-graining and its hierarchical relationship with $L_{m}$ can be written as $L_{N}=L_{m}+L_{\text{emergent}}$, where
\begin{itemize}
\item $L_{m}$ is the nanoscale loss coefficient (introduced earlier).
\item $L_{\text{emergent}}$ captures additional losses that become significant.
only at larger scales and longer times, including:
\begin{enumerate}
\item Boundary effects as heat reaches system edges.
\item Enhanced heat exchange when temperature gradients span the entire
assembly.
\item Collective modes of environmental coupling.
\end{enumerate}
\end{itemize}
Typical values for the systems studied here are:
\begin{align*}
\text{PMMA nanocomposite: } & L_{N}\sim10^{4}-10^{5}~\text{W}/(\text{K.m}^{3})\\
\text{Aqueous ferrofluid: } & L_{N}\sim10^{5}-10^{6}~\text{W}/(\text{K.m}^{3})
\end{align*}
The ratio $L_{N}/L_{m}\sim10-100$ quantifies the emergent losses that develop as heat diffuses beyond the nanoparticle vicinity.


Similarly to the nanoscale, we introduce the relative temperature elevation
\begin{equation}
\Theta(\bm{r},t)=\frac{T_{\mathrm{cg}}(\bm{r},t)-T_{0}}{T_{0}},
\end{equation}
the characteristic leakage time
\begin{equation}
t_{s}=\frac{\rho_{\mathrm{eff}}c_{v,\mathrm{eff}}}{L_{N}},\label{eq:ts}
\end{equation}
and the dimensionless coordinate
\begin{equation}
\bm{\xi}=\frac{\bm{r}}{\ell_{c}},\qquad \ell_{c}=L,
\end{equation}
so that $\nabla^{2}=\ell_{c}^{-2}\nabla_{\xi}^{2}$. Note that now $\ell_{c}$ is the full chain's length.

Then, defining the dimensionless time $\taus=t/t_{s}$, we obtain the heat equation at the assembly scale
\begin{equation}
\frac{\partial\Theta}{\partial\taus}=\zeta\nabla_{\xi}^{2}\Theta-\Theta+\Xi_{\mathrm{cg}}(\bm{\xi},\taus),\label{eq:HE_coarse_dimless}
\end{equation}
with
\begin{equation}
\zeta=\frac{\kappa_{\mathrm{eff}}}{L_{N}\ell_{c}^{2}}=\frac{\kappa_{\mathrm{eff}}}{L_{N}L^{2}},\qquad
\Xi_{\mathrm{cg}}=\frac{Q_{\mathrm{eff}}}{L_{N}T_{0}}.\label{eq:zeta_xi_def}
\end{equation}

We again refer to the work in Ref. \onlinecite{dejkac_as24} for a discussion of this formulation applied to magnetite ferrofluids.


At this scale, the effective source term $\Xi_{\mathrm{cg}}$ is directly related to the \emph{time-averaged} SLP of the nanomagnet, as derived
in Appendix~\ref{app:sar_coeffs}. After coarse-graining, it may be expressed in the linearized form
\begin{equation}
\Xi_{\mathrm{cg}}(\bm{r},\tau_s)\simeq a_{\mathrm{cg}}+b_{\mathrm{cg}}\,\Theta(\bm{r},\tau_s),\label{eq:SAR_coarse}
\end{equation}
where the coefficients $a_{\mathrm{cg}}$ and $b_{\mathrm{cg}}$ are obtained from the SLP expressions by averaging over the nanomagnets within
the coarse-graining volume and normalizing by $L_{N}T_{0}$.


For consistency with the nanoscale analysis, we restrict attention
to a one-dimensional domain $0\le\xi\le\Lambda$ and impose Dirichlet
boundary conditions,
\begin{equation}
\Theta(0,\tau_s)=\Theta(\Lambda,\tau_s)=0,\label{eq:Dirichlet_cg}
\end{equation}
corresponding to an idealized thermal bath at the system boundaries. As at the
nanoscale, the modal solution structure carries over directly to NBC with the
substitution of the corresponding eigenfunctions; the two cases are compared
in Sections~(\ref{sec:DBC_NBC_summary}, \ref{sec:Results}) and Appendix~\ref{app:BC_comparison}.

Under these conditions, Eq.~\eqref{eq:HE_coarse_dimless} admits
the same modal structure as the nanoscale problem, with the crucial
difference that the source term is now spatially smooth rather than
localized at discrete particle sites. As a result, only long-wavelength
modes contribute significantly to the temperature field, reflecting
the collective nature of heating at the assembly scale.


The assembly-scale equation \eqref{eq:HE_coarse_dimless} provides a macroscopic description
valid after sufficient temporal and spatial averaging. Its main role
in the present work is to connect the nanoscale heating physics to
previously studied collective thermal transport models \cite{rosensweig02j3m, Hergt_etal_JPMC2006, RodriguezEtAl_ijn11, Mehdaoui_AFM2011, mehdaouietal12apl, Haase_Nowak_PRB85_2012, OGRADY_APL2013, condeetal15jpcc, Ruta_ScientificReports_2015, RytovEtAl_sr22, DejKac2022, dejkac_as24} and to identify
the conditions under which localized heating crosses over to global,
homogenized behavior.

Now, we discuss the mathematical and physical connection between the nanoscale and assembly-scale equations.

\subsection{Relation between nanomagnet-scale and assembly-scale descriptions}
\label{subsec:scale_link}
Sections~\ref{sec:nanoscale} and~\ref{sec:HE_assembly} describe thermal transport at two distinct but complementary space-time scales. Here we derive how the coarse-grained heat equation emerges from the nanomagnet-scale formulation through controlled averaging, and establish quantitative relations between the parameters at each scale.

\subsubsection{Separation of space and time scales}

The key assumption underlying the two-scale approach is a strong separation of characteristic time and length scales:
\begin{equation}
t_d \ll T_{\rm AC} \ll t_s, \quad R \ll d \ll L,
\label{eq:scale_sep}
\end{equation}
where $t_d \sim d^2/\alpha$ is the thermal diffusion time across one interparticle spacing [Eq.~\eqref{eq:xi-tau-td}], $T_{\rm AC} = 2\pi/\omega$ is the AC-field period, and $t_s$ is the macroscopic (assembly-level) thermal leakage time [Eq.~\eqref{eq:ts}]; $R$ is the nanomagnet radius, $d$ the interparticle spacing, and $L$ the system size.

For the systems considered here (magnetite in PMMA or water):
\begin{itemize}
\item $t_d \sim 10-15$~ns (diffusion across $d \sim 30$~nm)
\item $T_{\rm AC} \sim 5~\mu$s (for $f = 194$~kHz)
\item $t_s \sim 0.1$--1~s (depending on $L_N$)
\item $R \sim 10$~nm, $d \sim 30$--50~nm, $L \sim 2~\mu$m
\end{itemize}
The hierarchy~(\ref{eq:scale_sep}) is thus satisfied with comfortable margins: $t_d/T_{\rm AC} \sim 10^{-3}$, $T_{\rm AC}/t_s \sim 10^{-5}$, and $d/L \sim 10^{-2}$.


At the nanomagnet scale, the magnetic power loss is first averaged over one AC cycle, yielding the time-averaged power $P_n(T)$ used in Eq.~\eqref{eq:HE_nm_dim}. The next averaging step is performed over times $\Delta t$ such that
\begin{equation}
T_{\rm AC} \ll \Delta t \ll t_s,
\label{eq:time_avg}
\end{equation}
which filters out nanomagnet-scale transients while preserving the slow evolution of the macroscopic temperature field.

Averaging the source term in Eq.~\eqref{eq:HE_nm_dim} over $\Delta t$ gives:
\begin{equation}
Q_{\rm eff}(x,t) = \frac{1}{\Delta t} \int_{t}^{t+\Delta t} \sum_{n=0}^{\mathcal{N}-1} P_n[T(x_n,t')] \delta(x - x_n) \, dt'.
\label{eq:temporal_avg}
\end{equation}
Because $\Delta t \gg t_d$, the temperature $T(x_n,t')$ varies slowly over the interval $[t, t+\Delta t]$, and we can approximate:
\begin{equation}
Q_{\rm eff}(x,t) \simeq \sum_{n=0}^{\mathcal{N}-1} \overline{P_n}[T(x_n,t)] \delta(x - x_n),
\end{equation}
where $\overline{P_n}$ denotes the time-averaged power at the quasi-steady temperature $T(x_n,t)$, see Section~\ref{sec:nanoscale}.


The next coarse-graining step consists in averaging over spatial regions of size $\ell_{\rm cg}$ satisfying
\begin{equation}
d \ll \ell_{\rm cg} \ll L.
\label{eq:spatial_avg}
\end{equation}
We define the coarse-grained temperature as:
\begin{equation}
T_{\rm cg}(x,t) = \frac{1}{\ell_{\rm cg}} \int_{x-\ell_{\rm cg}/2}^{x+\ell_{\rm cg}/2} T(x',t) \, dx',\label{eq:Tcg}
\end{equation}
and similarly for the source term:
\begin{equation}
Q_{\rm cg}(x,t) = \frac{1}{\ell_{\rm cg}} \int_{x-\ell_{\rm cg}/2}^{x+\ell_{\rm cg}/2} \sum_{n} \overline{P_n} \delta(x' - x_n) \, dx'\label{eq:Qcg}.
\end{equation}

For a uniform distribution of nanomagnets with concentration $n_{\rm NM} = 1/d$ (one nanomagnet per interparticle spacing), the number of nanomagnets within the averaging window is $N_{\rm win} = \ell_{\rm cg}/d \gg 1$. The discrete sum over delta functions is then replaced by:
\begin{equation}
Q_{\rm cg}(x,t) = n_{\rm NM} \langle \overline{P_n} \rangle = \frac{1}{d} \langle \overline{P_n} \rangle,
\label{eq:volumetric_source}
\end{equation}
where $\langle \overline{P_n} \rangle$ is the average power per nanomagnet in the coarse-graining volume.


Using the linearized form $\overline{P_n} = (V_p/A_p)[a_p + b_p \theta_i^{(n)}]$ from Eq.~\eqref{eq:SLP_linear}, and noting that after coarse-graining $\theta_i^{(n)} \simeq \Theta(x,t)$ (the assembly-scale relative temperature), we obtain:
\begin{equation}
Q_{\rm cg}(x,t) = \frac{V_p}{A_p d} [a_p + b_p \Theta(x,t)].
\end{equation}

Dividing by $L_N T_0$ to obtain the dimensionless source term $\Xi_{\rm cg} = Q_{\rm cg}/(L_N T_0)$ as in Eq.~\eqref{eq:HE_coarse_dimless}, we find:
\begin{align}
a_{\rm cg} &= \frac{V_p a_p}{A_p d L_N T_0}, \label{eq:acg}\\
b_{\rm cg} &= \frac{V_p b_p}{A_p d L_N T_0}. \label{eq:bcg}
\end{align}

We see that the coarse-grained coefficients scale inversely with the interparticle spacing $d$ (denser packing $\to$ higher effective heating) and inversely with the assembly-scale loss coefficient $L_N$. Note that the ratio $a_{\rm cg}/b_{\rm cg} = a_p/b_p$ is independent of the coarse-graining procedure.


The nanoscale loss term $-\beta \theta$ in Eq.~\eqref{eq:HE_nm_final} represents direct environmental coupling at short length scales ($\sim d$) and short times ($\sim t_d$). After spatial and temporal averaging, this term contributes to the assembly-scale loss $-\Theta$ in Eq.~\eqref{eq:HE_coarse_dimless}, but with a renormalized coefficient.

To derive the relationship explicitly, consider the average loss rate over the coarse-graining volume:
\begin{equation}
\langle \text{Loss rate} \rangle = \frac{1}{\ell_{\rm cg}} \int L_m [T(x) - T_0] \, dx = L_m [T_{\rm cg} - T_0].
\end{equation}

However, at larger scales, additional loss mechanisms become active:
\begin{enumerate}
\item \textbf{Boundary losses:} When the thermal field extends to the system edges ($x \sim 0, L$), heat leaks through the boundaries at a rate $\sim \kappa_{\rm eff} \nabla T / L$.
\item \textbf{Enhanced convective coupling:} At larger scales, collective thermal gradients drive more efficient heat transfer to the environment.
\item \textbf{Radiation losses:} For elevated temperatures (not relevant here), radiative losses $\propto T^4$ become significant.
\end{enumerate}

These emergent mechanisms contribute an additional loss coefficient $L_{\rm emergent}$, so that the total assembly-scale loss is:
\begin{equation}
L_N = L_m + L_{\rm emergent}.
\label{eq:LN_hierarchy}
\end{equation}

\textbf{Order-of-magnitude estimate:} For the systems studied here, $L_m \sim 10^4$--$10^5$~W~m$^{-3}$~K$^{-1}$ (nanoscale direct losses), while $L_N \sim 10^5$--$10^7$~W~m$^{-3}$~K$^{-1}$ (assembly-scale total losses), giving $L_N/L_m \sim 10$--100, which means that the emergent losses dominate at larger scales and longer times.

%
%
We can easily check that the coarse-graining procedure preserves the dimensional consistency.


In summary, the coarse-grained assembly-scale description emerges from the nanoscale formulation through:
\begin{enumerate}
\item \textbf{Temporal averaging} over $\Delta t$ (Eq.~\ref{eq:temporal_avg}), filtering out AC oscillations and nanomagnet-scale transients.
\item \textbf{Spatial averaging} over $\ell_{\rm cg}$ (Eq.~\ref{eq:Qcg}), replacing discrete sources by volumetric densities.
\item \textbf{Parameter mapping:} Nanoscale coefficients $(a_p, b_p, L_m)$ are related to assembly-scale coefficients $(a_{\rm cg}, b_{\rm cg}, L_N)$ via Eqs.~(\ref{eq:acg})--(\ref{eq:LN_hierarchy}).
\item \textbf{Emergent losses:} The assembly-scale loss $L_N$ incorporates both direct nanoscale losses $L_m$ and additional losses $L_{\rm emergent}$ that arise at larger scales.
\end{enumerate}

This explicit derivation clarifies that the two-scale framework is not merely a heuristic separation, but a systematic coarse-graining procedure with well-defined approximations. In Section~\ref{sec:ass-scale}, we quantify the accuracy of this approximation by comparing full nanoscale computations with assembly-scale predictions.

\subsection{Local-to-global heating crossover (L2GHC): indicators and conditions}

\label{subsec:crossover_indicators}

The transition from local to global heating is a central feature of
thermal transport in nanomagnet assemblies under alternating magnetic
fields. At short times and small scales, heating is confined to the
immediate vicinity of each nanomagnet, creating localized hot spots
ideal for targeted processes such as nanoscale catalysis or site-specific
activation. As time progresses and thermal diffusion couples neighboring
sources, the temperature field evolves toward spatial homogenization,
resulting in collective heating suitable for applications such as
magnetic hyperthermia or bulk thermal actuation.

The efficiency and controllability of such systems depend critically
on the competition between heat injection, diffusion, and thermal
leakage---quantified by parameters such as particle spacing $d$,
interfacial coupling $\gamma$, matrix conductivity $\kappa_{m}$,
and the thermo-magnetic feedback coefficient $b$. Understanding and
mapping the crossover conditions allows for the rational design of
materials in which heating can be tuned from highly localized to uniformly
global, optimizing therapeutic efficacy while minimizing unwanted
thermal spread.

\subsubsection{Analytical stability criterion}
\label{sec:stability_criterion}

The crossover from local to global heating is fundamentally a stability transition of the temperature field.  In the linearized regime, the
dynamics of the modal amplitudes $\mathbf{c}(\tau)$ are governed by the matrix $\mathbf{B}$ defined in Eqs.~(\ref{eq:modal_nm_full}, \ref{eq:B_matrix}).  The eigenvalues of $\mathbf{B}$ determine the growth or decay of thermal modes.

The system crosses from localized to global heating when the most
unstable mode becomes marginally stable:
\begin{equation}
  \max\bigl[\operatorname{Re}\bigl(\operatorname{eig}(\mathbf{B})\bigr)\bigr]
  = 0.
  \label{eq:eigen_criterion}
\end{equation}
For the fundamental Dirichlet mode ($r=1$), the diagonal element of $\mathbf{B}$ reads\footnote{See Appendix \ref{app:BC_comparison}  for NBC.}
\begin{equation}
  B_{11} = -(\lambda_1 + \beta)
           + \tilde{b}\;\frac{2}{\Lambda}
             \sum_{n=0}^{\mathcal{N}-1}\sin^{2}\!\!\left(\frac{\pi\xi_n}{\Lambda}\right),
  \qquad \lambda_1 = \left(\frac{\pi}{\Lambda}\right)^{2}.
  \label{eq:B11}
\end{equation}
Setting $B_{11}=0$ yields the critical feedback coefficient
\begin{equation}
  \tilde{b}_{c}
  = \frac{\lambda_1 + \beta}
         {\displaystyle \frac{2}{\Lambda}
           \sum_{n=0}^{\mathcal{N}-1}\sin^{2}\!\!\left(\frac{\pi\xi_n}{\Lambda}\right)}.
  \label{eq:bc_analytical}
\end{equation}

The sign of $\tilde{b}_c$ is strictly positive. Indeed, diffusion losses ($\lambda_1$) and Newton losses ($\beta$) must be compensated by a
positive thermo-magnetic feedback before the uniform mode can grow.  The crossover is therefore accessible only when the renormalized coefficient $\tilde{b}$ is itself positive.

The sign of $\tilde{b}$ is not, however, determined solely by the bare SLP coefficient~$b_p$ whose sign in turn depends on several system parameters [see Appendix \ref{app:sar_coeffs} and Fig. \ref{fig:SLP_coefficients}].
From the renormalization~\eqref{eq:renorm_coeffs},
\begin{equation}
  \tilde{b} = \frac{\gamma}{\gamma-\varepsilon}b = \Upsilon_0\frac{\gamma_s\,b_p}{\gamma_s - b_p},
  \label{eq:btilde_sign}
\end{equation}
where $\gamma_s = h_s A_p T_0/V_p > 0$ is the interfacial coupling
coefficient and $\Upsilon_0 > 0$ [see Eq.~\eqref{eq:Upsilon}].
Since $\Upsilon_0$ is strictly positive, the sign of $\tilde{b}$ is
determined entirely by the ratio $\gamma_s b_p/(\gamma_s - b_p)$,
i.e.\ by the relative magnitude of the interfacial coupling
$\gamma_s$ and the bare feedback coefficient~$b_p$.
Equivalently, in terms of the dimensionless quantities
$\gamma = \gamma_s/a_p$ and $\varepsilon = b_p/a_p$, the sign is
controlled by $\gamma$ relative to~$\varepsilon$.
Then, three cases may arise:
\begin{enumerate}
\item $b_p < 0$ (self-limiting bare SLP): $\tilde{b} < 0$.
  The sign is preserved by renormalization.

\item $b_p > 0$ and $\gamma_s > b_p$ (self-amplifying SLP,
  strong interfacial coupling): $\tilde{b} > 0$.  The positive
  feedback couples effectively into the matrix.

\item $b_p > 0$ and $\gamma_s < b_p$ (self-amplifying SLP,
  weak interfacial coupling): $\tilde{b} < 0$.  Physically, the interfacial thermal resistance is so
  large that the self-amplifying feedback cannot couple efficiently
  into the matrix temperature field; at the matrix level the effective
  feedback appears self-limiting.
\end{enumerate}

Three limiting values of $\gamma_s$ clarify the role of the nanomagnet--matrix interface.

\paragraph{No interfacial resistance (baseline model).}
The most direct formulation of the nanomagnet-chain heat equation is
obtained by writing the source term with the bare SLP coefficients,
$[a_p + b_p\,\theta(\xi_n)]\,\delta(\xi-\xi_n)$, without introducing
the interfacial boundary condition~\eqref{eq:Newton_NM} at all.
In this case the renormalization~\eqref{eq:renorm_coeffs} is not
needed and the entire stability analysis of the present section
applies with $b_p$ in place of~$\tilde{b}$.  The crossover criterion
\eqref{eq:bc_analytical} becomes simply $b_p > \tilde{b}_c$.
Within the renormalized theory, this baseline model is recovered
formally in the perfect-contact limit $h_s\to\infty$
($\gamma_s\to\infty$), where
Eqs.~\eqref{eq:renorm_coeffs} give $\tilde{a}\to a$ and
$\tilde{b}\to b$.
The interfacial boundary condition~\eqref{eq:Newton_NM} and the
associated renormalization are thus a refinement of this baseline,
introducing the richer structure discussed in the following two
limiting cases.

\paragraph{Decoupled limit ($\gamma_s\to 0$).}
In the opposite extreme, $h_s\to 0$ and the nanomagnet--matrix interface becomes thermally insulating. Equations~\eqref{eq:renorm_coeffs} give $\tilde{a}\to 0$ and $\tilde{b}\to 0$, so the source term in the matrix heat equation~\eqref{eq:HE_nm_final} vanishes identically.
The nanomagnet still dissipates power internally, but the infinite interfacial resistance traps all heat inside the nanomagnet; the matrix temperature field remains at ambient regardless of the values of $a_p$ and~$b_p$. No heating---neither local nor global---can develop in the embedding medium in this limit.

\paragraph{Resonant limit ($\gamma_s\to b_p^+$, for $b_p>0$).}
When $\gamma_s$ approaches~$b_p$ from above, the denominator $\gamma_s - b_p\to 0^+$ and both $\tilde{a}$ and $\tilde{b}$ diverge. Physically, the rate at which the interface extracts heat from the nanomagnet exactly matches the self-amplifying feedback: the nanomagnet temperature grows without bound in the linearized model, signaling the breakdown of the first-order Taylor expansion of the SLP before any collective matrix-level instability is reached.
For $\gamma_s < b_p$ the renormalized coefficients change sign (case~3 above), but the linearized framework is only quantitatively reliable
for $\gamma_s$ sufficiently far from~$b_p$. In practice, the divergence at $\gamma_s = b_p$ is regularized by higher-order (nonlinear) terms in the SLP, which saturate the nanomagnet temperature at a finite value.

\medskip

The interplay between $b_p$ and $\gamma_s$ thus governs not merely the magnitude of $\tilde{b}$ but its sign, and thereby determines whether
the local-to-global crossover is accessible at all.

Finally, combining these observations with the stability criterion \eqref{eq:bc_analytical}, we identify three physical regimes for the
temperature field:
\begin{itemize}
\item \textbf{Unconditionally stable (localized) regime} ($\tilde{b} \le 0$, i.e.\ $b_p < 0$, or $b_p > 0$ with  $\gamma_s < b_p$).  Both terms in $B_{11}$ are non-positive;  $\mathbf{B}$ is negative definite for all $|\tilde{b}|$.  The temperature field remains localized and spatially heterogeneous. No crossover to global heating is possible regardless of nanomagnet density or geometry.
\item \textbf{Sub-critical amplifying regime}
  ($0 < \tilde{b} < \tilde{b}_c$).  The positive feedback partially
  compensates diffusion and Newton losses.  The system is still
  stable, but as $\tilde{b}\to\tilde{b}_c^-$ the fundamental-mode
  decay rate $|B_{11}|\to 0$: thermal modes become long-lived, the
  temperature grows quasi-linearly, and spatial correlations extend
  over multiple interparticle spacings.

\item \textbf{Super-critical regime}
  ($\tilde{b} > \tilde{b}_c$).  The positive feedback overcomes all
  loss channels.  The fundamental mode grows exponentially, driving
  the temperature field toward spatially homogeneous, collective
  heating at the assembly scale.  In practice, nonlinear saturation
  of the SLP (higher-order terms in the Taylor expansion) arrests the
  growth and sets the final steady-state temperature.
\end{itemize}

Equation~\eqref{eq:bc_analytical} therefore encodes a \emph{dual}
condition for the local-to-global crossover: i) the bare SLP must be
self-amplifying ($b_p > 0$, controlled by the nanomagnet size~$D$, field
amplitude~$h_0$, frequency~$f$, and interparticle spacing; cf.\
Fig. \ref{fig:SLP_coefficients} in Appendix \ref{app:sar_coeffs}), and ii) the interfacial coupling must
be strong enough ($\gamma_s > b_p$) for the positive feedback to survive
renormalization into the matrix.
In the perfect-contact limit ($\gamma_s\to\infty$), or equivalently in
the baseline model where the interfacial boundary
condition~\eqref{eq:Newton_NM} is not imposed, the second condition
is automatically satisfied and the crossover is controlled by $b_p$
alone.
When either condition fails---whether because $b_p < 0$, owing to
$\gamma_s < b_p$, or in the decoupled limit $\gamma_s\to 0$ where the
matrix receives no heat at all---the system is locked into the
localized-heating regime.

\subsubsection{Physical indicators of the crossover}

To quantify the transition in practice, we introduce complementary
scalar and spatial indicators derived from the temperature field $\theta(\xi,\tau)$.
\begin{itemize}
\item \textbf{Mean temperature} (global indicator):
\begin{equation}
\langle\theta\rangle(\tau)=\frac{1}{\mathcal{N}}\sum_{n=0}^{\mathcal{N}-1}\theta(\xi_{n},\tau),\label{eq:theta_mean}
\end{equation}
which in steady state ($\tau\to\infty$) is given by
\[
\langle\theta\rangle_{\mathrm{ss}}=\frac{\tilde{a}\,\mathbf{1}^{\!\top}\mathbf{K}_{R}(0)\bigl[\mathbf{I}-\tilde{b}\,\mathbf{K}_{R}(0)\bigr]^{-1}\mathbf{1}}{\mathcal{N}}.
\]
A sharp rise in $\langle\theta\rangle_{\mathrm{ss}}$ as $\tilde{b}$
approaches $1/\lambda_{\max}(\mathbf{K}_{R}(0))$ signals the onset
of collective heating.
\item \textbf{Temperature variance} (local heterogeneity indicator):
\begin{equation}
\operatorname{Var}_{\tau}(\theta)=\frac{1}{\mathcal{N}}\sum_{n=0}^{\mathcal{N}-1}\bigl[\theta(\xi_{n},\tau)-\langle\theta\rangle(\tau)\bigr]^{2},\label{eq:theta_variance}
\end{equation}
which measures spatial inhomogeneity. In the local-heating regime,
the variance is large due to distinct hot spots; in the global-heating
regime, diffusion and boundary conditions homogenize the temperature,
driving the variance toward zero.
\item \textbf{Thermal correlation length} (spatial coherence indicator):
The two-point spatial correlator
\[
C(\Delta\xi,\tau)=\bigl\langle\theta(\xi,\tau)\,\theta(\xi+\Delta\xi,\tau)\bigr\rangle_{\xi}
\]
can be expanded in the modal basis as
\begin{equation}
C(\Delta\xi,\tau)=\sum_{r=1}^{\infty}|c_{r}(\tau)|^{2}\,\phi_{r}(0)\,\phi_{r}(\Delta\xi).\label{eq:correlator_expansion}
\end{equation}
From this, a thermal correlation length is defined via the first moment:
\[
\xi_{\mathrm{corr}}(\tau)=\frac{{\displaystyle \int_{0}^{\Lambda}\Delta\xi\,C(\Delta\xi,\tau)\,d\Delta\xi}}{{\displaystyle \int_{0}^{\Lambda}C(\Delta\xi,\tau)\,d\Delta\xi}}.
\]
Growth of $\xi_{\mathrm{corr}}$ beyond the interparticle spacing
($\xi_{\mathrm{corr}}\gtrsim1$) signals the emergence of collective,
assembly-scale heating.
\end{itemize}
Together, these indicators provide a multifaceted view of the local-to-global
crossover, linking the analytical stability criterion to observable
thermal statistics. Their behavior across parameter space is illustrated
numerically in Sec.~\ref{sec:Results}.

%

\subsubsection{Role of boundary conditions: Dirichlet versus Neumann}
\label{sec:DBC_NBC_summary}

The stability analysis and crossover indicators discussed above assume Dirichlet boundary conditions (DBC), $\theta(0,\tau)=\theta(\Lambda,\tau)=0$, modelling a chain terminated by ideal thermal baths.  However, alternative boundary conditions alter the modal spectrum and can qualitatively change both the absolute temperature levels and the spatial structure of the temperature field~\cite{MorseFeschbach_mgh53, duffy2015green}.  We summarize here the main results for the opposite limiting case---homogeneous Neumann (no-flux) conditions (NBC),
$\partial_\xi\theta|_{\xi=0,\Lambda}=0$---which models thermally insulated chain ends.  Full derivations are given in Appendix~\ref{app:BC_comparison}.

The Neumann eigenfunctions are cosines, $\phi_r^{(\mathrm{N})}(\xi)=\sqrt{2/\Lambda}\,\cos(r\pi\xi/\Lambda)$ for $r\ge1$, supplemented by the spatially uniform zero mode $\phi_0^{(\mathrm{N})}=1/\sqrt{\Lambda}$ with eigenvalue $\lambda_0^{(\mathrm{N})}=0$.  This zero mode---absent from the Dirichlet spectrum, where $\lambda_1^{(\mathrm{D})}=(\pi/\Lambda)^2>0$---is the structural feature responsible for all qualitative differences between the two boundary conditions, see Section~\ref{sec:Results}.

\paragraph{Stability.}
The zero-mode growth rate in the diagonal approximation is $-\beta + \mathcal{N}\,\tilde{b}/\Lambda$.
The corresponding critical feedback coefficient,
\begin{equation}
  \tilde{b}_c^{(0)}
  = \frac{\beta\,\Lambda}{\mathcal{N}}\,,
  \label{eq:bc_Neumann_main}
\end{equation}
vanishes as $\beta\to0$. So, with insulated boundaries (NBC) and negligible
volumetric losses, \emph{any} positive feedback---however
small---destabilises the uniform temperature channel.  By contrast,
the Dirichlet critical value~\eqref{eq:bc_analytical} remains finite
even for $\beta=0$, because the boundary heat sinks provide diffusive
stabilisation through $\lambda_1^{(\mathrm{D})}>0$.  The Neumann
system is therefore generically less stable.  In the physically
relevant regime ($\tilde{b}<0$, $\beta>0$), both systems are
unconditionally stable, but their steady-state properties differ
markedly.

\paragraph{Steady state and the background effect.}
The zero mode contributes a spatially uniform temperature offset to the Neumann steady state [see Appendix~\ref{app:BC_comparison}],
\begin{equation}
  \theta_{\mathrm{ss}}^{(\mathrm{N})}\big|_{r=0}
  = \frac{\tilde{a}\,\mathcal{N}/\Lambda}
         {\beta - \tilde{b}\,\mathcal{N}/\Lambda}\,,
  \label{eq:NBC_pedestal}
\end{equation}
which, for $\beta\ll|\tilde{b}|\,\mathcal{N}/\Lambda$, simplifies to $\tilde{a}/|\tilde{b}|$---independent of system size and loss
parameter.
No analogous contribution exists under DBC, where $\theta$ vanishes at the boundaries.  Consequently, the absolute NBC temperature
exceeds the DBC value at every interior point; the difference is dominated by the zero-mode term $1/(\Lambda\beta)$ in the diagonal
Green's function [Eq.~\eqref{eq:GN_minus_GD} in Appendix~\ref{app:BC_comparison}].

\paragraph{Spatial heterogeneity.}
Because the zero mode is spatially uniform, it cancels exactly in
the hotspot contrast $\Delta\theta_n$ (defined as the difference
between a nanomagnet-site temperature and its inter-site baseline),
which is governed entirely by the $r\ge1$ modes common to both
spectra.  Interior hotspot contrasts are therefore comparable under
DBC and NBC.
However, the \emph{relative} spatial variance
$\mathcal{V}_{\mathrm{rel}}=\mathrm{Var}(\theta_{\mathrm{ss}})/\langle\theta_{\mathrm{ss}}\rangle^2$
is strongly suppressed under NBC, because the zero-mode pedestal
inflates the mean while leaving the variance (set by $r\ge1$ modes)
essentially unchanged:
$\mathcal{V}_{\mathrm{rel}}^{(\mathrm{N})}\ll\mathcal{V}_{\mathrm{rel}}^{(\mathrm{D})}$
when $\beta\ll|\tilde{b}|\,\mathcal{N}/\Lambda$.
This is the central trade-off: \emph{NBC maximises absolute temperature at
the cost of washing out spatial heterogeneity, while DBC preserves
strong spatial contrast at the cost of lower absolute temperatures}.

\paragraph{Boundary nanomagnet activity.}
Under DBC, the two boundary nanomagnets ($n=0$ and $n=\mathcal{N}-1$) are thermally invisible: the sine eigenfunctions vanish at $\xi=0$ and $\xi=\Lambda$, so these particles do not contribute to the modal source vector or coupling matrix.  Under NBC, the cosine eigenfunctions attain their extremal values at the boundaries, so all $\mathcal{N}$ particles are thermally active.


\paragraph{Design implications.}
Real systems are likely intermediate between these two limits,
described by Robin boundary conditions with an effective Biot number
$\mathrm{Bi}_b = h_b d/\kappa_m$ interpolating between
$\mathrm{Bi}_b\to\infty$ (Dirichlet) and $\mathrm{Bi}_b\to0$
(Neumann).  The analysis shows that conducting boundaries
($\mathrm{Bi}_b\gg1$) favour applications requiring maximum spatial
contrast (e.g.\ site-selective activation), while insulating
boundaries ($\mathrm{Bi}_b\ll1$) favour those requiring maximum
absolute temperature (e.g.\ bulk hyperthermia).
These conclusions apply equally to the crossover indicators
introduced above: under NBC, the correlation length
$\xi_{\mathrm{corr}}$ saturates rapidly at the system size once the zero mode dominates, and the variance-to-mean ratio drops considerably compared to DBC. Numerical illustrations for both boundary conditions are presented
in the Results Sec.~\ref{sec:Results}.

\section{Physical parameters and scaling}
\label{sec:parameters}

For numerical convenience, we introduced in Sec.~\ref{sec:framework} the dimensionless formulation of the thermal problem in terms of a compact set of control parameters.  To connect this formulation to real experimental systems, we provide explicit scaling relations between these
dimensionless parameters and physical material properties, geometric arrangements, and excitation conditions.  This mapping is essential
for interpreting the results in Sec.~\ref{sec:Results} and for designing experiments aimed at observing the local-to-global heating crossover.

\subsection{Parameter hierarchy}

We distinguish three levels of parameters:
(i)~\emph{raw input parameters}, directly measurable or independently
tunable quantities that define the experimental system;
(ii)~\emph{inferred physical parameters}, derived from raw inputs via
constitutive relations (magnetic response, thermal transport); and
(iii)~\emph{dimensionless control parameters}, the final scaled
variables governing the heat equations.

From the raw inputs, the specific loss power (SLP) is obtained from
linear-response theory including dipolar interactions, as detailed
in Appendix~\ref{app:sar_coeffs}.  The key outputs are the
dimensional coefficients $a_p$ and~$b_p$ appearing in the linearized
heating law~\eqref{eq:SLP_linear}, which depend on the reduced
anisotropy barrier $\sigma_0 = KV_p/(k_BT_0)$, the reduced
frequency $\varpi_0 = \omega\tau_0$, the dipolar coupling strength (renormalized by $2KV_p$)
\begin{equation}
  \lambda
  = \frac{\mu_0}{4\pi}\,\frac{M_s^2\,V_p^2}{d^3}\,\frac{1}{2KV_p}\,,
\end{equation}
and the field amplitude~$h_0$.  Their explicit expressions are given in Eqs.~\eqref{eq:a_sar_app}--\eqref{eq:b_sar_app}.


The dimensionless parameters entering the nanomagnet-scale and
assembly-scale heat equations were introduced in
Sec.~\ref{sec:framework}; here we collect the definitions and
explain how they are evaluated from the raw inputs above.

The nanomagnet-scale coefficients $a = \Upsilon_0\,a_p$ and
$b = \Upsilon_0\,b_p$, with the scaling prefactor~$\Upsilon_0$
defined in Eq.~\eqref{eq:Upsilon}, set the baseline heating
strength and the thermo-magnetic feedback.  The dimensionless
interfacial coupling~$\gamma$ and ratio~$\varepsilon$
[Eq.~\eqref{gameps}], together with the interfacial coupling
strength $\gamma_s = A_p h_s T_0/V_p$, determine the renormalized
coefficients $\tilde{a}$ and~$\tilde{b}$
[Eq.~\eqref{eq:renorm_coeffs}], which reduce to $a$ and~$b$ in
the strong interfacial coupling limit $\gamma \gg |\varepsilon|$.
The nanoscale loss parameter~$\beta = L_m\,t_d/(\rho_m c_{v,m})$
captures environmental coupling at the nanomagnet level, while
the diffusive coupling~$\zeta$ [Eq.~\eqref{eq:zeta_xi_def}] and the system
size~$\Lambda = L/d$ complete the nanomagnet-scale description.

At the assembly scale, the discrete delta-function sources are
replaced by an effective volumetric heating density after spatial
averaging over $\ell_{\mathrm{cg}} \gg d$
[Eq.~\eqref{eq:HE_coarse_dimless}].  For a uniform chain of identical
nanomagnets, the coarse-grained source density is
$(V_p/A_p)\,a_p/d$, and the linearized source
term~\eqref{eq:SAR_coarse} yields the coarse-grained coefficients
\begin{equation}
  a_{\mathrm{cg}}
  = \frac{a}{\beta_N}\,,
  \qquad
  b_{\mathrm{cg}}
  = \frac{b}{\beta_N}\,,
  \qquad
  \beta_N
  \equiv \frac{L_N\,t_d}{\rho_m c_{v,m}}\,=\frac{L_N\,d^2}{\kappa_m}.
  \label{eq:acg_bcg}
\end{equation}

Here $\beta_N$ is the assembly-scale analogue of~$\beta$: it measures the ratio of the diffusion time~$t_d$ to the
assembly-scale leakage time $t_s \simeq \rho_m c_{v,m}/L_N$.  Because $L_N \gg L_m$ (the hierarchical relation discussed in
Sec.~\ref{sec:HE_assembly}), we have $\beta_N/\beta = L_N/L_m \gg 1$, so $a_{\mathrm{cg}}$ and $b_{\mathrm{cg}}$ are parametrically smaller than $a/\beta$ and $b/\beta$, respectively, but typically of order unity or larger.

\subsection{Reference parameter set}
\label{sec:ref_params}

Table~\ref{table:reference_params} lists the reference values used
for the magnetite--PMMA system, by way of illustration, throughout
Sec.~\ref{sec:Results}.  These values are representative of typical
nanocomposites used in magnetic hyperthermia studies.  For the
reference set, $\sigma_0 \simeq 6.6$ places the NM in the
intermediate barrier regime, with $\eta_0 \simeq 0.37$ and
$b_p < 0$ (self-limiting heating, cf.\ Appendix~\ref{app:sar_coeffs}).
The interfacial coupling is moderate ($\gamma \simeq 2.5$), so the
renormalization~\eqref{eq:renorm_coeffs} is significant: $\tilde{a}$
is reduced by about a factor of 3 relative to~$a$.  The system lies
in the unconditionally stable regime
($\tilde{b} < 0$, cf.\ Sec.~\ref{sec:stability_criterion}).

The coarse-grained coefficients $a_{\mathrm{cg}} \simeq 11$ and
$b_{\mathrm{cg}} \simeq -66$ reflect the large ratio of source
power to assembly-scale losses:
$\langle a_p \rangle_{\mathrm{cg}} / (L_N T_0) \gg 1$.

\begin{table}[t]
\centering
\caption{Reference physical and dimensionless parameters for the
         magnetite--PMMA nanocomposite.}
\label{table:reference_params}
\begin{tabular}{@{}llc@{}}
\toprule
\textbf{Parameter} & \textbf{Symbol} & \textbf{Value} \\
\midrule
\multicolumn{3}{c}{\textbf{Nanoparticle properties}} \\
\midrule
Diameter             & $D$           & 12.0~nm \\
Volume               & $V_p$         & $9.05\times10^{-25}$~m$^3$ \\
Density              & $\rho_p$      & 5200~kg/m$^3$ \\
Saturation magn.     & $M_s$         & 480~kA/m \\
Anisotropy constant  & $K$           & 30~kJ/m$^3$ \\
Attempt time         & $\tau_0$      & $1.0\times10^{-9}$~s \\
\midrule
\multicolumn{3}{c}{\textbf{Geometric arrangement}} \\
\midrule
Interparticle spacing & $d$          & 36.0~nm \\
System length         & $L$          & 1.98~$\mu$m \\
Number of NMs         & $\mathcal{N}$ & 56 \\
\midrule
\multicolumn{3}{c}{\textbf{Thermal properties}} \\
\midrule
Matrix conductivity   & $\kappa_m$           & 0.2~W/(m$\cdot$K) \\
Matrix heat capacity  & $\rho_m c_{v,m}$     & $2.21\times10^{6}$~J/(m$^3\cdot$K) \\
Base temperature      & $T_0$                & 300~K \\
Interfacial coeff.    & $h_s$                & 0.33~W/(m$^2\cdot$K) \\
Nanoscale loss coeff. & $L_m$                & $3.0\times10^{3}$~W/(K$\cdot$m$^3$) \\
Assembly-scale loss   & $L_N$                & $3.3\times10^{5}$~W/(K$\cdot$m$^3$) \\
\midrule
\multicolumn{3}{c}{\textbf{AC magnetic field}} \\
\midrule
Amplitude            & $h_0$         & 38.2~kA/m \\
Frequency            & $f$           & 194~kHz \\
\midrule
\multicolumn{3}{c}{\textbf{Inferred magnetic parameters}} \\
\midrule
Reduced anisotropy   & $\sigma_0$    & 6.55 \\
Reduced frequency    & $\varpi_0$    & $1.22\times10^{-3}$ \\
Dipolar strength     & $\lambda$     & 0.0074 \\
\midrule
\multicolumn{3}{c}{\textbf{Dimensionless control parameters}} \\
\midrule
Baseline heating     & $a$              & $2.38\times10^{-8}$ \\
Feedback coefficient & $b$              & $-1.41\times10^{-7}$ \\
Interfacial coupling & $\gamma$         & 2.50 \\
Renormalized baseline& $\tilde{a}$      & $7.05\times10^{-9}$ \\
Renormalized feedback& $\tilde{b}$      & $-4.18\times10^{-8}$ \\
Nanoscale loss       & $\beta$          & $1.94\times10^{-11}$ \\
Diffusive coupling   & $\zeta$          & $1.55\times10^{5}$ \\
System size          & $\Lambda$        & 55 \\
CG baseline          & $a_{\mathrm{cg}}$ & 11.1 \\
CG feedback          & $b_{\mathrm{cg}}$ & $-65.8$ \\
\bottomrule
\end{tabular}
\end{table}

\paragraph{Case studies.}
Table~\ref{table:case_studies} contrasts two representative configurations,
both using the same nanomagnet ($D = 12$~nm magnetite) but differing
in matrix environment and geometric packing.
\begin{itemize}
 \item The aqueous ferrofluid ($d/D = 5$) has high matrix
conductivity, strong interfacial coupling
($\gamma \simeq 4.4 \times 10^4$, so $\tilde{a} \simeq a$), and
large environmental losses ($\beta$ an order of magnitude above the
reference, $L_N/L_m = 50$).  The dilute packing and efficient heat
removal favor the localized-heating regime.
 \item The reference PMMA nanocomposite ($d/D = 3$) reproduces the
parameter set of Table~\ref{table:reference_params}.  Here the
interfacial coupling is moderate ($\gamma \simeq 2.5$), making
the renormalization significant: $\tilde{a}$ is reduced by about a
factor of 3 relative to~$a$.  The denser packing and lower matrix
conductivity yield $a_{\mathrm{cg}} \simeq 11$, promoting collective
heating.
\end{itemize}

\begin{table}[t]
\centering
\caption{Case studies: aqueous ferrofluid vs.\ reference PMMA
         nanocomposite (Table~\ref{table:reference_params}).
         Both use $D = 12$~nm magnetite, $L = 1.98$~$\mu$m,
         $h_0 = 38.2$~kA/m, $f = 194$~kHz.}
\label{table:case_studies}
\begin{tabular}{@{}lcc@{}}
\toprule
\textbf{Parameter}
  & \textbf{Ferrofluid}
  & \textbf{Reference PMMA} \\
\midrule
\multicolumn{3}{c}{\textbf{Physical parameters}} \\
\midrule
$d/D$                                    & 5          & 3 \\
$\kappa_m$ (W/mK)                        & 0.6        & 0.2 \\
$\rho_m c_{v,m}$ ($10^6$ J/m$^3$K)      & 4.19       & 2.21 \\
$h_s$ (W/m$^2$K)                         & 5000       & 0.33 \\
$L_m$ ($10^3$ W/Km$^3$)                  & 20.0       & 3.0 \\
$L_N$ ($10^5$ W/Km$^3$)                  & 10.0       & 3.3 \\
\midrule
\multicolumn{3}{c}{\textbf{Dimensionless parameters}} \\
\midrule
$a$                        & $1.14\times10^{-8}$  & $2.38\times10^{-8}$ \\
$b$                        & $-7.05\times10^{-8}$ & $-1.41\times10^{-7}$ \\
$\gamma$                   & $4.4\times10^{4}$    & $2.50$ \\
$\tilde{a}$                & $1.14\times10^{-8}$  & $7.05\times10^{-9}$ \\
$\tilde{b}$                & $-7.04\times10^{-8}$ & $-4.18\times10^{-8}$ \\
$\beta$                    & $1.20\times10^{-10}$ & $1.94\times10^{-11}$ \\
$\zeta$                    & $1.50\times10^{5}$   & $1.55\times10^{5}$ \\
$\Lambda$                  & 33                   & 55 \\
$a_{\mathrm{cg}}$          & 1.89                 & 11.1 \\
$b_{\mathrm{cg}}$          & $-11.7$              & $-65.8$ \\
\bottomrule
\end{tabular}
\end{table}

\subsection{Physical versus illustrative parameters}
\label{sec:nano_to_chain_peaks}

As discussed in the introduction, a single nanomagnet driven by an AC magnetic field dissipates heat, but
the associated temperature perturbation is both \emph{weak} (limited
power per particle) and \emph{rapidly homogenized} by diffusion in the
embedding medium.
The renormalized coefficients $\tilde{a}_{\mathrm{phys}}$ and
$\tilde{b}_{\mathrm{phys}}$ for the reference magnetite--PMMA system are
of order $10^{-8}$--$10^{-9}$, reflecting this extreme weakness of
individual nanomagnet heating relative to thermal diffusion.
Under these physical conditions local temperature peaks at source sites (hotspots)
are unresolvable: the dimensionless temperature field is dominated by the
lowest spatial modes and thermal heterogeneity is suppressed by factors
of $10^8$ or more.

\paragraph{Single-particle scales (energy, diffusion time, AC cycles).}
We recall from the introduction that, for a spherical nanomagnet of radius $R \sim 10$~nm, the energy needed to raise
its temperature by $\Delta T \sim 1$~K is
\begin{equation}
E_{1\mathrm{K}} \sim \rho_{\mathrm{NP}}c_{p,\mathrm{NP}}\,
\frac{4\pi R^3}{3}\,(1\,\mathrm{K})\,\simeq\,10^{-17}\text{--}10^{-16}\,\mathrm{J}.
\end{equation}
Yet the surrounding medium (water/polymers, $\alpha\sim 10^{-7}$~m$^2$/s)
diffuses heat away on the nanoscale time
\begin{equation}
\tau_{\mathrm{diff}}\sim \frac{R^2}{\alpha}\sim 10^{-9}\,\mathrm{s}.
\end{equation}
Comparing $\tau_{\mathrm{diff}}$ with the AC period $T = 1/f$ gives
\begin{equation}
N_{\mathrm{cyc}}\sim f\tau_{\mathrm{diff}}\ll 1
\qquad (f\sim 10^5\,\mathrm{Hz}\Rightarrow N_{\mathrm{cyc}}\sim10^{-4}),
\end{equation}
so nanoscale gradients relax within a small fraction of a single cycle.
In the time-averaged (quasi-steady) picture a localized power $P$ yields
the far-field scaling $\Delta T(r)\sim P/(4\pi\kappa r)$ for $r\gtrsim R$\footnote{%
  In steady state and with spherical symmetry, the temperature field
  around a point source of power $P$ embedded in an infinite medium of
  thermal conductivity $\kappa$ satisfies $\kappa\nabla^{2}T +
  P\,\delta^{(3)}(\mathbf{r})=0$.  For $r>0$ this reduces to
  $(1/r^{2})\,\mathrm{d}(r^{2}\,\mathrm{d}T/\mathrm{d}r)/\mathrm{d}r=0$,
  whose general solution vanishing at infinity is $T=C/r$.  The
  constant $C$ is fixed by energy conservation: the total power
  conducted outward through any sphere of radius $r$ must equal $P$,
  i.e.\ $P = -\kappa(\mathrm{d}T/\mathrm{d}r)\,4\pi r^{2} = 4\pi\kappa C$,
  giving $\Delta T(r)=P/(4\pi\kappa r)$.%
},
so local contrasts decay rapidly with distance; conversely, maintaining
$\Delta T(a)\sim 1$~K would require $P\sim 4\pi k a\,\Delta T\sim
10^{-7}$~W for $k\sim 0.6$~W/(m\,K), corresponding to unrealistically
large per-particle SLP.
The key message is therefore not that local peaks are absent, but that
under realistic conditions they are \emph{too small and too fast} to
appear as resolved hotspots in mesoscale profiles.

\paragraph{Orders-of-magnitude gap and design criteria.}
To render the analytical structure of the solution visible---the
cusp-like peaks at source sites, the edge--bulk asymmetry imposed by
boundary conditions, and the localized-to-diffuse heating crossover---we
employ an illustrative parameter set elevated by $\sim\!7$--9 orders of
magnitude relative to the physical values
(Table~\ref{table:illustrative_params}).
These illustrative values are chosen so that (i)~the dimensionless temperature field
is of order unity, (ii)~the system lies in the stable localized heating
regime ($\tilde{b} < \tilde{b}_c$), and (iii)~the interplay between
boundary conditions, interfacial coupling, and loss hierarchy is clearly
legible.
All qualitative features of the solution---including the crossover
indicators and the DBC/NBC contrast---are independent of the absolute
magnitude of the source coefficients.

\begin{table}[t]
\centering
\caption{Illustrative parameter set used in the local-profile
         figures (Sec.~\ref{sec:Results}), compared with the physical
         values from Table~\ref{table:reference_params}.}
\label{table:illustrative_params}
\begin{tabular}{@{}lccc@{}}
\toprule
\textbf{Parameter}
  & \textbf{Physical}
  & \textbf{Illustrative}
  & \textbf{Ratio} \\
\midrule
$\tilde{a}$
  & $7.05\times10^{-9}$
  & 0.4
  & $5.7\times10^{7}$ \\
$\tilde{b}$
  & $-4.18\times10^{-8}$
  & $-0.5$
  & $1.2\times10^{7}$ \\
$\beta$
  & $1.94\times10^{-11}$
  & 0.015
  & $7.7\times10^{8}$ \\
\bottomrule
\end{tabular}
\end{table}

\paragraph{Assembly-scale perspective.}
Real nanomagnet assemblies in water/polymer are expected to exhibit
mainly a smooth, collective temperature rise governed by boundary
exchange and low spatial modes\cite{DejKac2022, DejKac2024}.
The use of illustrative parameters should therefore best be viewed as a
pedagogical step: they expose how localized heating imprints peaks at
source locations before those features are washed out by diffusion and
by averaging across many nanomagnets at the assembly scale.
This amplification is a deliberate visualization device and does not
affect the qualitative mechanism by which global heating emerges from
many weak nanoscale sources.

In practice, in the following section we will apply \textit{baseline-subtraction (or detrending) procedure} to remove the zero-mode pedestal (or background) and low-mode structure so as to expose the cusp-like peaks.
Indeed, under the physical magnetite--PMMA parameters the dimensionless temperature field is dominated by the lowest spatial modes, which produce a smooth, slowly varying background of order $10^{-6}$; the source-localised cusps sit on top of this background with amplitudes of order $10^{-8}$--$10^{-9}$, rendering them invisible on a linear scale.
Hence, to expose the local peak structure, \textit{e.g.,} in Fig.~\ref{fig:local_profiles_phys}, a local linear detrend is applied independently for each panel and each time snapshot. More precisely, within the plotting window centred on source~$\xi_s$, a least-squares linear fit $\hat\theta(\xi) = a_0 + a_1\xi$ is computed and subtracted, yielding the residual $\delta\theta(\xi,\tau) = \theta(\xi,\tau)-\hat\theta(\xi)$.
This removes the contribution of all modes whose wavelength exceeds the window width (a few interparticle spacings) while preserving the cusp-like structure at~$\xi_s$.


\section{Results}
\label{sec:Results}

Unless otherwise stated, all results presented in this section are obtained using the reference physical and dimensionless parameters
summarized in Sec.~\ref{sec:parameters} and Table~\ref{table:reference_params}, corresponding to a one-dimensional chain of magnetite nanoparticles embedded in a PMMA matrix.

The local-to-global heating crossover in nanomagnet assemblies is
governed by the competition between heat injection, thermal diffusion,
and environmental losses across multiple scales.  At the nanomagnet
scale, the renormalized feedback coefficient $\tilde{b}$ (incorporating
both magnetic feedback $b$ and interfacial resistance via $\gamma$)
and the nanoscale loss parameter $\beta$ determine the stability
of localized hotspots.  The analytical condition $\tilde{b}=\tilde{b}_{c}$,
where $\tilde{b}_{c}$ depends on both diffusion ($\lambda_{1}$)
and nanoscale losses ($\beta$) as given in Eq.~\eqref{eq:bc_analytical},
provides a compact criterion for the onset of collective heating.
However, how this transition manifests in space and time depends on
the spatial arrangement of the heat sources, the efficiency of thermal
coupling between them, and the loss mechanisms operating
at different scales.  In the following, we also vary the interparticle
distance $d$ as an organizing geometric parameter, as it directly reflects nanoparticle
concentration and simultaneously controls thermal diffusion, dipolar
interactions, and the dimensionless loss parameters.  For any fixed geometry, equivalent transitions can also be induced
by varying magnetic or excitation parameters through their impact
on $a$, $b$, and thereby $\tilde{a}$, $\tilde{b}$; the role
of these additional control parameters is highlighted throughout the section.

\paragraph{Raw versus inferred parameters.}

We emphasize that only physically independent quantities are specified
as raw input parameters in the numerical implementation.  These include
the nanomagnet properties ($D$, $K$, $M_{s}$), the interparticle
spacing $d$, the matrix thermal properties ($\kappa_{m}$, $\rho_{m}c_{v,m}$),
the interfacial coefficient $h_{s}$, the nanoscale loss coefficient
$L_{m}$, the assembly-scale loss coefficient $L_{N}$, and the AC-field
parameters ($h_{0}$, $f_{0}$).  All secondary quantities---in particular
the dimensionless parameters $\tilde{a}$, $\tilde{b}$, $\beta$,
$\gamma$, $\zeta$, and the characteristic times $t_{d}$ and $t_{s}$---are
then inferred from these raw parameters using the scaling relations
defined in Sec.~\ref{sec:parameters}.

As detailed in Sec.~\ref{sec:nano_to_chain_peaks}, we employ two parameter sets throughout this section: the physical magnetite, with PMMA coefficients, and an illustrative set elevated by several orders of magnitude, the latter being designed to make the source-localized structure of the temperature field, namely the cusp-like peaks, edge-bulk asymmetry, and the localized-to-diffuse crossover, directly visible in the figures.

\subsection{Local temperature profiles at source sites}
\label{sec:results_local}
\begin{figure*}[!ht]
  \centering
  \includegraphics[width=\textwidth]{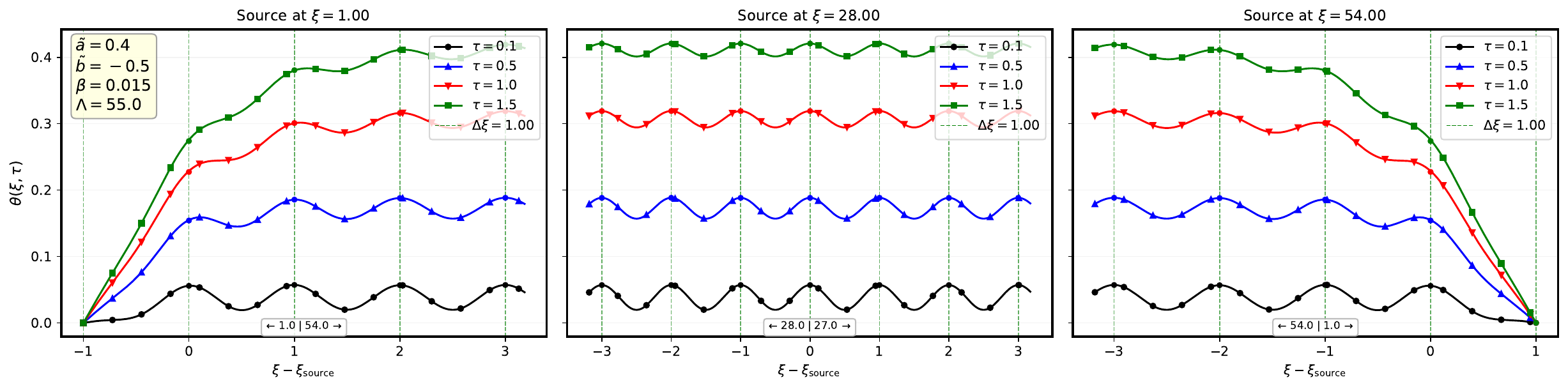}\\[2pt]
  \includegraphics[width=\textwidth]{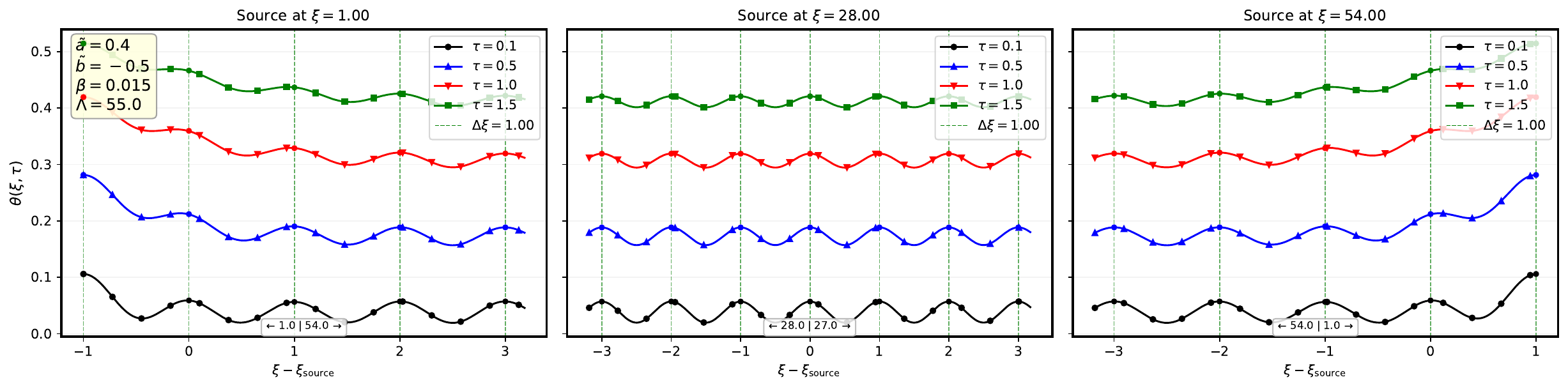}
  \caption{Local temperature profiles $\theta(\xi,\tau)$ around
           three source sites ($\xi=1$, $28$, $54$) at instants
           $\tau=0.1$, $0.5$, $1.0$, $1.5$ for the illustrative
           parameter set (Table~\ref{table:illustrative_params}).
           Vertical dashed lines mark the interparticle spacing $\Delta\xi=1$.
           (Upper panel)~Dirichlet BCs.
           (Lower panel)~Neumann BCs.}
  \label{fig:local_profiles_ill}
\end{figure*}

\begin{figure*}[!ht]
  \centering
  \includegraphics[width=\textwidth]{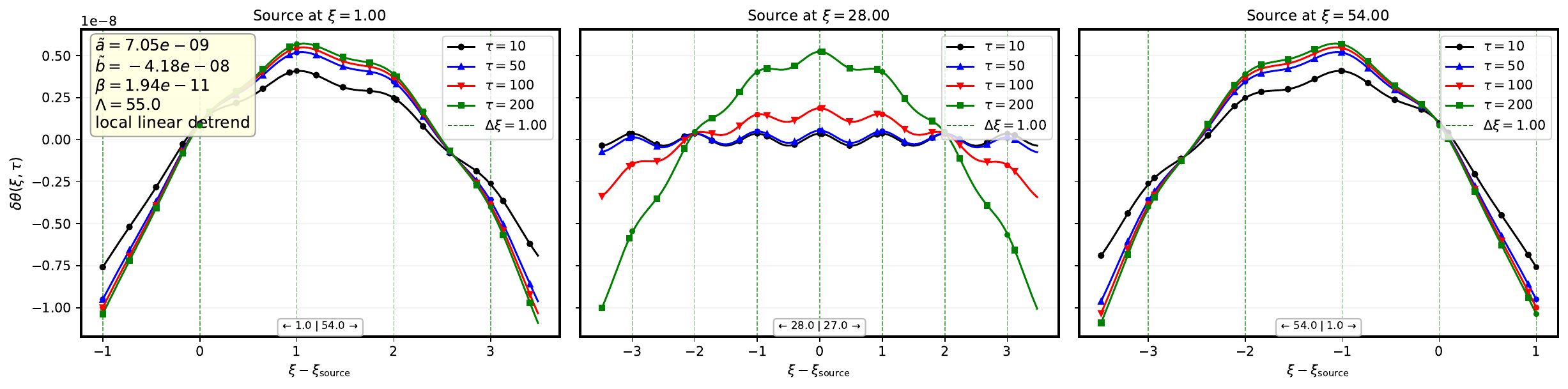}\\[2pt]
  \includegraphics[width=\textwidth]{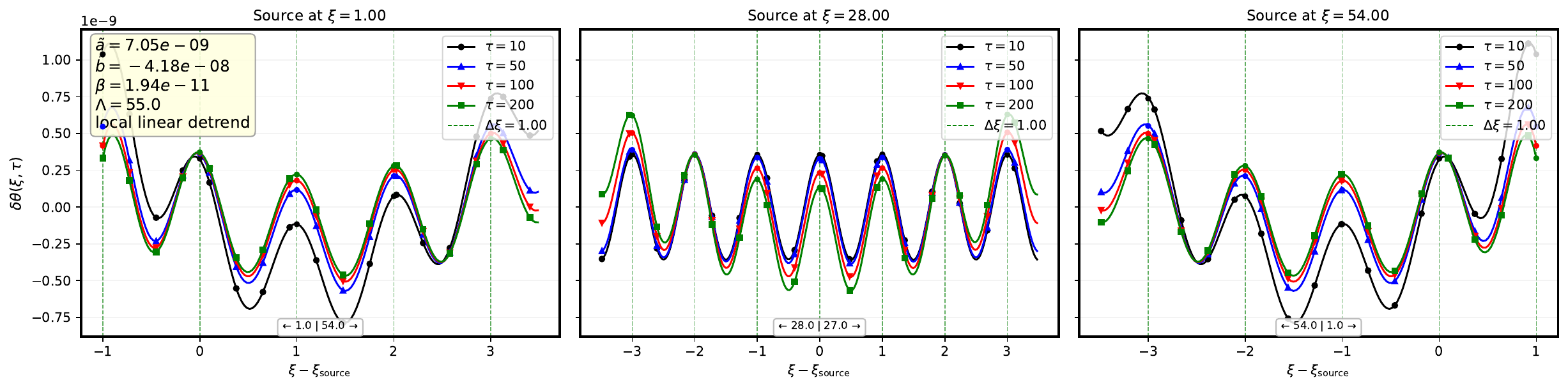}
  \caption{Detrended local profiles $\delta\theta(\xi,\tau)$ around
           three source sites ($\xi=1$, $28$, $54$) for the physical
           magnetite--PMMA parameter set (Table~\ref{table:reference_params})
           at $\tau=10$, $50$, $100$, $200$.
           (Upper panel)~Dirichlet BCs.
           (Lower panel)~Neumann BCs.}
  \label{fig:local_profiles_phys}
\end{figure*}

In Fig.~\ref{fig:local_profiles_ill} and~\ref{fig:local_profiles_phys} we display the temperature field in the neighbourhood of three selected source sites, near the left boundary ($\xi=1$), at the chain centre ($\xi=28$), and near the right boundary ($\xi=54$), under DBC and NBC.
Both figure pairs use the reference magnetite--PMMA chain geometry ($\Lambda=55$, $\mathcal{N}=56$, $\Delta\xi=1$); the illustrative set (Fig.~\ref{fig:local_profiles_ill}) replaces the physical SLP coefficients by $\tilde{a}=0.4$, $\tilde{b}=-0.5$, $\beta=0.015$ while keeping everything else identical, so that the only difference between the two figures is the magnitude of the source and loss terms.

As mentioned earlier, in Fig.~\ref{fig:local_profiles_phys}, we used a baseline-subtraction (or detrending) procedure to remove the zero-mode pedestal (or background) and low-mode structure so as to expose the cusp-like peaks.
On the other hand, we have applied no detrending to the illustrative profiles (Fig.~\ref{fig:local_profiles_ill}), where the source peaks are directly visible without post-processing. Even after detrending, the peak amplitudes in  Fig.~\ref{fig:local_profiles_phys} are of order $10^{-8}$ (DBC) and $10^{-9}$ (NBC), corresponding to absolute temperature excursions $\delta T\sim 10^{-8}T_0 \sim 3\;\mu$K and $\sim 0.3\;\mu$K, respectively.
Note that since both figure pairs share identical chain geometry, the comparison isolates the sole effect of the 7--9 order-of-magnitude coefficient gap given in Table~\ref{table:illustrative_params} of Sec.~\ref{sec:parameters}: under the illustrative set (Fig.~\ref{fig:local_profiles_ill}), $\theta$ is of order unity and no detrending is needed.
The fact that the NBC detrended amplitude is one order of magnitude \emph{smaller} than the DBC value is due to the fact that NBC produces a higher temperature field, but the zero-mode pedestal inflates the local linear trend, so that the residual peak-above-trend shrinks.
Despite the dramatic scale difference, the qualitative features are identical across both parameter sets: cusp-like peaks at source sites,
edge--bulk asymmetry under DBC, near-symmetric profiles under NBC, and monotonic growth with $\tau$.  The physical profiles thus confirm the
conclusion of Sec.~\ref{sec:nano_to_chain_peaks}, namely that nanoscale thermal localisation, while mathematically exact, is physically unresolvable for individual nanomagnets under realistic conditions; the observable temperature rise emerges only through the collective action of the full assembly at the coarse-grained scale.

\paragraph{Effects of the boundary condition.}
Under DBC [Figs.~\ref{fig:local_profiles_ill} and~\ref{fig:local_profiles_phys}, upper panels], the temperature vanishes at the chain
ends and the peak envelope is approximately uniform across the interior, while the edge source ($\xi=1$) is visibly suppressed and
asymmetrically truncated on the boundary side.  We see that each source generates a cusp-like peak that decays within one or two interparticle spacings, though in the bulk the overlap of many such contributions builds a smooth background that grows with $\tau$.
On the other hand, under NBC [Figs.~\ref{fig:local_profiles_ill} and~\ref{fig:local_profiles_phys}, lower panels], the edge--bulk hierarchy is reversed. Indeed,  the boundary sources produce the highest peaks, because the zero-flux condition reflects all diffusing heat inward, effectively doubling the local source strength at the chain ends.  In addition, the inter-peak baseline is lifted by the growing zero-mode background $\phi_0^{(\mathrm{N})}=1/\sqrt{\Lambda}$, which accumulates heat without decay (Appendix~\ref{app:BC_comparison}).
In the illustrative case, the global peak temperature reaches $\theta\simeq 0.51$ under NBC versus $\simeq 0.42$ under DBC; however, the
raised baseline reduces the peak-to-valley contrast, which is consistent with the suppression of relative spatial variance
$\mathcal{V}_{\mathrm{rel}}^{(\mathrm{N})}\ll\mathcal{V}_{\mathrm{rel}}^{(\mathrm{D})}$ derived in Appendix~\ref{app:BC_comparison}.
The bulk source profiles ($\xi=28$) display nearly identical cusp shapes under both boundary conditions, obviously confirming that the DBC/NBC distinction is primarily an edge and global-mode effect.

\paragraph{Effect of interparticle separation.}
\begin{figure*}[!ht]
  \centering
  \includegraphics[width=\textwidth]{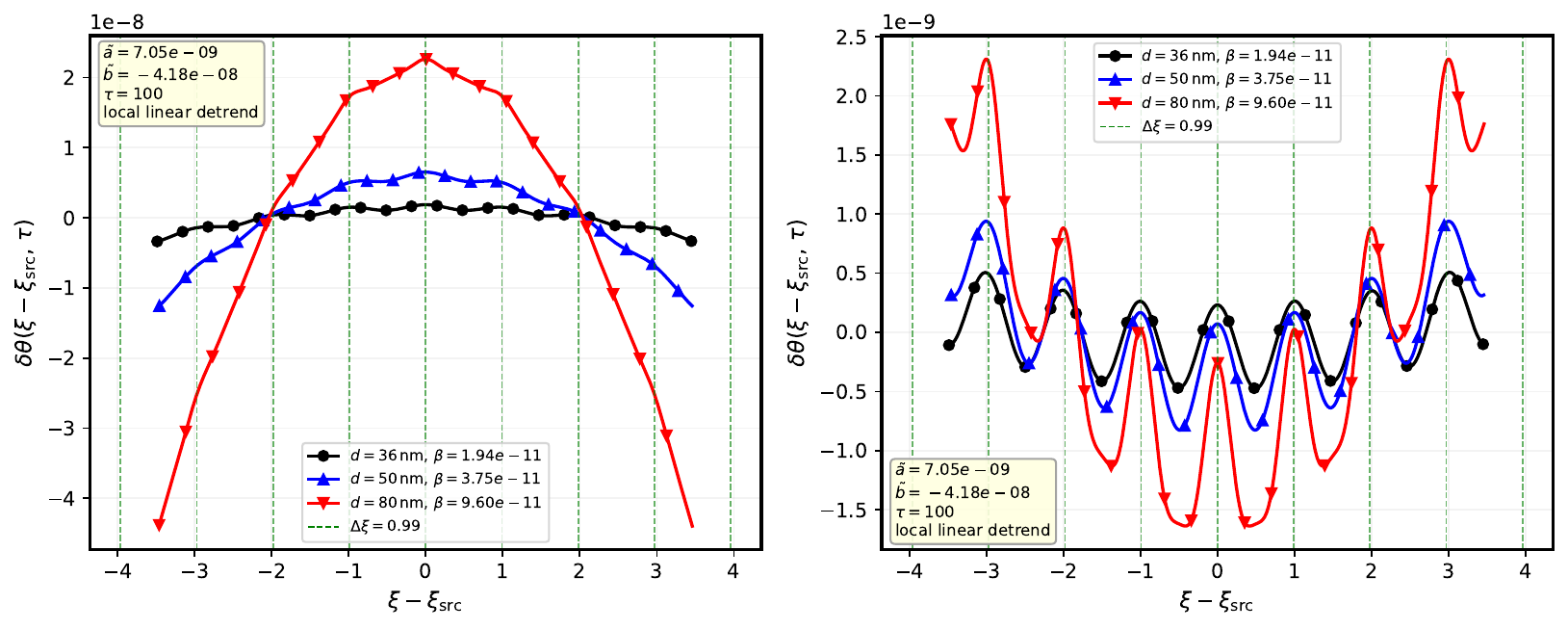}
  \caption{Local profiles at the central source vs.\ interparticle
           separation $d$.
           Detrended profile $\delta\theta(\xi - \xi_{\mathrm{src}},\tau)$
           for the physical magnetite--PMMA parameter set
           (same as Fig.~\ref{fig:local_profiles_phys})
           at $\tau = 100$, for $d = 36$, 50, 80~nm.
           Note that $\beta\propto d^2$ and $\Lambda = L/d$ vary
           with~$d$; the annotation box shows the reference values
           at $d = 36$~nm.
           (Left panel)~Dirichlet BCs.
           (Right panel)~Neumann BCs.}
  \label{fig:local_profiles_sep}
\end{figure*}

In Fig.~\ref{fig:local_profiles_sep}, we examine the dependence of the local profile shape on the interparticle spacing $d$, using the physical
magnetite--PMMA parameter set at the central source ($\xi_{\mathrm{src}} = \Lambda/2$) and late time $\tau = 100$ for three spacings:
$d = 36$, 50, and 80~nm.
Since $\beta = L_m t_d / (\rho_m c_{v,m})$ and $t_d = d^2/\alpha_m$, the Newton loss parameter scales as $\beta \propto d^2$, ranging from
$1.94\times10^{-11}$ at $d = 36$~nm to $9.60\times10^{-11}$ at $d = 80$~nm.
Simultaneously, the chain length in dimensionless units $\Lambda = L/d$ decreases from $55.0$ to $25.0$ as $d$ increases, thus reducing the number of sources from $\mathcal{N} = 56$ to $26$, while the chain physical length $L$ remains fixed.

We see that, under DBC [Fig.~\ref{fig:local_profiles_sep}, left panel], the detrended profile at $d = 80$~nm (red curve) exhibits the largest amplitude, with a broad, dome-shaped peak extending over several interparticle spacings.
As $d$ decreases, both the amplitude and the spatial extent of the peak shrink: at $d = 36$~nm (black curve), the detrended structure is
barely discernible.
This behavior reflects two effects:
(i)~at smaller~$d$ more sources contribute to the smooth background that the detrending removes, and
(ii)~the thermal decay length $\ell_T \sim 1/\sqrt{\beta + |\tilde{b}|}$ in units of $d$ increases (since $\beta$ shrinks faster than
$|\tilde{b}|$), causing the tails of adjacent cusps to overlap and merge into the background.

On the other hand, uder NBC [Fig.~\ref{fig:local_profiles_sep}, right panel], the same ordering persists ($d = 80$~nm largest, $d = 36$~nm smallest), but the amplitudes are uniformly one order of magnitude lower.
The profiles also display resolved satellite peaks at neighbouring source positions, particularly visible for $d = 80$~nm where the oscillatory structure of the Neumann Green's function is most prominent.  These satellites arise because the cosine eigenfunctions couple neighbouring sources more effectively than the sine eigenfunctions (of thhe DBC), and the zero-mode background is removed by the detrending.

\subsection{Transient dynamics at individual source sites}
\label{sec:results_traces}

\begin{figure*}[!ht]
  \centering
  \begin{subfigure}[b]{0.5\textwidth}
    \includegraphics[width=\textwidth]{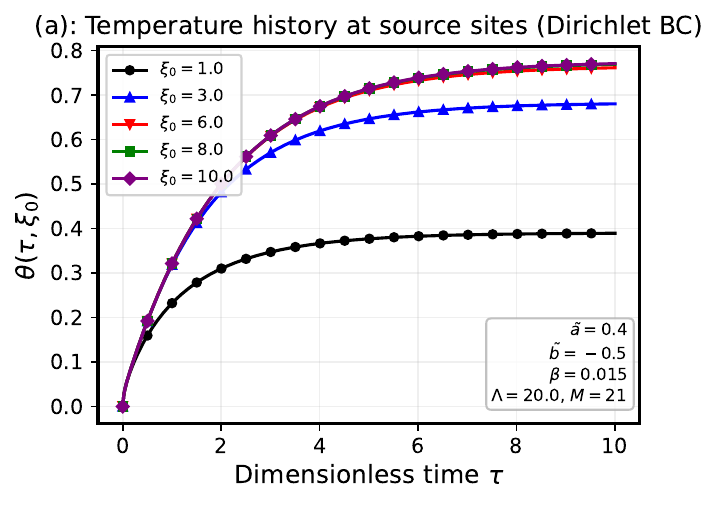}
  \end{subfigure}\hfill
  \begin{subfigure}[b]{0.5\textwidth}
    \includegraphics[width=\textwidth]{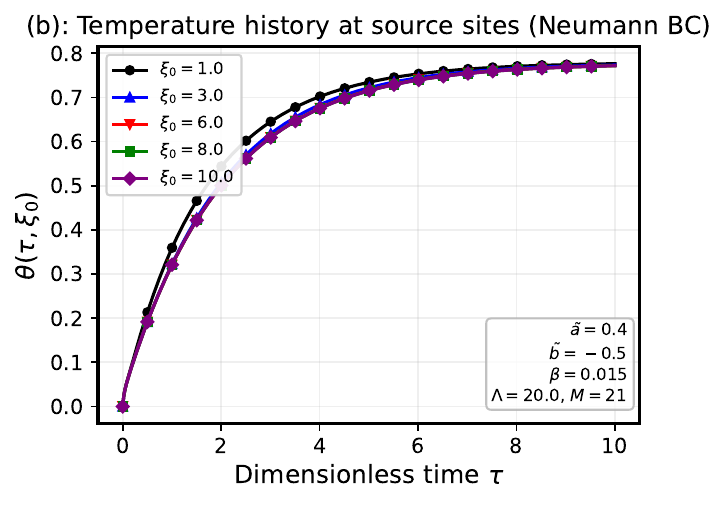}
  \end{subfigure}
  \caption{Temperature history $\theta(\tau)$ at five source positions
           from near-edge ($\xi_0=1$) to chain centre ($\xi_0=10$)
           ($\Lambda=20$, $\mathcal{N}=21$, $\tilde{a}=0.4$, $\tilde{b}=-0.5$,
           $\beta=0.015$).}
  \label{fig:theta_vs_tau}
\end{figure*}

Now, we examine the temperature time profile $\theta(\tau)$ at selected source positions. Figure~\ref{fig:theta_vs_tau} shows temperature temporal evolution at five probe positions, from near-edge ($\xi_0 = 1$) to chain centre ($\xi_0 = 10$), for a chain with $\Lambda = 20$, $\mathcal{N} = 21$ (illustrative parameters: $\tilde{a} = 0.4$, $\tilde{b} = -0.5$, $\beta = 0.015$).

Under DBC [Fig.~\ref{fig:theta_vs_tau}(a)], we may organize the five probe positions into three distinct groups:
the near-edge source ($\xi_0 = 1$) saturates at $\theta_{\mathrm{ss}} \simeq 0.39$, the intermediate source
($\xi_0 = 3$) at $\simeq 0.68$, and the three bulk sources ($\xi_0 = 6, 8, 10$) cluster tightly at $\simeq 0.77$.
We see that the edge--bulk contrast ratio is roughly $1{:}2$ and persists at all times. This reflects the permanent cooling imposed by the boundary thermal baths, or in other words, sources near the boundaries are cooled more efficiently than interior ones, producing a graded temperature distribution across the chain.

The picture is qualitatively different under NBC [Fig.~\ref{fig:theta_vs_tau}(b)] since all five curves converge to a narrow band at $\theta_{\mathrm{ss}} \simeq 0.77$--$0.78$, corresponding to the spatially uniform steady state governed by the zero-mode balance~\eqref{eq:NBC_pedestal}.

Regarding the saturation value of $\theta$, we may show that the saturation time is given by $\tau_{\mathrm{sat}} = 1/|\mu_{\min}| \simeq 2$, where $\mu_{\min}$ is the least negative eigenvalue of the evolution matrix $\mathbf{B}$, and we note that 95\% saturation requires $\tau \simeq 3\tau_{\mathrm{sat}} \simeq 6$.
This timescale is set by the competition between the source strength $\tilde{a}$ and the effective decay rate $|\tilde{b}| + \beta + \lambda_1$ of the fundamental mode [see Eqs.~(\ref{eq:BD}, \ref{eq:BN})], and is independent of the boundary condition for interior sources.
\subsection{Steady-state temperature profiles}

\begin{figure*}[!ht]
  \centering
  \begin{subfigure}[b]{0.33\textwidth}
    \includegraphics[width=\textwidth]{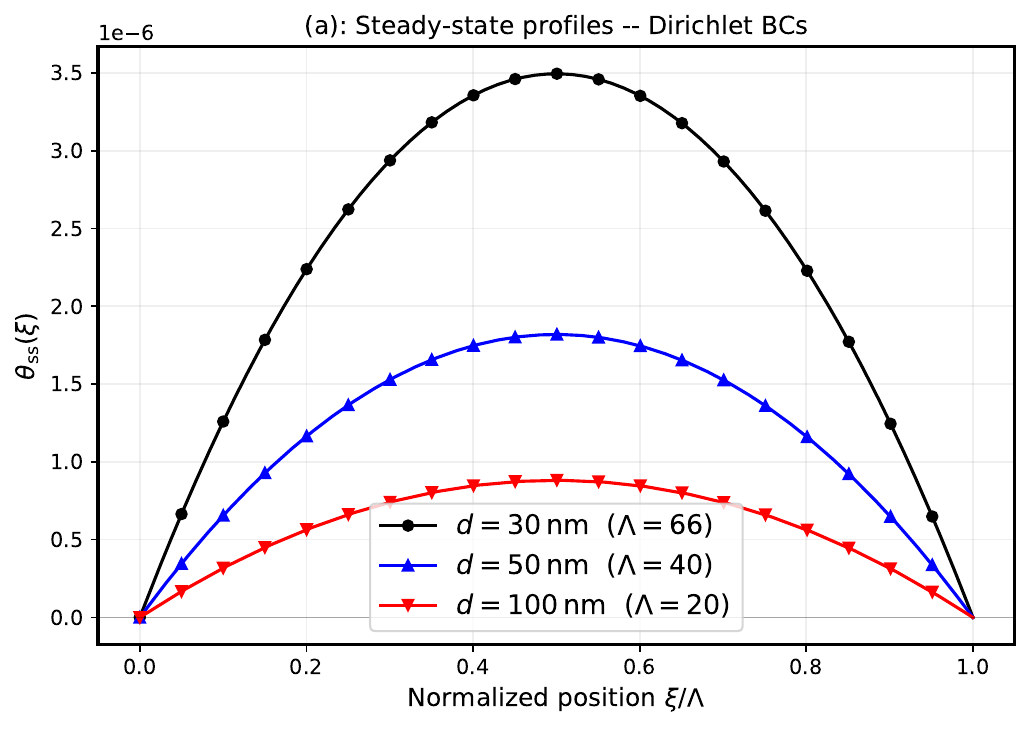}
  \end{subfigure}\hfill
  \begin{subfigure}[b]{0.33\textwidth}
    \includegraphics[width=\textwidth]{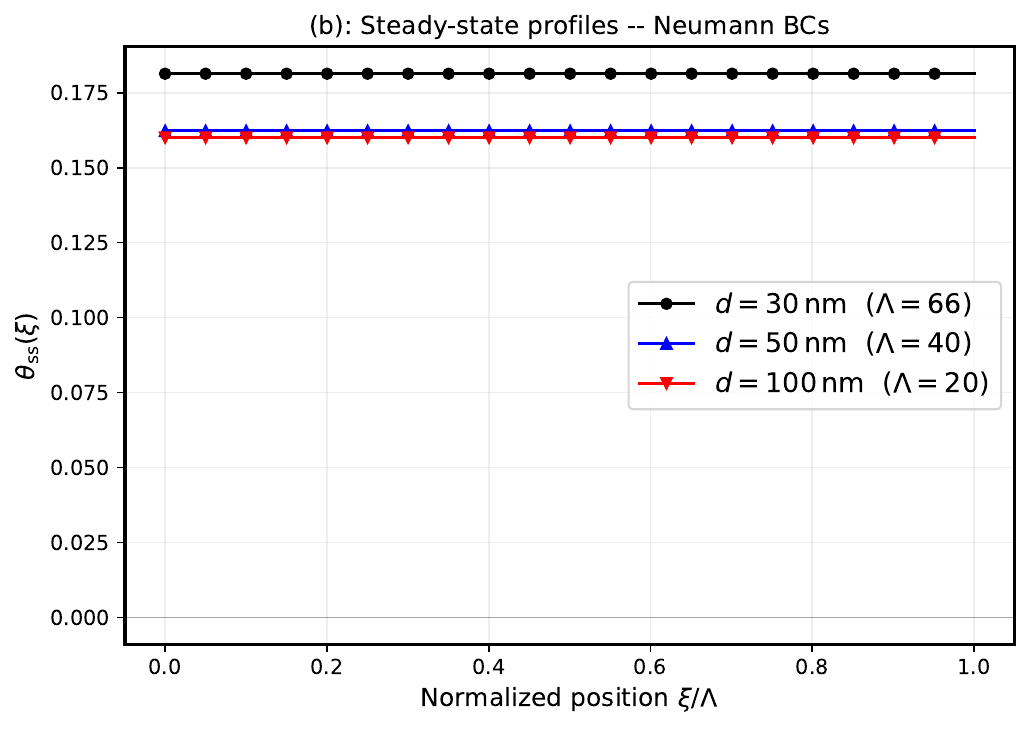}
  \end{subfigure}\hfill
  \begin{subfigure}[b]{0.33\textwidth}
    \includegraphics[width=\textwidth]{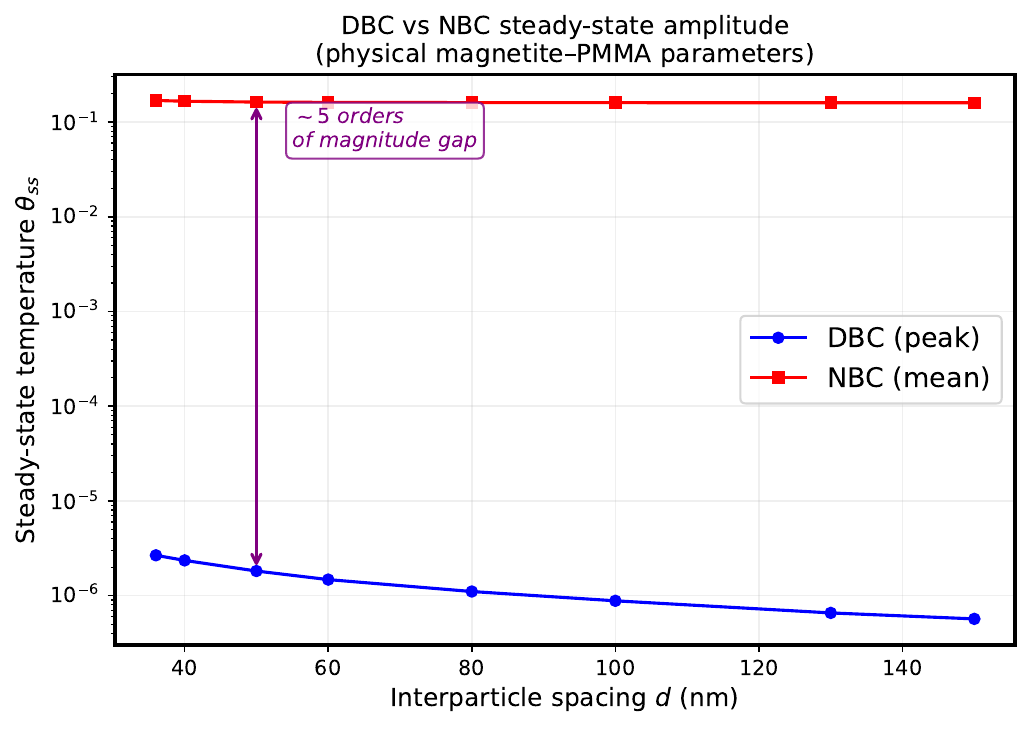}
  \end{subfigure}
  \caption{Steady-state profiles for $d=30$, $50$, $100$\,nm
           (physical magnetite--PMMA parameters).
           (a)~DBC: $\theta_{\mathrm{ss}}(\xi)$. (b)~NBC: $\theta_{\mathrm{ss}}(\xi)$.
           (c)~Peak amplitudes $\theta_{\mathrm{ss}}^{(\mathrm{D})}$ (circles)
           and $\theta_{\mathrm{ss}}^{(\mathrm{N})}$ (squares) vs $d$ (log scale).}
  \label{fig:steady_state}
\end{figure*}

The most direct illustration of the DBC/NBC dichotomy is provided by the steady-state temperature field $\theta_{\mathrm{ss}}(\xi)$, obtained in the limit $\tau\to\infty$ as $\mathbf{c}_{\mathrm{ss}}=-\mathbf{B}^{-1}\mathbf{d}$ [cf.~Eq.~\eqref{eq:exact_modal_solution_full}].
Accordingly, Figure~\ref{fig:steady_state} displays $\theta_{\mathrm{ss}}(\xi)$ for three interparticle spacings $d=30, 50, 100$\,nm, using the physical magnetite--PMMA parameter set (Table~\ref{table:reference_params}).

As expected, under DBC [Fig.~\ref{fig:steady_state}(a)], $\theta_{\mathrm{ss}}(\xi)$ forms a smooth dome that vanishes at both chain ends, with a broad maximum in the interior. Then, smaller spacing $d$ (i.e.\ larger $\Lambda$ and more sources) yields higher peak temperatures.
At this scale, individual nanomagnet cusps are entirely invisible and the temperature field is dominated by the lowest-order modal contributions, which produce a smooth, macroscopic envelope. Therefore, the characteristic parabolic-to-dome shape results from the balance between uniform internal heating and diffusive losses to the boundary baths, with the $\sin(\pi\xi/\Lambda)$ fundamental mode providing the dominant contribution.

NBC [Fig.~\ref{fig:steady_state}(b)] leads to a strikingly different situation. More precisely, the profile is essentially flat for all three spacings, with amplitudes of order $0.16$--$0.18$, \textit{i.e.} roughly five orders of magnitude larger than the DBC peak.
This dramatic difference arises directly from the zero-mode accumulation: because $\lambda_0=0$, the zero mode cannot decay by diffusion to the boundaries and only the volumetric loss $\beta\theta$ and the negative feedback $\tilde{b}<0$ can drain it.
In the hypothetical $\tilde{b}\to0$ limit, the zero-mode balance [Eq.~\eqref{eq:NBC_pedestal}] would give
$\theta_{\mathrm{ss}}\simeq\tilde{a}\,\mathcal{N}/(\Lambda\beta)$. For the physical parameters, $\beta_{\mathrm{phys}}\simeq 2\times10^{-11}$ is negligible compared with $|\tilde{b}|\,\mathcal{N}/\Lambda\simeq 4\times10^{-8}$, so the denominator of Eq.~\eqref{eq:NBC_pedestal} is dominated by the feedback term and the pedestal reduces to $\theta_{\mathrm{ss}}\simeq\tilde{a}/|\tilde{b}|\approx 0.17$, which is still five orders of magnitude above the DBC peak.

A word is in order regarding the weak $d$-dependence of the NBC profiles: the three curves in Fig.~\ref{fig:steady_state}(b) are nearly coincident, making it difficult to read off the variation quantitatively from a spatial plot. Figure~\ref{fig:steady_state}(c) addresses this directly by displaying both the NBC uniform amplitude and the DBC peak temperature as functions of $d$ on a logarithmic scale.

Consequently, three observations follow from Fig.~\ref{fig:steady_state}(c).
First, the DBC peak falls monotonically with $d$, reflecting the decrease in the total number of sources $\mathcal{N}\propto d^{-1}$ at fixed
system length $L$.
Second, the NBC amplitude is almost perfectly flat. Indeed, to leading order it equals $\tilde{a}/|\tilde{b}|$ [Eq.~\eqref{eq:NBC_pedestal}], which is $d$-independent, and the sub-leading correction $\propto\beta/|\tilde{b}|\mathcal{N}/\Lambda$ is tiny for $\beta_{\mathrm{phys}}\ll|\tilde{b}|\mathcal{N}/\Lambda$.
Third, the gap between the two curves, roughly five orders of magnitude across the full range, encapsulates the central physical trade-off
identified in Sec.~\ref{sec:DBC_NBC_summary}, namely: insulating (Neumann-like) boundaries maximise absolute temperature, while conducting (Dirichlet-like) boundaries preserve spatial heterogeneity.

\subsection{Spatial profiles: localized regime and crossover}

To visualise the full spatio-temporal dynamics, we display spatial profiles $\theta(\xi,\tau)$ at selected times for both the stable localized regime (Fig.~\ref{fig:profiles_loc}) and the critical crossover regime (Fig.~\ref{fig:profiles_cross}).
This representation conveys the same information as a space-time map but makes the growth of individual source peaks, the edge--bulk hierarchy, and the onset of spatial homogenisation directly readable on a common scale.

\subsubsection{Localized heating regime}

\begin{figure*}[!ht]
  \centering
  \includegraphics[width=\textwidth]{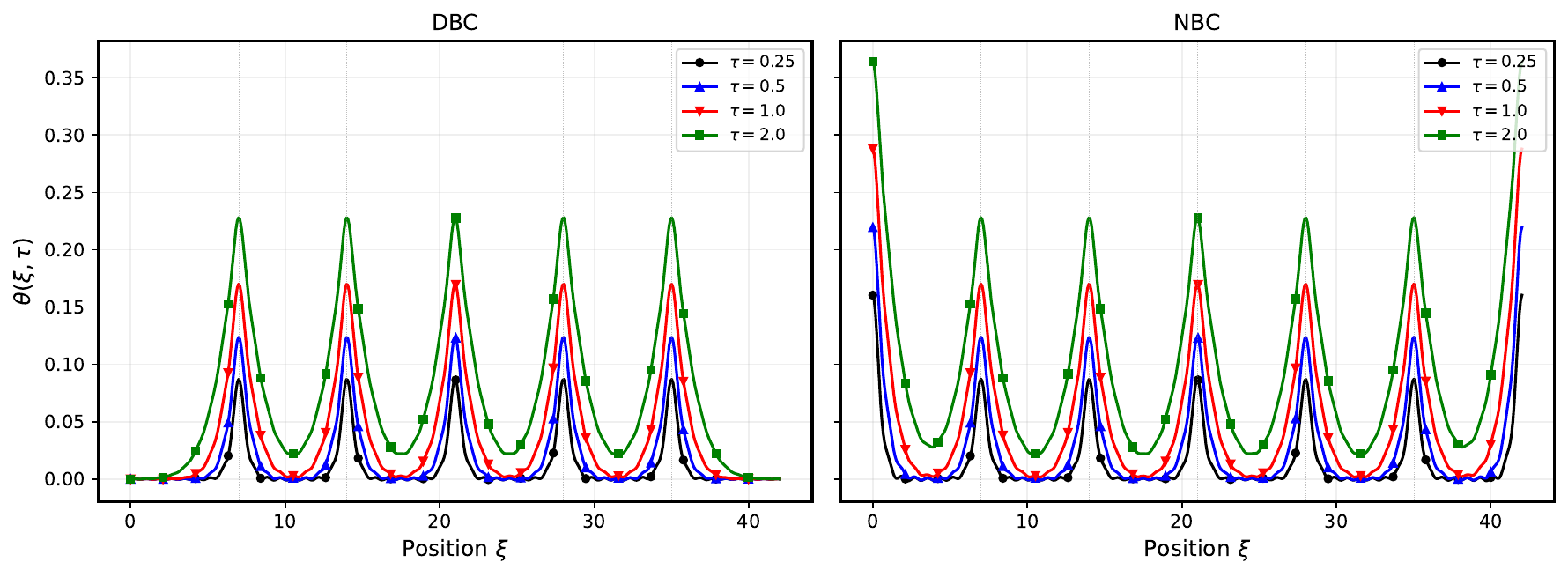}
  \caption{Spatial temperature profiles $\theta(\xi,\tau)$ at
           $\tau=0.25$, $0.5$, $1.0$, $2.0$ in the localized regime
           ($\Lambda=42$, $\mathcal{N}=7$, $\Delta\xi=7$, $\tilde{b}=-0.52$, $\beta=0.017$).
           Faint dashed lines mark the source positions.}
  \label{fig:profiles_loc}
\end{figure*}

Figure~\ref{fig:profiles_loc} shows $\theta(\xi,\tau)$ at $\tau=0.25, 0.5, 1.0, 2.0$ for $\Lambda=42$, $\mathcal{N}=7$, $\Delta\xi=7$, $\tilde{b}=-0.52$, $\beta=0.017$ (localized regime, well below $\tilde{b}_c$). Both panels share the same set of source positions; the faint vertical dashed lines mark each source site for orientation.

Under DBC [Fig.~\ref{fig:profiles_loc}, left], the profiles grow uniformly across the interior of the chain, with the edge sources visibly suppressed relative to the bulk.
We see that the envelope of the peaks is nearly flat in the interior at each time, thus confirming that the fundamental Dirichlet mode
$\sin(\pi\xi/\Lambda)$ spreads the heating broadly. On the other hand, individual source peaks are resolved. Indeed, since the inter-particle spacing $\Delta\xi=7$ exceeds the thermal decay length $\ell_T=1/\sqrt{\beta+|\tilde{b}|}\approx 1.3$, the tails of adjacent peaks do not overlap significantly.
As $\tau$ increases from $0.25$ to $2.0$, the profiles grow quasi-proportionally, approaching the dome-shaped steady state of
Fig.~\ref{fig:steady_state}(a).

In the case of NBC [Fig.~\ref{fig:profiles_loc}, right], two distinct features appear. First, the baseline rises monotonically with $\tau$: even the inter-peak minima increase in time, reflecting the accumulation of heat in the spatially uniform zero mode $\phi_0=1/\sqrt{\Lambda}$.
So, by $\tau=2.0$, the baseline has risen to a level comparable to the individual peak amplitudes, indicating that the zero-mode background
is already substantial.
Second, the edge sources ($\xi=0$ and $\xi=\Lambda$) produce the largest peaks at every time, reversing the DBC hierarchy. Indeed, as already mentioned earlier, the no-flux boundary reflects all diffusing heat inward, effectively doubling the local source strength.
The peak-to-valley contrast, \textit{i.e.}, the direct measure of spatial localisation, decreases visibly with $\tau$ as the zero mode inflates the baseline, consistent with the suppression of the relative variance $\mathcal{V}_{\mathrm{rel}}^{(\mathrm{N})}$ established analytically in Appendix~\ref{app:BC_comparison}.

Comparing the two panels at the same time and on the same scale makes explicit the fact that the overall temperature level is uniformly higher under NBC, but the spatial contrast (peak height above local baseline) is substantially lower. This is the real-space counterpart of the variance-mean ratio diagnostic we discussed in Sec.~\ref{sec:results_varmean}\,E.

\subsubsection{Crossover regime}

\begin{figure*}[!ht]
  \centering
  \includegraphics[width=\textwidth]{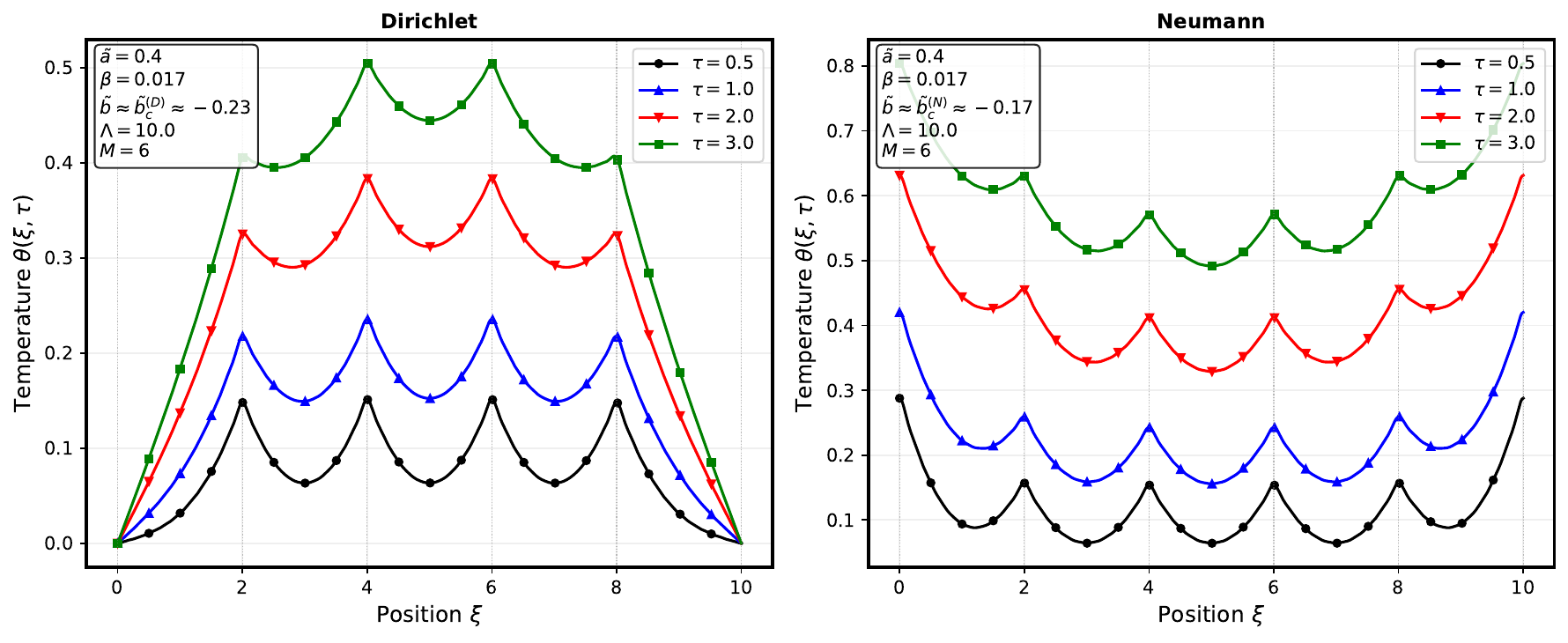}
  \caption{Spatial temperature profiles $\theta(\xi,\tau)$ at
           $\tau=0.5$, $1.0$, $2.0$, $3.0$ near criticality
           ($\Lambda=10$, $\mathcal{N}=6$, $\Delta\xi=2$, $\beta=0.017$).
           Left:~DBC at $\tilde{b}=\tilde{b}_c^{(\mathrm{D})}=-0.23$.
           Right:~NBC at $\tilde{b}=\tilde{b}_c^{(0)}=-0.17$.}
  \label{fig:profiles_cross}
\end{figure*}

In Figure~\ref{fig:profiles_cross}, we plot the same profile representation for the crossover regime: $\Lambda=10$, $\mathcal{N}=6$, $\Delta\xi=2$, $\beta=0.017$, $\tilde{b}\simeq\tilde{b}_c$. The DBC panel uses $\tilde{b}\simeq\tilde{b}_c^{(\mathrm{D})}=-0.23$ and the NBC panel uses $\tilde{b}\simeq\tilde{b}_c^{(0)}=-0.17$ (the respective critical values).

We note that near the critical feedback coefficient $\tilde{b}_c$, the least-negative eigenvalue of $\mathbf{B}$ approaches zero and thereby the temperature grows quasi-linearly in time. The profile plots reveal the spatial structure of this growth in detail.

Under DBC [Fig.~\ref{fig:profiles_cross}, left], the profiles retain a clearly resolved, sine-shaped envelope at all four times, while individual source peaks sit on top of the fundamental-mode background. Again, the temperature vanishes at both edges at all $\tau$,  maintaining the edge--bulk hierarchy even as the amplitude grows. This growth is nearly proportional in time up to $\tau=3$: the ratio of the $\tau=3$ and $\tau=0.5$ profiles is approximately constant across $\xi$, thus confirming the dominance of a single slow mode ($r=1$) whose decay rate $|B_{11}|$ is small but finite.

Under NBC [Fig.~\ref{fig:profiles_cross}, right], the temporal evolution is qualitatively different. At early times ($\tau=0.5$), individual source peaks are still resolved and the edge sources are enhanced relative to the interior, essentially the same pattern seen in Fig.~\ref{fig:profiles_loc}.
By $\tau=1.0$, the zero-mode pedestal starts to dominate the overall amplitude while individual source peaks remain resolved above it. By $\tau=2.0$--$3.0$, the pedestal continues to grow steeply and the peak-to-baseline contrast decreases markedly, though discrete source peaks remain discernible at all times shown.
This is the direct visualisation of the near-critical NBC regime: the spatially uniform zero mode drives the dominant temperature rise, thus progressively reducing the relative spatial heterogeneity without fully suppressing the individual source peaks.

\subsection{Statistical indicators of the crossover: variance and mean}
\label{sec:results_varmean}

\begin{figure*}[!ht]
  \centering
  \begin{subfigure}[b]{0.48\textwidth}
    \includegraphics[width=\textwidth]{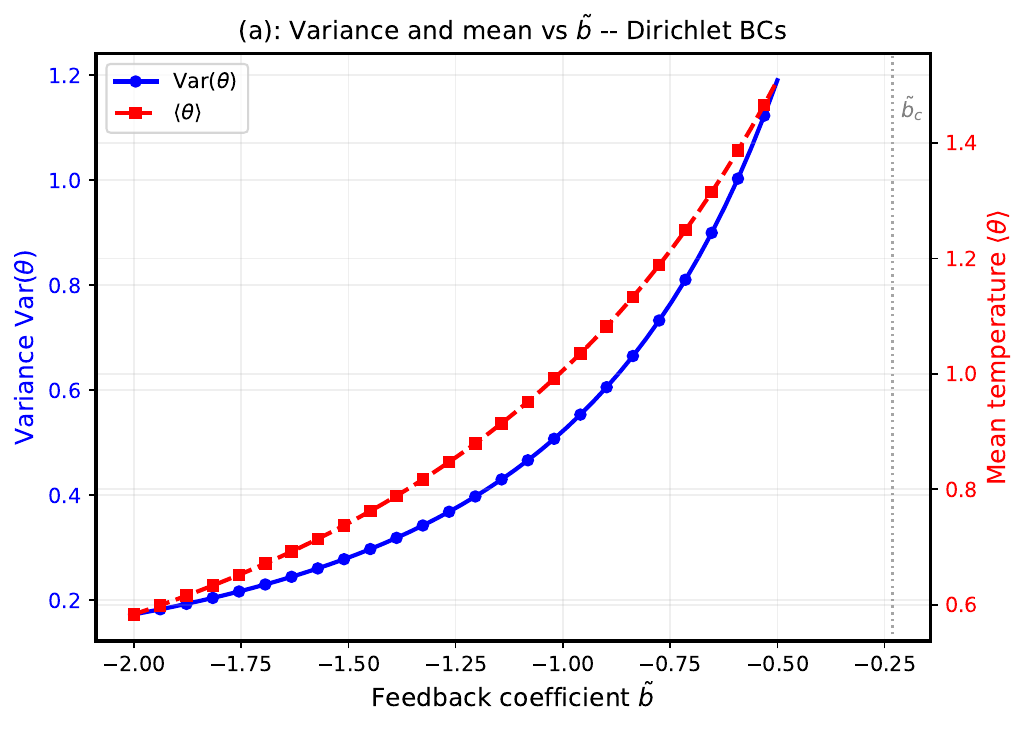}
  \end{subfigure}\hfill
  \begin{subfigure}[b]{0.48\textwidth}
    \includegraphics[width=\textwidth]{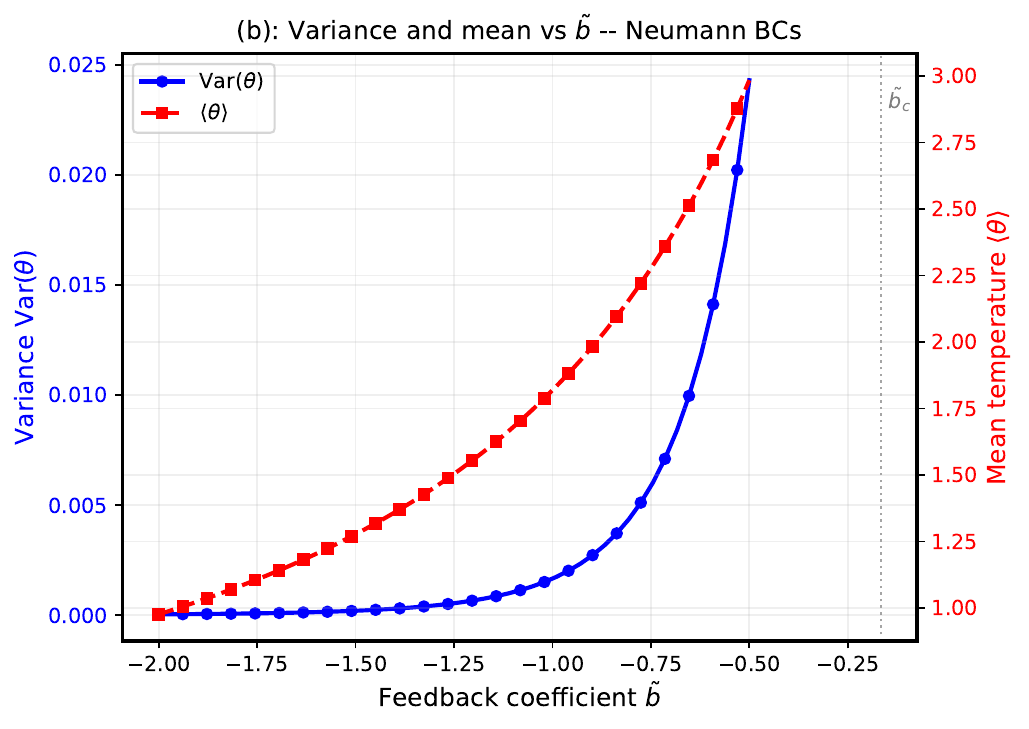}
  \end{subfigure}
  \caption{Spatial variance $\mathrm{Var}(\theta)$ and mean
           $\langle\theta\rangle$ at late time vs feedback coefficient
           $\tilde{b}$. Dotted line: $\tilde{b}_c$.
           }
  \label{fig:varmean}
\end{figure*}

The physical indicators introduced in Sec.~\ref{subsec:crossover_indicators}, namely the mean temperature $\langle\theta\rangle$ [Eq.~\eqref{eq:theta_mean}] and spatial variance $\mathrm{Var}(\theta)$ [Eq.~\eqref{eq:theta_variance}], provide us with scalar diagnostics that track the local-to-global crossover as $\tilde{b}$ varies.
Accordingly, Fig.~\ref{fig:varmean} displays both quantities at late time as a function of~$\tilde{b}$, with the vertical dotted line marking the analytical critical value~$\tilde{b}_c$.

As we can see, both quantities diverge as $\tilde{b} \to \tilde{b}_c^-$. This behavior signals the approach to marginal stability: the fundamental-mode decay rate $|\mu_1| \to 0$, the modal amplitudes grow without bound, and the temperature field fails to reach a finite steady state within the simulated time window.

Under DBC [Fig.~\ref{fig:varmean}(a)], the variance and mean are of the same order of magnitude (both $\sim 0.2$--$1.2$ in the pre-critical range), confirming that the temperature field retains significant spatial structure. The peaks at source sites contribute comparably to both the mean (through their absolute height) and the spread (through the peak-to-valley contrast), so the ratio $\mathrm{Var}(\theta)/\langle\theta\rangle^2 = \mathcal{O}(1)$.

Under NBC [Fig.~\ref{fig:varmean}(b)], we see that the mean temperature is much larger ($\langle\theta\rangle \sim 1$--$3$), while the variance
remains very small ($\mathrm{Var}(\theta) \sim 0$--$0.025$), roughly two orders of magnitude below the DBC values. This quantitatively confirms the homogenising effect of the zero mode. More precisely, NBC produces a large but spatially uniform temperature, whereas DBC preserves strong spatial contrast.

This figure provides the most direct numerical verification of the central trade-off identified analytically in
Sec.~\ref{sec:DBC_NBC_summary}, namely the relative spatial variance $\mathcal{V}_{\mathrm{rel}}^{(\mathrm{N})} \ll
\mathcal{V}_{\mathrm{rel}}^{(\mathrm{D})}$ whenever $\beta \ll|\tilde{b}|\,\mathcal{N}/\Lambda$.
Therefore, for applications requiring spatially targeted heating (e.g.\ site-selective drug release), the DBC regime with its preserved
heterogeneity is preferable; for applications requiring maximal bulk temperature rise (e.g.\ whole-tumour hyperthermia), the NBC regime is
more effective.

\subsection{Assembly-scale temperature evolution}
\label{sec:ass-scale}

\begin{figure*}[!ht]
  \centering
  \begin{subfigure}[b]{0.33\textwidth}
    \includegraphics[width=\textwidth]{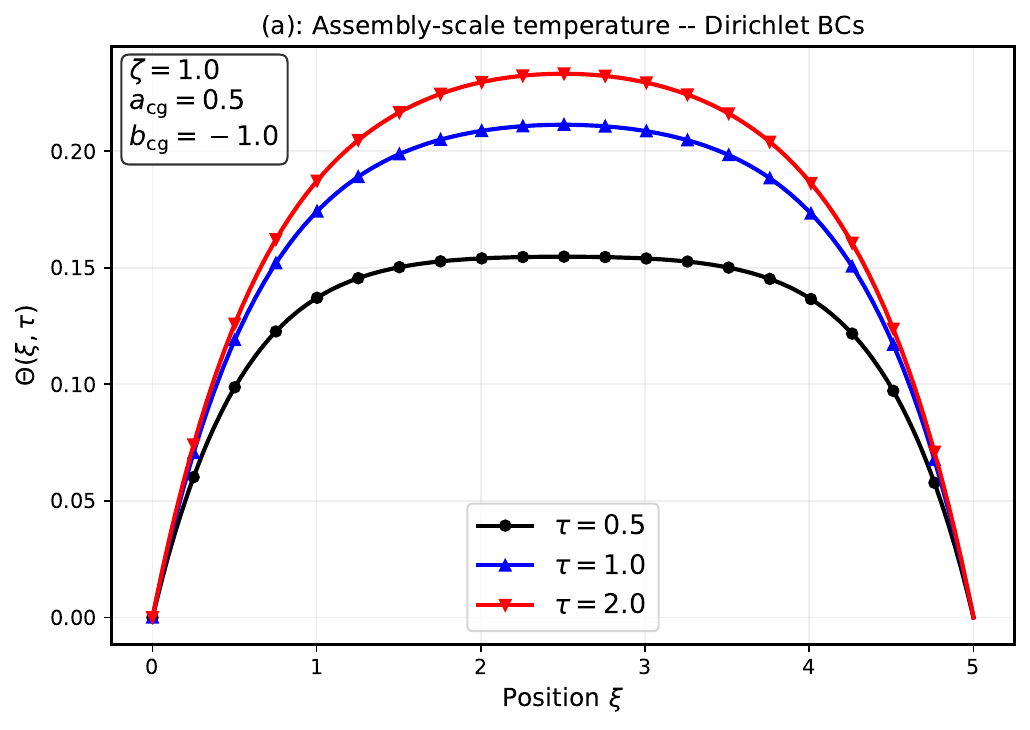}
  \end{subfigure}\hfill
  \begin{subfigure}[b]{0.33\textwidth}
    \includegraphics[width=\textwidth]{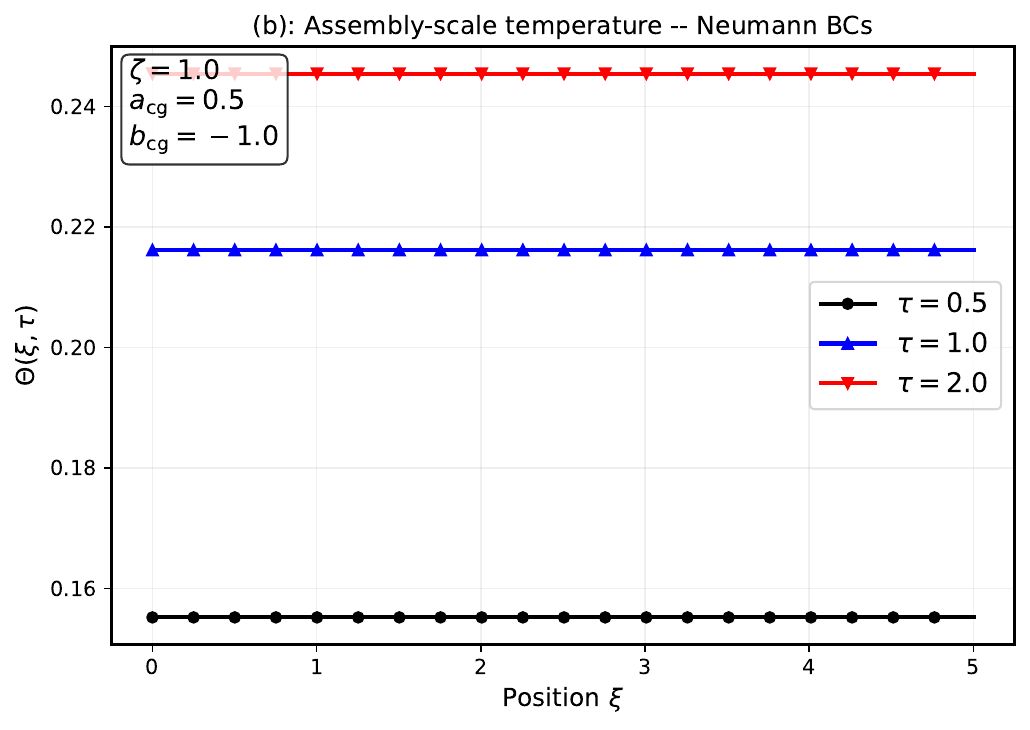}
  \end{subfigure}\hfill
  \begin{subfigure}[b]{0.33\textwidth}
    \includegraphics[width=\textwidth]{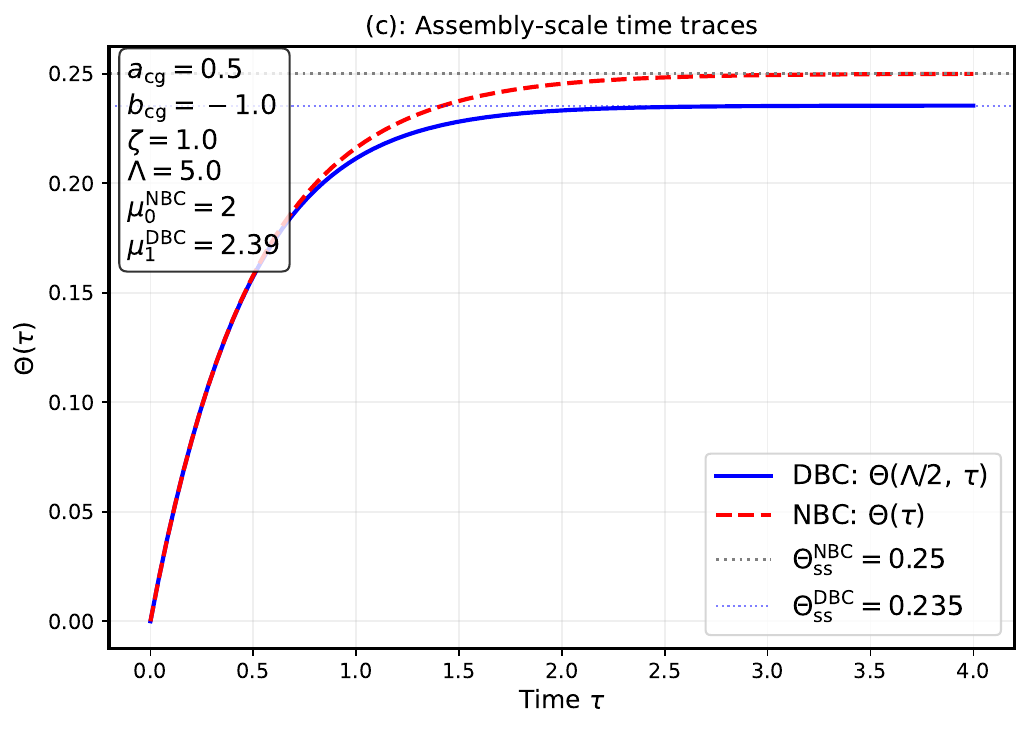}
  \end{subfigure}
  \caption{Coarse-grained assembly-scale temperature
           ($a_{\mathrm{cg}}=0.5$, $b_{\mathrm{cg}}=-1.0$, $\zeta=1$, $\Lambda=5$).
           (a)~DBC spatial profiles at $\tau=0.5$, $1.0$, $2.0$.
           (b)~NBC spatial profiles at the same times.
           (c)~DBC midpoint $\Theta(\Lambda/2,\tau)$ (blue solid) and
           NBC uniform temperature $\Theta(\tau)$ (red dashed).
           Dotted lines: respective long-time asymptotes (see text).}
  \label{fig:assembly}
\end{figure*}

We recall that the coarse-grained assembly-scale heat equation~\eqref{eq:HE_coarse_dimless}, with the linearized source term~\eqref{eq:SAR_coarse}, is designed to describe the macroscopic temperature field $\Theta(\xi,\tau_s)$ that emerges from spatial averaging of the nanoscale sources. Accordingly, Figure~\ref{fig:assembly} shows $\Theta(\xi,\tau)$ at three time snapshots ($\tau=0.5, 1.0, 2.0$), using the coarse-grained coefficients $a_{\mathrm{cg}}=0.5$, $b_{\mathrm{cg}}=-1.0$ (illustrative values) and $\zeta=1$.

Under DBC [Fig.~\ref{fig:assembly}(a)], $\Theta$ vanishes at both ends and develops a broad, flat-topped profile in the interior, with the plateau rising as $\tau$ increases. The boundary layers sharpen with time as the interior approaches the steady-state value $\Theta_{\mathrm{ss}}=a_{\mathrm{cg}}/(1-b_{\mathrm{cg}})$. This dome-shaped profile is the assembly-scale analogue of the nanoscale steady-state profile [Fig.~\ref{fig:steady_state}(a)], thus confirming the consistency of the two-scale framework.

In contrast, under NBC [Fig.~\ref{fig:assembly}(b)], $\Theta$ is perfectly spatially uniform at all times, because the zero-flux boundaries eliminate all spatial modes, and only the spatially uniform zero mode survives. So, the three horizontal lines in Fig.~\ref{fig:assembly}(b) make explicit what the modal analysis already shows analytically, \textit{i.e.} the fact that the assembly-scale NBC problem reduces to the first-order ODE $\dot{\Theta}=a_{\mathrm{cg}}+(b_{\mathrm{cg}}-1)\Theta$, whose solution is a pure exponential approach to $\Theta_{\mathrm{ss}}$.
Since the spatial profiles provide no additional information beyond the amplitude at each time, Fig.~\ref{fig:assembly}(c) presents the more informative comparison as a time profile. These plots make quantitative two features, which follow directly from the coarse-grained equation~\eqref{eq:HE_coarse_dimless}.

\paragraph{Asymptotes.}
Setting $\partial\Theta/\partial\tau_s=0$ in Eq.~\eqref{eq:HE_coarse_dimless} yields the steady-state ODE
\begin{equation}
  \zeta\Theta'' - k^2\zeta\,\Theta = -a_{\mathrm{cg}},
  \qquad k=\sqrt{(1-b_{\mathrm{cg}})/\zeta}.
  \label{eq:ss_ode}
\end{equation}
Under NBC the zero-mode condition $\Theta''=0$ gives the spatially uniform solution
$\Theta_{\mathrm{ss}}=a_{\mathrm{cg}}/(1-b_{\mathrm{cg}})=0.25$.
Under DBC the general even solution (the source and domain are both symmetric)
is $\Theta(\tilde\xi)=\Theta_{\mathrm{ss}}+C\cosh(k\tilde\xi)$,
where $\tilde\xi=\xi-\Lambda/2$.
Imposing $\Theta(\pm\Lambda/2)=0$ fixes $C=-\Theta_{\mathrm{ss}}/\cosh(k\Lambda/2)$,
so the midpoint value is
\begin{equation}
  \Theta_{\mathrm{ss}}^{\mathrm{DBC}}(\xi)
    = \Theta_{\mathrm{ss}}\!\left[1-\frac{1}{\cosh(k\Lambda/2)}\right].
  \label{eq:asymp_dbc}
\end{equation}
For the chosen parameters ($k\Lambda\approx 7.07$) this evaluates to
$\Theta_{\mathrm{ss}}^{\mathrm{DBC}}(\Lambda/2)\approx 0.235$,
lying $\sim$6\% below $\Theta_{\mathrm{ss}}^{\mathrm{NBC}}=0.25$,
a correction of order $e^{-k\Lambda/2}$ that vanishes for $k\Lambda\gg 1$.
Both asymptotes are therefore governed by the source--loss balance alone,
not by the boundary conditions.

\paragraph{Decay rates.}
Expanding $\Theta$ in the eigenbasis of $-\nabla^2$ (eigenvalues $\lambda_r$),
the linear operator $\zeta\nabla^2+(b_{\mathrm{cg}}-1)$ in Eq.~\eqref{eq:HE_coarse_dimless}
is diagonal with eigenvalues $-({\zeta\lambda_r+1-b_{\mathrm{cg}}})$,
so each modal amplitude decays as $e^{-\mu_r\tau}$ with the rate
\begin{equation}
  \mu_r = \zeta\lambda_r + (1-b_{\mathrm{cg}}).
  \label{eq:decay_rate}
\end{equation}
For NBC the only excited mode is the zero mode ($\lambda_0=0$), giving
$\mu_0^{\mathrm{NBC}}=1-b_{\mathrm{cg}}=2$.
For DBC the dominant late-time mode is the fundamental mode
($\lambda_1=\pi^2/\Lambda^2$), giving
$\mu_1^{\mathrm{DBC}}=\zeta(\pi/\Lambda)^2+(1-b_{\mathrm{cg}})\approx 2.39$.
Since $\zeta(\pi/\Lambda)^2>0$, the DBC midpoint always saturates \emph{faster}
than NBC by the diffusive excess $\zeta(\pi/\Lambda)^2$ [see Fig. \ref{fig:assembly}(c)], which encodes the
additional spatial relaxation of the DBC eigenfunction basis.
This faster DBC saturation is the assembly-scale counterpart
of the higher critical threshold
$|\tilde{b}_c^{(\mathrm{D})}|>|\tilde{b}_c^{(0)}|$
identified at the nanomagnet scale: the extra spatial mode
structure of the DBC solution provides an additional
dissipation channel absent under NBC.

\subsection{Two-scale consistency: nanoscale mean versus assembly-scale prediction}
\label{sec:results_twoscale}

\begin{figure*}[!ht]
  \centering
  \begin{subfigure}[b]{0.48\textwidth}
    \includegraphics[width=\textwidth]{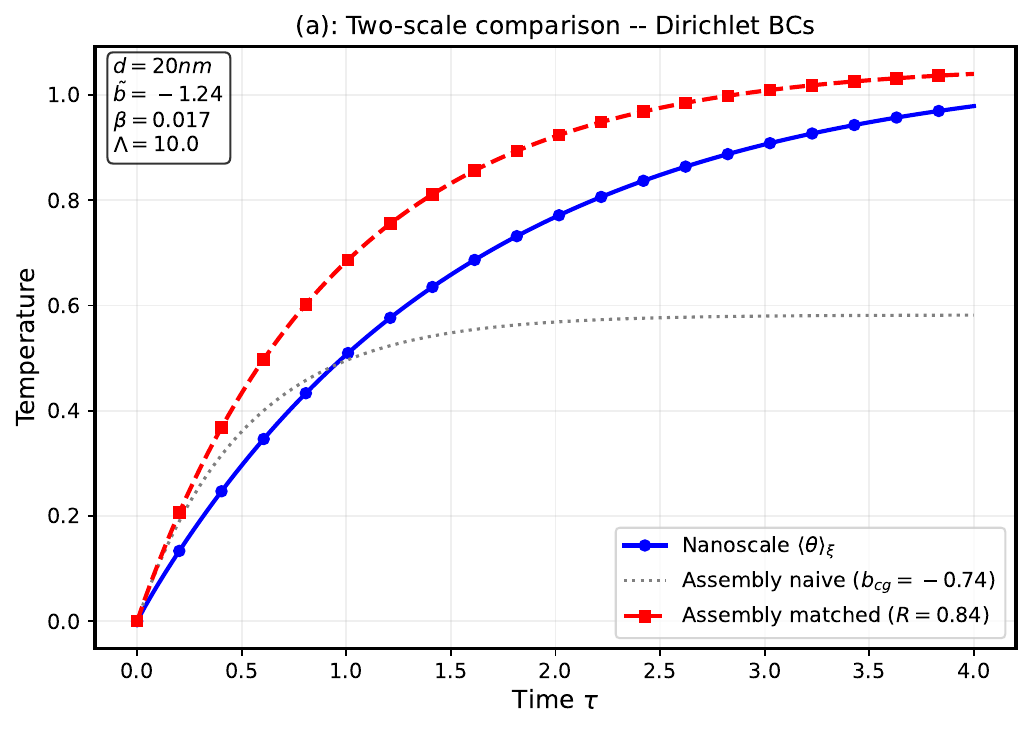}
  \end{subfigure}\hfill
  \begin{subfigure}[b]{0.48\textwidth}
    \includegraphics[width=\textwidth]{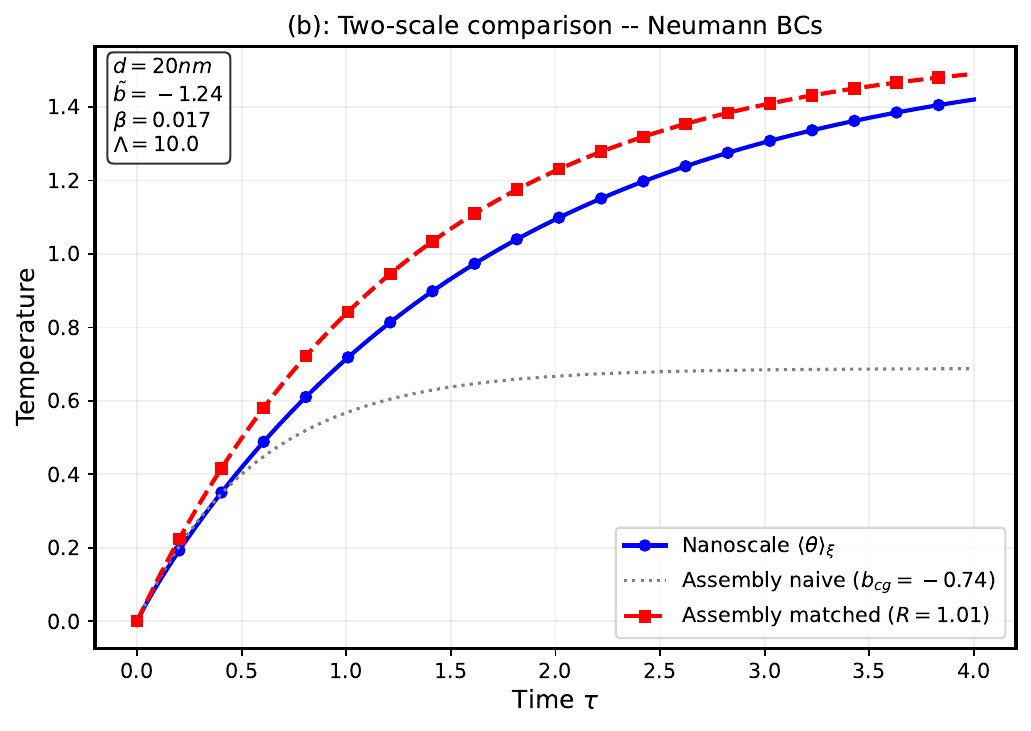}
  \end{subfigure}
  \caption{Nanoscale spatial mean $\langle\theta\rangle_\xi$ (blue solid)
           vs coarse-grained spatial mean $\langle\Theta\rangle_\xi$
           with naive density coefficients
           $b_{\mathrm{cg}}^{(0)} = \mathcal{N}\tilde{b}/\Lambda = -0.74$
           (gray dotted) and analytically matched coefficients
           (red dashed; $b_{\mathrm{cg}}^{\mathrm{DBC}}=+0.11$,
           $b_{\mathrm{cg}}^{\mathrm{NBC}}=+0.23$).
           Left: DBC. Right: NBC.
           \textsc{crossover} parameter set
           ($\tilde{b}=-1.24$, $\beta=0.017$, $\Lambda=10$, $\mathcal{N}=6$, $\Delta\xi=2$).}
  \label{fig:twoscale}
\end{figure*}

The two-scale framework we develop here rests on the assumption that the spatial mean of the nanoscale temperature field can be faithfully represented by the coarse-grained assembly-scale equation with effective coefficients $a_{\mathrm{cg}}$, $b_{\mathrm{cg}}$.
We test this assumption by comparing, in Figure~\ref{fig:twoscale}, the nanoscale spatial mean $\langle\theta\rangle_\xi(\tau)$ with the
coarse-grained spatial mean $\langle\Theta\rangle_\xi(\tau)$, for two choices of coefficients: the analytically matched coefficients derived
below (red dashed) and the naive density mapping included for comparison (gray dotted)
\footnote{The assembly-scale equation~\eqref{eq:HE_coarse_dimless} under NBC (spatially uniform zero mode) reads:
\begin{equation}
  \frac{d\Theta}{d\tau_s} = \acg - (1-\bcg)\Theta.
\end{equation}
Changing to nanoscale time via $d/d\tau = (t_d/t_s)\,d/d\tau_s$, and defining the effective coefficients to absorb the time-scale factor,
the (naive) coarse format in $\tau$ is:
\begin{equation}\label{eq:coarse_tau}
  \frac{d\Theta}{d\tau} = \acg - (1-\bcg)\Theta,
\end{equation}
where the coefficients $\acg$, $\bcg$ are understood as matched to nanoscale time.}.

\paragraph{Derivation of the effective coarse-grained coefficients.}
To obtain the coefficients that faithfully reproduce the nanoscale spatial mean, we integrate Eq.~\eqref{eq:HE_nm_final} over the domain
$[0,\Lambda]$.
Under NBC (where the boundary flux vanishes), this yields an exact ODE for the spatial mean $\langle\theta\rangle$,
\begin{equation}\label{eq:mean_theta_ode}
  \frac{d\langle\theta\rangle}{d\tau}
  = a_{\mathrm{cg}}^{(0)}
    - \bigl(\beta - R\,b_{\mathrm{cg}}^{(0)}\bigr)\,\langle\theta\rangle,
\end{equation}
where $a_{\mathrm{cg}}^{(0)} = \mathcal{N}\tilde{a}/\Lambda$ and
$b_{\mathrm{cg}}^{(0)} = \mathcal{N}\tilde{b}/\Lambda$ are the
source densities, and
$R = \langle\theta\rangle_{\mathrm{sites}}/\langle\theta\rangle_\xi = \left(\mathcal{N}^{-1}\sum_n\theta(\xi_n) \right)/\langle\theta\rangle_\xi$
is the source-site enhancement factor.
Under NBC the spatially uniform zero mode ($\lambda_0=0$) contributes
equally to $\langle\theta\rangle_{\mathrm{sites}}$ and
$\langle\theta\rangle_\xi$, so $R=1$ exactly in the zero-mode limit;
the correction $R-1$ arises from the higher cosine modes ($r\ge 1$),
which peak at the source positions but integrate to zero over
$[0,\Lambda]$, making $R$ only slightly above unity
($R_{\mathrm{NBC}}\approx 1.01$ for the crossover set).
$R$ is evaluated numerically from the steady-state modal solution
and can also be expressed analytically via the static Green's function
of the operator $-\partial_\xi^2+\beta$
(Appendix~\ref{app:GF_nanoscale}).
The two corrections encoded in Eq.~\eqref{eq:mean_theta_ode} relative
to a direct density substitution are: (i)~the feedback term
$\tilde{b}\,\theta(\xi_n,\tau)$ in the nanoscale equation is
evaluated at the source sites, not at the domain average, introducing
the factor $R$; and (ii)~the physical nanoscale loss rate is
$\beta \ll 1$, whereas the assembly-scale equation~\eqref{eq:HE_coarse_dimless}
normalises its loss coefficient to unity.
The true nanoscale effective loss rate is therefore
$\mu_{\mathrm{nano}} = \beta + R\,|b_{\mathrm{cg}}^{(0)}| \simeq 0.76$
for the crossover parameter set.

\paragraph{Matched coarse-grained coefficients.}
Matching the effective loss rate of Eq.~\eqref{eq:mean_theta_ode} to the coarse format $d\Theta/d\tau = a_{\mathrm{cg}} - (1-b_{\mathrm{cg}})\Theta$ yields the NBC correction
\begin{equation}\label{eq:bcg_nbc}
  b_{\mathrm{cg}}^{\mathrm{NBC}} = 1 - \beta + R\,b_{\mathrm{cg}}^{(0)},
\end{equation}
which simultaneously matches both the decay rate and the long-time asymptote of the spatial mean. For the crossover set: $R_{\mathrm{NBC}} \simeq 1.01$, giving $b_{\mathrm{cg}}^{\mathrm{NBC}} = +0.232$, with steady-state $\langle\Theta\rangle_\infty^{\mathrm{NBC}}
= a_{\mathrm{cg}}^{(0)}/(1-b_{\mathrm{cg}}^{\mathrm{NBC}}) = 1.56$.
By contrast, the naive density substitution $b_{\mathrm{cg}}^{(0)} = \mathcal{N}\tilde{b}/\Lambda = -0.744$ omits both corrections (i) and (ii) above, yielding an effective loss rate $1 - b_{\mathrm{cg}}^{(0)} = 1.74$, a factor $2.3$ larger than $\mu_{\mathrm{nano}}$, and should be avoided.

For DBC, the escape of heat through the boundaries introduces an additional effective loss channel absent from the NBC case. Thus, the coarse DBC spatial-mean steady state has the closed form
\begin{equation}\label{eq:Theta_ss_dbc_mean}
  \langle\Theta\rangle_\infty^{\mathrm{DBC}}
  = \frac{a_{\mathrm{cg}}}{1-b_{\mathrm{cg}}}
    \left[1 - \frac{\tanh(k\Lambda/2)}{k\Lambda/2}\right], \quad
  k = \sqrt{\frac{1-b_{\mathrm{cg}}}{\zeta}},
\end{equation}
and $b_{\mathrm{cg}}^{\mathrm{DBC}}$ is found by equating this expression to the exact nanoscale steady-state spatial mean $\langle\theta_{\mathrm{ss}}\rangle^{\mathrm{DBC}}$ (obtained from the modal linear solution $\boldsymbol{\theta}_{\mathrm{ss}} = -\mathbf{B}^{-1}\mathbf{d}$). For the crossover set,  $\langle\theta_{\mathrm{ss}}\rangle^{\mathrm{DBC}} = 1.06$,
giving $b_{\mathrm{cg}}^{\mathrm{DBC}} = +0.110$.

\paragraph{Numerical results.}
We have found that the matched coefficients reduce the RMS error between $\langle\theta\rangle_\xi$ and $\langle\Theta\rangle_\xi$ at $\tau \in [0, 4]$ by a factor of $\sim 4$ under NBC (from $0.46$ to $0.11$) and by a factor of $\sim 2$ under DBC  (from $0.24$ to $0.13$).
The residual discrepancy reflects multi-mode effects: the nanoscale solution involves many decay modes of $\mathbf{B}$ with different rates, while the coarse equation has a single effective mode. We believe that retaining higher modes should further reduce this residual discrepancy.

\subsection{Correlation length, interfacial coupling, and TC--DC interplay}
\label{sec:results_coupling}

To complete the picture of the crossover and its dependence on the two coupling channels, namely the interfacial thermal conductance $h_s$ (encoded in $\gamma_s$) and the dipolar coupling (encoded in $\tilde{\lambda}$), we have considered the following three complementary observables.

\begin{figure*}[!ht]
  \centering
  \begin{subfigure}[b]{0.48\textwidth}
    \includegraphics[width=\textwidth]{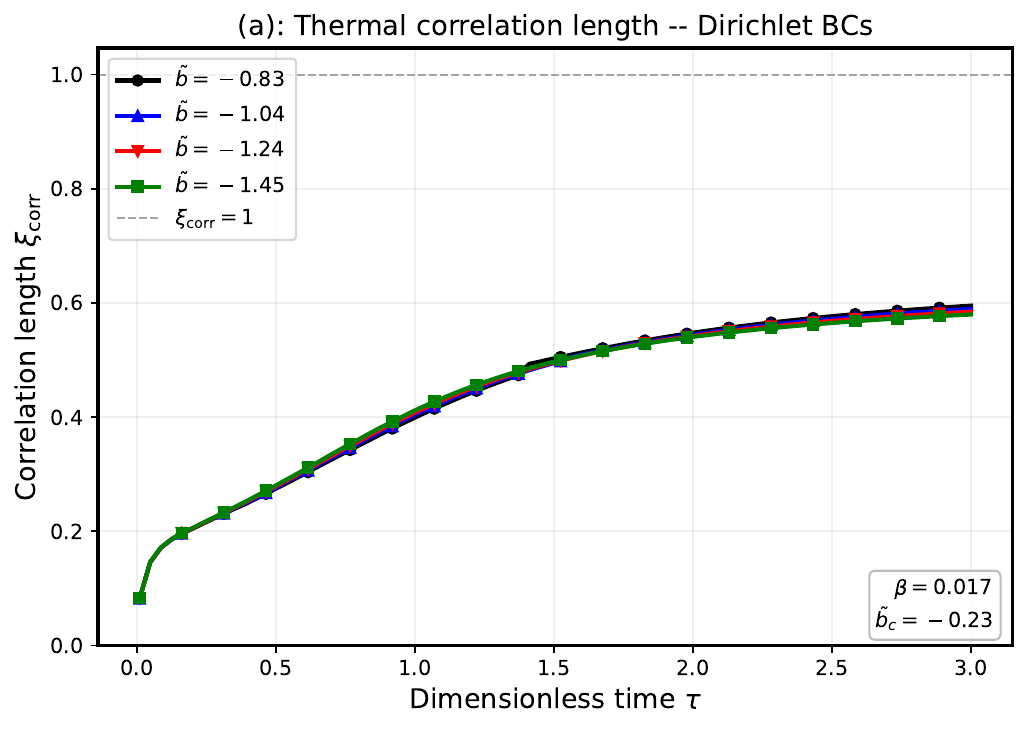}
  \end{subfigure}\hfill
  \begin{subfigure}[b]{0.48\textwidth}
    \includegraphics[width=\textwidth]{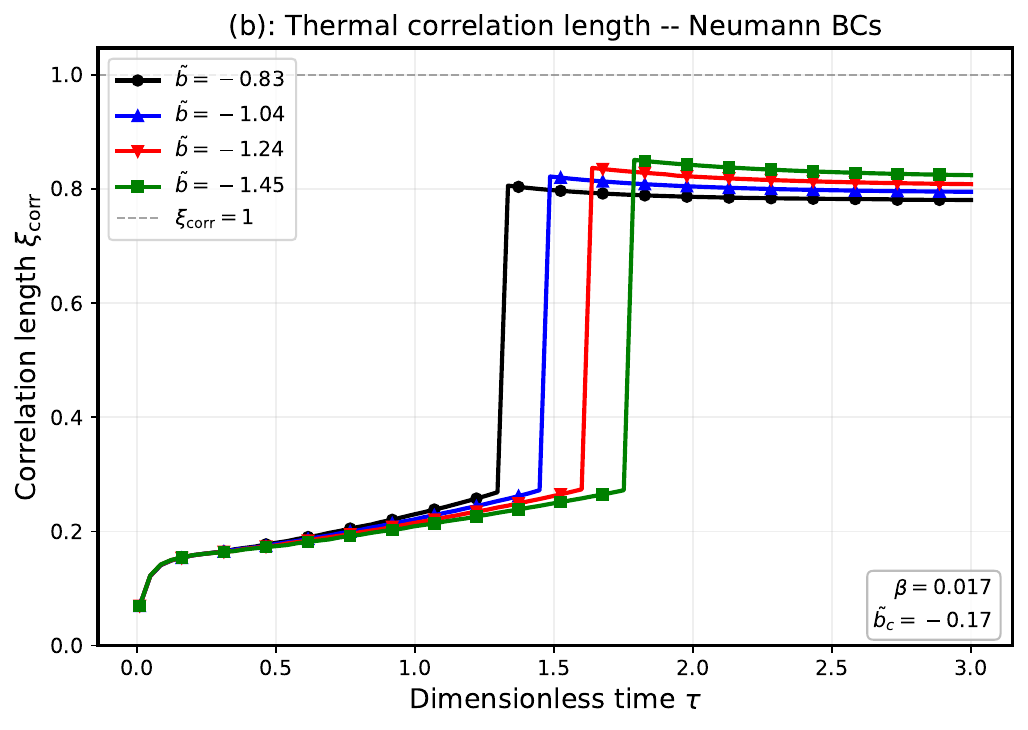}
  \end{subfigure}
  \caption{Thermal correlation length $\xi_{\mathrm{corr}}(\tau)$ for
           four values of $\tilde{b}$ bracketing $\tilde{b}_c$
           (\textsc{crossover} set: $\beta=0.017$, $\Lambda=10$, $\mathcal{N}=6$, $\Delta\xi=2$).
           DBC~(a): all curves grow slowly and remain $\lesssim 1$, with
           little sensitivity to $\tilde{b}$.
           NBC~(b): each curve undergoes a sharp jump at a time
           $\tau^*(\tilde{b})$ that decreases as $|\tilde{b}|\to|\tilde{b}_c|$,
           the direct signature of zero-mode dominance.}
  \label{fig:treat_corr_length}
\end{figure*}

\paragraph{Thermal correlation length.}
In Fig.~\ref{fig:treat_corr_length} we show the first-moment autocorrelation length $\xi_{\mathrm{corr}}(\tau)$ for four values of $\tilde{b}$
bracketing the critical value $\tilde{b}_c$, using the crossover parameter set ($\tilde{a}=2.0$, $\beta=0.017$, $\Lambda=10$, $\mathcal{N}=6$, $\Delta\xi=2$).
We see that under DBC, all curves grow slowly and remain near unity, since the boundary conditions suppress any qualitative dependence on $\tilde{b}$. On the other hand, under NBC, each curve undergoes a sharp jump to $\xi_{\mathrm{corr}}\approx\Lambda/2$ at a time $\tau^*$ that
decreases as $|\tilde{b}|$ approaches $|\tilde{b}_c|$: this is the direct signature of the crossover, \textit{i.e.} the spatially uniform
NBC zero mode abruptly dominates the profile.

\begin{figure*}[!ht]
  \centering
  \includegraphics[width=\textwidth]{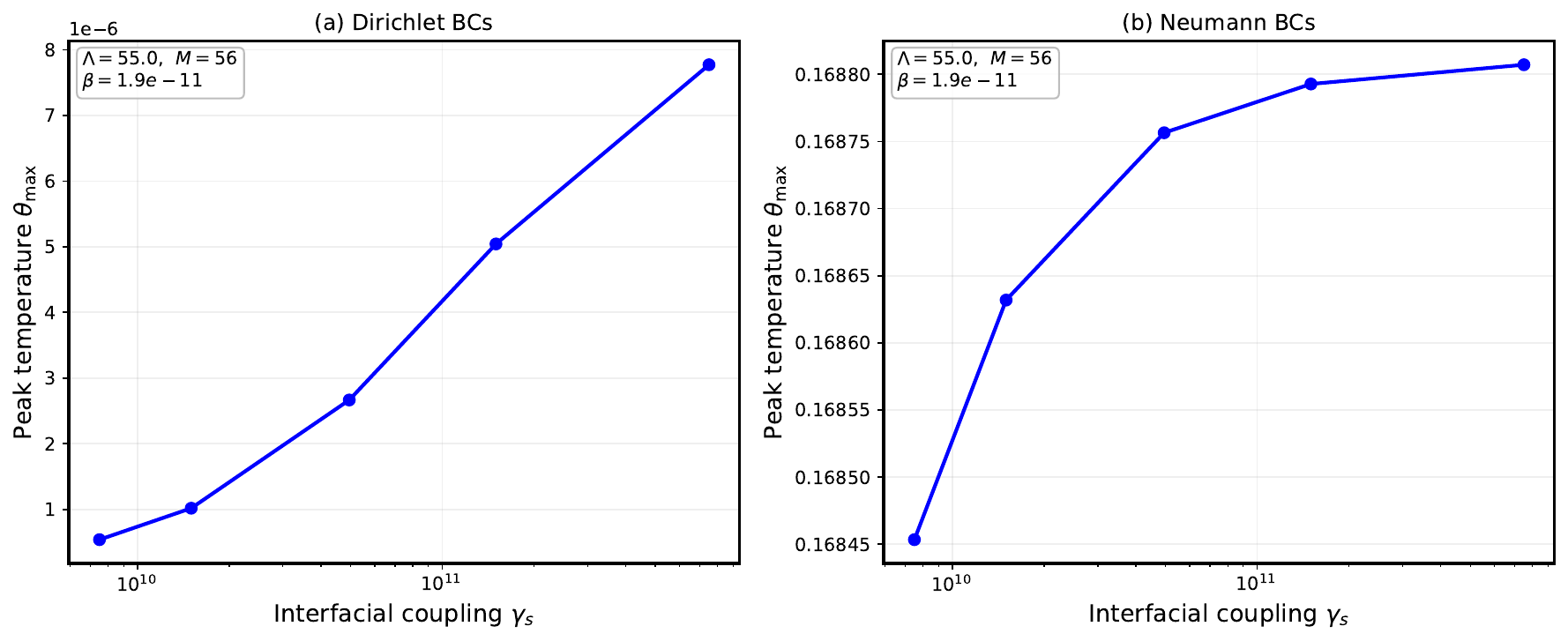}
  \caption{Peak steady-state temperature $\theta_{\max}$ vs.\
           dimensionless interfacial coupling $\gamma_s$ for
           (a)~DBC and (b)~NBC (physical magnetite--PMMA:
           $\Lambda=55$, $\mathcal{N}=56$, $\beta=1.9\times10^{-11}$).
           DBC: strong monotonic growth spanning nearly one decade.
           NBC: negligible variation
           ($\Delta\theta/\theta\sim2\times10^{-3}$) because the
           ratio $\tilde{a}/\tilde{b}$ is $\gamma_s$-independent.}
  \label{fig:treat_gamma_effect}
\end{figure*}

\paragraph{Sensitivity to the interfacial coupling $\gamma_s$.}
In Fig.~\ref{fig:treat_gamma_effect} we plot the peak steady-state temperature $\theta_{\max}$ as a function of $\gamma_s$ for both
boundary conditions, using the physical magnetite--PMMA parameter set ($\Lambda=55$, $\mathcal{N}=56$, $\beta=1.9\times10^{-11}$).
Under DBC, $\theta_{\max}$ rises by nearly one order of magnitude over two decades of $\gamma_s$, driven by the SAR renormalisation
$\tilde{a}\propto\gamma_s/(\gamma_s-\varepsilon)$.
Under NBC, $\theta_{\max}$ is essentially constant ($\Delta\theta/\theta\sim 2\times10^{-3}$): since the NBC steady state is set by the ratio $\tilde{a}/|\tilde{b}|$ and both coefficients scale identically with $\gamma_s$, the peak temperature is, to leading order, $\gamma_s$-independent.

\begin{figure*}[!ht]
  \centering
  \includegraphics[width=\textwidth]{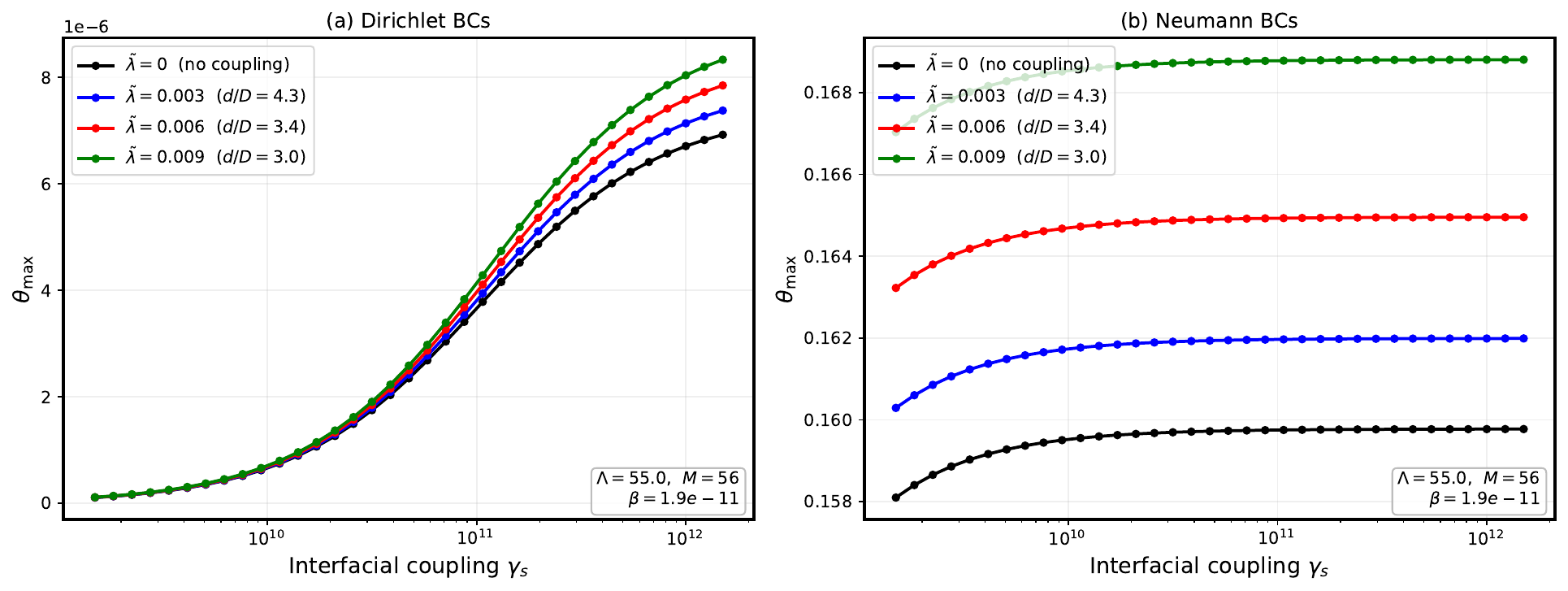}
  \caption{Peak steady-state temperature $\theta_{\max}$ vs.\ $\gamma_s$
           for four physically valid dipolar coupling values
           $\tilde{\lambda}\in\{0,\,0.003,\,0.006,\,0.009\}$
           ($d/D\in\{\infty,\,4.8,\,3.8,\,3.0\}$), for
           (a)~DBC and (b)~NBC.
           DBC: curves fan out with increasing $\gamma_s$ (non-trivial
           TC--DC interplay; enhancement $6\%\to18\%$).
           NBC: curves are parallel (multiplicative independence of the
           two coupling channels).}
  \label{fig:treat_tc_dc}
\end{figure*}

\paragraph{TC--DC interplay.}
Finally, figure~\ref{fig:treat_tc_dc} adds dipolar coupling $\tilde{\lambda}$ to the picture. The constraint $d\geq 3D$ (validity of the dipole approximation) restricts $\tilde{\lambda}$ to the range $[0,\,\tilde{\lambda}_{\mathrm{nat}}]\approx[0,\,0.009]$, achieved  at centre-to-centre separations $d/D\in[3.0,\infty)$.
We see that under DBC, the dipolar coupling enhances $\theta_{\max}$ and this enhancement is itself $\gamma_s$-dependent: it grows from
${\sim}\,6\%$ at small $\gamma_s$ to ${\sim}\,18\%$ at large $\gamma_s$, causing the four curves to fan out, and this is a non-trivial
interplay between the two coupling channels.
Under NBC, the ratio $\theta_{\max}(\tilde{\lambda})/\theta_{\max}(0)$ is constant to numerical precision across the full $\gamma_s$ sweep:
the two effects enter multiplicatively and independently, a direct consequence of zero-mode dominance.

\subsection{Summary of DBC versus NBC effects across all observables}
\label{sec:results_summary}

After this lengthy comparison of the two boundary conditions, we summarize in Table~\ref{tab:DBC_NBC_summary} the principal differences
between them across all observables presented above. The entries summarise the results of Figs.~\ref{fig:local_profiles_ill}--\ref{fig:treat_tc_dc} and provide a compact reference for the DBC/NBC dichotomy.

\begin{table}[H]
\centering
\small
\begin{tabular}{lcc}
\toprule
\textbf{Observable} & \textbf{DBC} & \textbf{NBC} \\
 & \textbf{(ideal bath)} & \textbf{(insulated)} \\
\midrule
Edge source $\theta$    & Suppressed               & Enhanced (reflected heat) \\
Absolute $\theta_{\max}$   & Lower                 & Higher \\
Spatial contrast        & Preserved                & Washed out by zero mode \\
Zero mode               & Absent                   & Present ($\lambda_0=0$) \\
$\tilde{b}_c$ threshold & $|\tilde{b}_c|$ larger   & $|\tilde{b}_c|$ smaller \\
$\mathrm{Var}/\langle\theta\rangle$ & $\mathcal{O}(1)$ & $\ll 1$ \\
Assembly profile        & Dome-shaped              & Perfectly flat \\
Steady-state scaling
  & $\theta_{\mathrm{ss}} \propto \Lambda^2$
  & \shortstack[c]{%
      $a_{\mathrm{cg}}\mathcal{N}/(\Lambda\beta)$\\
      {\footnotesize(loss-dominated)}\\[4pt]
      $|\tilde{a}|/|\tilde{b}|$\\
      {\footnotesize(feedback-dominated)}} \\
\bottomrule
\end{tabular}
\caption{Summary of the principal differences between Dirichlet and Neumann boundary conditions across all observables.}
\label{tab:DBC_NBC_summary}
\end{table}
\section{Conclusions}
\label{sec:conclusions}
We have developed a two-scale analytical framework, namely the Local-to-Global Heat Crossover, for studying the thermal physics of
one-dimensional chains of nanomagnets subjected to alternating magnetic fields.
This framework allows us to resolve individual nanomagnet sources as temperature-dependent heat sources and to connect the resulting nanoscale temperature field, through spatial and temporal coarse-graining, to the macroscopic heating observable in, \textit{e.g.}, magnetic hyperthermia applications.
We summarize our main findings as follows:
\paragraph{Nanomagnet-scale thermal structure.}
The dimensionless heat equation~\eqref{eq:HE_nm_final}, solved exactly via modal decomposition [Eq.~\eqref{eq:exact_modal_solution_full}], reveals that each nanomagnet generates a cusp-like temperature peak at its source position, with a spatial extent controlled by the thermal decay length $\ell_T = 1/\sqrt{\beta + |\tilde{b}|}$.
For realistic magnetite--PMMA parameters, the peak amplitudes are of order $10^{-8}$--$10^{-9}$ in dimensionless units (corresponding to
$\sim\!\mu$K temperature excursions), confirming that nanoscale thermal localisation, while mathematically exact, is physically unresolvable by any current experimental technique.
The source-localised structure is, however, well displayed using illustrative parameters elevated by 7--9 orders of magnitude
(Table~\ref{table:illustrative_params}), which preserve all qualitative features of the solution.

\paragraph{Role of boundary conditions.}
The extensive comparison of Dirichlet (ideal thermal bath) and Neumann (perfect insulation) boundary conditions reveals a fundamental
trade-off with direct consequences for applications. Dirichlet conditions preserve strong spatial heterogeneity: the relative variance $\mathcal{V}_{\mathrm{rel}}^{(\mathrm{D})}$ is of order unity, steady-state profiles are dome-shaped with a clear edge--bulk difference, and the temperature vanishes at the chain boundaries.
Neumann conditions, by contrast, activate a zero mode ($\lambda_0 = 0$) that accumulates heat globally, at the cost of washing out all spatial structure ($\mathcal{V}_{\mathrm{rel}}^{(\mathrm{N})} \ll\mathcal{V}_{\mathrm{rel}}^{(\mathrm{D})}$).
The critical feedback coefficient for the local-to-global crossover is smaller under NBC [$\tilde{b}_c^{(0)} = \beta\Lambda/\mathcal{N}$,
Eq.~\eqref{eq:bc_Neumann_main}] than under DBC [Eq.~\eqref{eq:bc_analytical}].
In fact, real nanomagnet assemblies will exhibit intermediate (Robin-type) boundary conditions, with the Biot number $\mathrm{Bi}_b$ interpolating between the two limits analysed here.

\paragraph{Local-to-global crossover.}
We have demonstrated that the analytical stability criterion~\eqref{eq:bc_analytical} provides a compact condition for the onset of collective heating: the renormalized feedback coefficient $\tilde{b}$ must be positive (\textit{i.e.}\ the bare SLP must be self-amplifying and the interfacial coupling strong enough to transmit the feedback to the matrix) and must exceed the critical threshold $\tilde{b}_c$, which depends on the eigenvalue spectrum, nanoscale losses, and source geometry.
In the physically prevalent regime ($\tilde{b} < 0$, self-limiting feedback), the system is unconditionally stable and heating remains localised at the nanomagnet sites; the transition to global heating requires engineering the system into the self-amplifying regime
(case~2 of Sec.~\ref{sec:stability_criterion}), for instance by tuning the nanoparticle diameter or field amplitude into a region
where $b_p > 0$.

\paragraph{Assembly-scale heating and two-scale consistency.}
The coarse-grained assembly-scale equation correctly captures the qualitative features of the macroscopic temperature field, such as the dome-shaped profiles under DBC, spatially uniform rise under NBC, and recovers the classical lumped-parameter hyperthermia model as a special case (NBC with uniform source).
We then show that the dominant quantitative mismatch between the naive coarse description and the full nanoscale computation originates from a time-scale normalisation: the assembly-scale loss coefficient in Eq.~\eqref{eq:HE_coarse_dimless} is set to unity, while the physical
nanoscale loss rate is $\beta \ll 1$.
Correcting this through the analytically matched coefficients of Eqs.~\eqref{eq:bcg_nbc}--\eqref{eq:Theta_ss_dbc_mean} allows us to reduce the
RMS error by a factor of $\sim 4$ for NBC and $\sim 2$ for DBC, attributing the residual to multi-mode dynamics that a single-effective-mode coarse description cannot capture.

\paragraph{Design implications.}
We believe that the present framework will help to identify several controllable design parameters for tuning the thermal response of nanomagnet assemblies:
(i)~the interparticle spacing $d$, which sets both the source density and the dimensionless thermal parameters via the scaling
$\beta \propto d^2$;
(ii)~the boundary thermal conductance, which interpolates between the DBC (maximum heterogeneity) and NBC (maximum temperature) limits;
(iii)~the interfacial coupling $\gamma_s$, whose interplay with the bare SLP feedback~$b_p$ determines the sign and magnitude of $\tilde{b}$ and thereby controls whether the local-to-global crossover is accessible; and
(iv)~the nanoscale loss coefficient $L_m$, which provides a dissipation channel that competes with the zero-mode accumulation under NBC.

For magnetic hyperthermia applications, our results suggest that thermally confined (Neumann-like) assemblies with dense packing maximise the absolute temperature rise, whereas well-thermalised (Dirichlet-like) assemblies with moderate packing preserve spatial selectivity.

The one-dimensional local-to-global heating crossover framework we have developed here can be extended in several directions.
First, the generalisation to two- and three-dimensional nanomagnet arrays is conceptually straightforward within the modal formalism, although the computational cost of the matrix exponential should increase.
Second, the linearized source term $\tilde{a} + \tilde{b}\,\theta$ can be replaced by the full nonlinear SLP dependence to capture saturation effects in the supercritical regime; we may proceed by solving the stochastic Landau-Lifshitz equation.
Third, the Robin boundary condition (combining Dirichlet and Neumann) can be treated via the transcendental eigenvalue equation, bridging the two limiting cases analysed here.
Finally, coupling this framework to realistic magnetic response models (including polydispersity, dipolar interactions beyond the perturbative approach, and Brown relaxation) would allow for a more direct comparison with experimental hyperthermia data, which naturally operate near the local-to-global crossover. Unfortunately, this particular extension can only be carried out using numerical simulations with a relatively high cost.


\begin{thebibliography}{66}%
\makeatletter
\providecommand \@ifxundefined [1]{%
 \@ifx{#1\undefined}
}%
\providecommand \@ifnum [1]{%
 \ifnum #1\expandafter \@firstoftwo
 \else \expandafter \@secondoftwo
 \fi
}%
\providecommand \@ifx [1]{%
 \ifx #1\expandafter \@firstoftwo
 \else \expandafter \@secondoftwo
 \fi
}%
\providecommand \natexlab [1]{#1}%
\providecommand \enquote  [1]{``#1''}%
\providecommand \bibnamefont  [1]{#1}%
\providecommand \bibfnamefont [1]{#1}%
\providecommand \citenamefont [1]{#1}%
\providecommand \href@noop [0]{\@secondoftwo}%
\providecommand \href [0]{\begingroup \@sanitize@url \@href}%
\providecommand \@href[1]{\@@startlink{#1}\@@href}%
\providecommand \@@href[1]{\endgroup#1\@@endlink}%
\providecommand \@sanitize@url [0]{\catcode `\\12\catcode `\$12\catcode
  `\&12\catcode `\#12\catcode `\^12\catcode `\_12\catcode `\%12\relax}%
\providecommand \@@startlink[1]{}%
\providecommand \@@endlink[0]{}%
\providecommand \url  [0]{\begingroup\@sanitize@url \@url }%
\providecommand \@url [1]{\endgroup\@href {#1}{\urlprefix }}%
\providecommand \urlprefix  [0]{URL }%
\providecommand \Eprint [0]{\href }%
\providecommand \doibase [0]{https://doi.org/}%
\providecommand \selectlanguage [0]{\@gobble}%
\providecommand \bibinfo  [0]{\@secondoftwo}%
\providecommand \bibfield  [0]{\@secondoftwo}%
\providecommand \translation [1]{[#1]}%
\providecommand \BibitemOpen [0]{}%
\providecommand \bibitemStop [0]{}%
\providecommand \bibitemNoStop [0]{.\EOS\space}%
\providecommand \EOS [0]{\spacefactor3000\relax}%
\providecommand \BibitemShut  [1]{\csname bibitem#1\endcsname}%
\let\auto@bib@innerbib\@empty
\bibitem [{\citenamefont {Pop}(2010)}]{Pop2010EnergyDissipation}%
  \BibitemOpen
  \bibfield  {author} {\bibinfo {author} {\bibfnamefont {E.}~\bibnamefont
  {Pop}},\ }\bibfield  {title} {\bibinfo {title} {{Energy dissipation and
  transport in nanoscale devices}},\ }\href
  {https://doi.org/10.1007/s12274-010-1019-z} {\bibfield  {journal} {\bibinfo
  {journal} {Nano Research}\ }\textbf {\bibinfo {volume} {3}},\ \bibinfo
  {pages} {147} (\bibinfo {year} {2010})}\BibitemShut {NoStop}%
\bibitem [{\citenamefont {Chen}(2005)}]{Chen2005}%
  \BibitemOpen
  \bibfield  {author} {\bibinfo {author} {\bibfnamefont {G.}~\bibnamefont
  {Chen}},\ }\href@noop {} {\emph {\bibinfo {title} {Nanoscale Energy Transport
  and Conversion: A Parallel Treatment of Electrons, Molecules, Phonons, and
  Photons}}}\ (\bibinfo  {publisher} {Oxford University Press},\ \bibinfo
  {address} {Oxford},\ \bibinfo {year} {2005})\BibitemShut {NoStop}%
\bibitem [{\citenamefont {Benenti}\ \emph {et~al.}(2023)\citenamefont
  {Benenti}, \citenamefont {Donadio}, \citenamefont {Lepri},\ and\
  \citenamefont {Livi}}]{benenti23rivi}%
  \BibitemOpen
  \bibfield  {author} {\bibinfo {author} {\bibfnamefont {G.}~\bibnamefont
  {Benenti}}, \bibinfo {author} {\bibfnamefont {D.}~\bibnamefont {Donadio}},
  \bibinfo {author} {\bibfnamefont {S.}~\bibnamefont {Lepri}},\ and\ \bibinfo
  {author} {\bibfnamefont {R.}~\bibnamefont {Livi}},\ }\bibfield  {title}
  {\bibinfo {title} {Non-fourier heat transport in nanosystems},\ }\href
  {https://doi.org/10.1007/s40766-023-00041-w} {\bibfield  {journal} {\bibinfo
  {journal} {La Rivista del Nuovo Cimento}\ }\textbf {\bibinfo {volume} {46}},\
  \bibinfo {pages} {105–161} (\bibinfo {year} {2023})}\BibitemShut {NoStop}%
\bibitem [{\citenamefont {Rosensweig}(2002)}]{rosensweig02j3m}%
  \BibitemOpen
  \bibfield  {author} {\bibinfo {author} {\bibfnamefont {R.~E.}\ \bibnamefont
  {Rosensweig}},\ }\bibfield  {title} {\bibinfo {title} {{Heating magnetic
  fluid with alternating magnetic field}},\ }\href
  {http://www.sciencedirect.com/science/article/pii/S0304885302007060}
  {\bibfield  {journal} {\bibinfo  {journal} {J. Magn. Magn. Mater.}\ }\textbf
  {\bibinfo {volume} {252}},\ \bibinfo {pages} {370} (\bibinfo {year}
  {2002})},\ \bibinfo {note} {proceedings of the 9th International Conference
  on Magnetic Fluids, 23-27 Jul. 2001}\BibitemShut {NoStop}%
\bibitem [{\citenamefont {Ortega}\ and\ \citenamefont
  {Pankhurst}(2013)}]{OrtegaPankhurst_nsc13}%
  \BibitemOpen
  \bibfield  {author} {\bibinfo {author} {\bibfnamefont {D.}~\bibnamefont
  {Ortega}}\ and\ \bibinfo {author} {\bibfnamefont {Q.}~\bibnamefont
  {Pankhurst}},\ }\bibinfo {title} {Magnetic hyperthermia}\ (\bibinfo {year}
  {2013})\ pp.\ \bibinfo {pages} {60--88}\BibitemShut {NoStop}%
\bibitem [{\citenamefont {{Hergt}}\ \emph {et~al.}(2006)\citenamefont
  {{Hergt}}, \citenamefont {{Dutz}}, \citenamefont {{M{\"u}ller}},\ and\
  \citenamefont {{Zeisberger}}}]{Hergt_etal_JPMC2006}%
  \BibitemOpen
  \bibfield  {author} {\bibinfo {author} {\bibfnamefont {R.}~\bibnamefont
  {{Hergt}}}, \bibinfo {author} {\bibfnamefont {S.}~\bibnamefont {{Dutz}}},
  \bibinfo {author} {\bibfnamefont {R.}~\bibnamefont {{M{\"u}ller}}},\ and\
  \bibinfo {author} {\bibfnamefont {M.}~\bibnamefont {{Zeisberger}}},\
  }\bibfield  {title} {\bibinfo {title} {{Magnetic particle hyperthermia:
  nanoparticle magnetism and materials development for cancer therapy}},\
  }\href {https://doi.org/10.1088/0953-8984/18/38/S26} {\bibfield  {journal}
  {\bibinfo  {journal} {J. Phys.: Condens. Mater.}\ }\textbf {\bibinfo {volume}
  {18}},\ \bibinfo {pages} {2919} (\bibinfo {year} {2006})}\BibitemShut
  {NoStop}%
\bibitem [{\citenamefont {Bell}(2003)}]{Bell2003}%
  \BibitemOpen
  \bibfield  {author} {\bibinfo {author} {\bibfnamefont {A.~T.}\ \bibnamefont
  {Bell}},\ }\bibfield  {title} {\bibinfo {title} {The impact of nanoscience on
  heterogeneous catalysis},\ }\href {https://doi.org/10.1126/science.1083671}
  {\bibfield  {journal} {\bibinfo  {journal} {Science}\ }\textbf {\bibinfo
  {volume} {299}},\ \bibinfo {pages} {1688} (\bibinfo {year}
  {2003})}\BibitemShut {NoStop}%
\bibitem [{\citenamefont {Haruta}(1997)}]{Haruta1997}%
  \BibitemOpen
  \bibfield  {author} {\bibinfo {author} {\bibfnamefont {M.}~\bibnamefont
  {Haruta}},\ }\bibfield  {title} {\bibinfo {title} {Size- and
  support-dependency in the catalytic activity of gold},\ }\href
  {https://doi.org/10.1016/S0920-5861(96)00208-8} {\bibfield  {journal}
  {\bibinfo  {journal} {Catalysis Today}\ }\textbf {\bibinfo {volume} {36}},\
  \bibinfo {pages} {153} (\bibinfo {year} {1997})}\BibitemShut {NoStop}%
\bibitem [{\citenamefont {N{\o}rskov}\ \emph {et~al.}(2009)\citenamefont
  {N{\o}rskov}, \citenamefont {Bligaard}, \citenamefont {Rossmeisl},\ and\
  \citenamefont {Christensen}}]{Norskov2009}%
  \BibitemOpen
  \bibfield  {author} {\bibinfo {author} {\bibfnamefont {J.~K.}\ \bibnamefont
  {N{\o}rskov}}, \bibinfo {author} {\bibfnamefont {T.}~\bibnamefont
  {Bligaard}}, \bibinfo {author} {\bibfnamefont {J.}~\bibnamefont
  {Rossmeisl}},\ and\ \bibinfo {author} {\bibfnamefont {C.~H.}\ \bibnamefont
  {Christensen}},\ }\bibfield  {title} {\bibinfo {title} {Towards the
  computational design of solid catalysts},\ }\href
  {https://doi.org/10.1038/nchem.121} {\bibfield  {journal} {\bibinfo
  {journal} {Nature Chemistry}\ }\textbf {\bibinfo {volume} {1}},\ \bibinfo
  {pages} {37} (\bibinfo {year} {2009})}\BibitemShut {NoStop}%
\bibitem [{\citenamefont {Bauer}\ \emph {et~al.}(2012)\citenamefont {Bauer},
  \citenamefont {Saitoh},\ and\ \citenamefont {van Wees}}]{Bauer2012}%
  \BibitemOpen
  \bibfield  {author} {\bibinfo {author} {\bibfnamefont {G.~E.~W.}\
  \bibnamefont {Bauer}}, \bibinfo {author} {\bibfnamefont {E.}~\bibnamefont
  {Saitoh}},\ and\ \bibinfo {author} {\bibfnamefont {B.~J.}\ \bibnamefont {van
  Wees}},\ }\bibfield  {title} {\bibinfo {title} {Spin caloritronics},\ }\href
  {https://doi.org/10.1038/nmat3301} {\bibfield  {journal} {\bibinfo  {journal}
  {Nature Materials}\ }\textbf {\bibinfo {volume} {11}},\ \bibinfo {pages}
  {391} (\bibinfo {year} {2012})}\BibitemShut {NoStop}%
\bibitem [{\citenamefont {Uchida}\ \emph {et~al.}(2008)\citenamefont {Uchida},
  \citenamefont {Takahashi}, \citenamefont {Harii}, \citenamefont {Ieda},
  \citenamefont {Koshibae}, \citenamefont {Ando}, \citenamefont {Maekawa},\
  and\ \citenamefont {Saitoh}}]{Uchida2008}%
  \BibitemOpen
  \bibfield  {author} {\bibinfo {author} {\bibfnamefont {K.}~\bibnamefont
  {Uchida}}, \bibinfo {author} {\bibfnamefont {S.}~\bibnamefont {Takahashi}},
  \bibinfo {author} {\bibfnamefont {K.}~\bibnamefont {Harii}}, \bibinfo
  {author} {\bibfnamefont {J.}~\bibnamefont {Ieda}}, \bibinfo {author}
  {\bibfnamefont {W.}~\bibnamefont {Koshibae}}, \bibinfo {author}
  {\bibfnamefont {K.}~\bibnamefont {Ando}}, \bibinfo {author} {\bibfnamefont
  {S.}~\bibnamefont {Maekawa}},\ and\ \bibinfo {author} {\bibfnamefont
  {E.}~\bibnamefont {Saitoh}},\ }\bibfield  {title} {\bibinfo {title}
  {Observation of the spin seebeck effect},\ }\href
  {https://doi.org/10.1038/nature07321} {\bibfield  {journal} {\bibinfo
  {journal} {Nature}\ }\textbf {\bibinfo {volume} {455}},\ \bibinfo {pages}
  {778} (\bibinfo {year} {2008})}\BibitemShut {NoStop}%
\bibitem [{\citenamefont {Adachi}\ \emph {et~al.}(2013)\citenamefont {Adachi},
  \citenamefont {Uchida}, \citenamefont {Saitoh},\ and\ \citenamefont
  {Maekawa}}]{Adachi2013}%
  \BibitemOpen
  \bibfield  {author} {\bibinfo {author} {\bibfnamefont {H.}~\bibnamefont
  {Adachi}}, \bibinfo {author} {\bibfnamefont {K.-i.}\ \bibnamefont {Uchida}},
  \bibinfo {author} {\bibfnamefont {E.}~\bibnamefont {Saitoh}},\ and\ \bibinfo
  {author} {\bibfnamefont {S.}~\bibnamefont {Maekawa}},\ }\bibfield  {title}
  {\bibinfo {title} {Theory of the spin seebeck effect},\ }\href
  {https://doi.org/10.1088/0034-4885/76/3/036501} {\bibfield  {journal}
  {\bibinfo  {journal} {Reports on Progress in Physics}\ }\textbf {\bibinfo
  {volume} {76}},\ \bibinfo {pages} {036501} (\bibinfo {year}
  {2013})}\BibitemShut {NoStop}%
\bibitem [{\citenamefont {El~Baraji}\ \emph {et~al.}(2009)\citenamefont
  {El~Baraji}, \citenamefont {Javerliac}, \citenamefont {Guo}, \citenamefont
  {Prenat},\ and\ \citenamefont {Dieny}}]{ElBarajietal2009}%
  \BibitemOpen
  \bibfield  {author} {\bibinfo {author} {\bibfnamefont {M.}~\bibnamefont
  {El~Baraji}}, \bibinfo {author} {\bibfnamefont {V.}~\bibnamefont
  {Javerliac}}, \bibinfo {author} {\bibfnamefont {W.}~\bibnamefont {Guo}},
  \bibinfo {author} {\bibfnamefont {G.}~\bibnamefont {Prenat}},\ and\ \bibinfo
  {author} {\bibfnamefont {B.}~\bibnamefont {Dieny}},\ }\bibfield  {title}
  {\bibinfo {title} {Dynamic compact model of thermally assisted switching
  magnetic tunnel junctions},\ }\bibfield  {journal} {\bibinfo  {journal}
  {Journal of Applied Physics}\ }\textbf {\bibinfo {volume} {106}},\ \href
  {https://doi.org/10.1063/1.3259373} {10.1063/1.3259373} (\bibinfo {year}
  {2009})\BibitemShut {NoStop}%
\bibitem [{\citenamefont {Herzog}\ \emph {et~al.}(2010)\citenamefont {Herzog},
  \citenamefont {Krause},\ and\ \citenamefont {Wiesendanger}}]{Herzogetal2010}%
  \BibitemOpen
  \bibfield  {author} {\bibinfo {author} {\bibfnamefont {G.}~\bibnamefont
  {Herzog}}, \bibinfo {author} {\bibfnamefont {S.}~\bibnamefont {Krause}},\
  and\ \bibinfo {author} {\bibfnamefont {R.}~\bibnamefont {Wiesendanger}},\
  }\bibfield  {title} {\bibinfo {title} {Heat assisted spin torque switching of
  quasistable nanomagnets across a vacuum gap},\ }\bibfield  {journal}
  {\bibinfo  {journal} {Applied Physics Letters}\ }\textbf {\bibinfo {volume}
  {96}},\ \href {https://doi.org/10.1063/1.3354023} {10.1063/1.3354023}
  (\bibinfo {year} {2010})\BibitemShut {NoStop}%
\bibitem [{\citenamefont {Granitzka}\ \emph {et~al.}(2017)\citenamefont
  {Granitzka}, \citenamefont {Jal}, \citenamefont {Guyader}, \citenamefont
  {Savoini}, \citenamefont {Higley}, \citenamefont {Liu}, \citenamefont {Chen},
  \citenamefont {Chase}, \citenamefont {Ohldag}, \citenamefont {Dakovsky},
  \citenamefont {Schlotter}, \citenamefont {Carron}, \citenamefont {Hoffman},
  \citenamefont {Shafer}, \citenamefont {Arenholz}, \citenamefont {Hellwig},
  \citenamefont {Mehta}, \citenamefont {Takahashi}, \citenamefont {Wang},
  \citenamefont {Fullerton}, \citenamefont {Stöhr}, \citenamefont {Reid},\
  and\ \citenamefont {Dürr}}]{Granitzkaetal2017}%
  \BibitemOpen
  \bibfield  {author} {\bibinfo {author} {\bibfnamefont {P.~W.}\ \bibnamefont
  {Granitzka}}, \bibinfo {author} {\bibfnamefont {E.}~\bibnamefont {Jal}},
  \bibinfo {author} {\bibfnamefont {L.~L.}\ \bibnamefont {Guyader}}, \bibinfo
  {author} {\bibfnamefont {M.}~\bibnamefont {Savoini}}, \bibinfo {author}
  {\bibfnamefont {D.~J.}\ \bibnamefont {Higley}}, \bibinfo {author}
  {\bibfnamefont {T.}~\bibnamefont {Liu}}, \bibinfo {author} {\bibfnamefont
  {Z.}~\bibnamefont {Chen}}, \bibinfo {author} {\bibfnamefont {T.}~\bibnamefont
  {Chase}}, \bibinfo {author} {\bibfnamefont {H.}~\bibnamefont {Ohldag}},
  \bibinfo {author} {\bibfnamefont {G.~L.}\ \bibnamefont {Dakovsky}}, \bibinfo
  {author} {\bibfnamefont {W.}~\bibnamefont {Schlotter}}, \bibinfo {author}
  {\bibfnamefont {S.}~\bibnamefont {Carron}}, \bibinfo {author} {\bibfnamefont
  {M.}~\bibnamefont {Hoffman}}, \bibinfo {author} {\bibfnamefont
  {P.}~\bibnamefont {Shafer}}, \bibinfo {author} {\bibfnamefont
  {E.}~\bibnamefont {Arenholz}}, \bibinfo {author} {\bibfnamefont
  {O.}~\bibnamefont {Hellwig}}, \bibinfo {author} {\bibfnamefont
  {V.}~\bibnamefont {Mehta}}, \bibinfo {author} {\bibfnamefont {Y.~K.}\
  \bibnamefont {Takahashi}}, \bibinfo {author} {\bibfnamefont {J.}~\bibnamefont
  {Wang}}, \bibinfo {author} {\bibfnamefont {E.~E.}\ \bibnamefont {Fullerton}},
  \bibinfo {author} {\bibfnamefont {J.}~\bibnamefont {Stöhr}}, \bibinfo
  {author} {\bibfnamefont {A.~H.}\ \bibnamefont {Reid}},\ and\ \bibinfo
  {author} {\bibfnamefont {H.~A.}\ \bibnamefont {Dürr}},\ }\bibfield  {title}
  {\bibinfo {title} {Magnetic switching in granular fept layers promoted by
  near-field laser enhancement},\ }\bibfield  {journal} {\bibinfo  {journal}
  {Nano Letters}\ }\href {https://doi.org/10.1021/acs.nanolett.7b00052}
  {10.1021/acs.nanolett.7b00052} (\bibinfo {year} {2017})\BibitemShut {NoStop}%
\bibitem [{\citenamefont {{Riedinger, A and Guardia, P and Curcio, A and
  Garcia, MA and Cingolani, R and Manna, L and Pellegrino,
  T}}(2013)}]{riedinger2013subnanometer}%
  \BibitemOpen
  \bibfield  {author} {\bibinfo {author} {\bibnamefont {{Riedinger, A and
  Guardia, P and Curcio, A and Garcia, MA and Cingolani, R and Manna, L and
  Pellegrino, T}}},\ }\bibfield  {title} {\bibinfo {title} {{Subnanometer local
  temperature probing and remotely controlled drug release based on
  azo-functionalized iron oxide nanoparticles}},\ }\href@noop {} {\bibfield
  {journal} {\bibinfo  {journal} {Nano letters}\ }\textbf {\bibinfo {volume}
  {13}},\ \bibinfo {pages} {2399 } (\bibinfo {year} {2013})}\BibitemShut
  {NoStop}%
\bibitem [{\citenamefont {Dong}\ and\ \citenamefont
  {Zink}(2014)}]{DongZink2014}%
  \BibitemOpen
  \bibfield  {author} {\bibinfo {author} {\bibfnamefont {J.}~\bibnamefont
  {Dong}}\ and\ \bibinfo {author} {\bibfnamefont {J.~I.}\ \bibnamefont
  {Zink}},\ }\bibfield  {title} {\bibinfo {title} {Taking the temperature of
  the interiors of magnetically heated nanoparticles},\ }\href
  {https://doi.org/10.1021/nn501250e} {\bibfield  {journal} {\bibinfo
  {journal} {ACS Nano}\ }\textbf {\bibinfo {volume} {8}},\ \bibinfo {pages}
  {5199} (\bibinfo {year} {2014})}\BibitemShut {NoStop}%
\bibitem [{\citenamefont {Serantes}\ \emph {et~al.}(2020)\citenamefont
  {Serantes}, \citenamefont {Nieves}, \citenamefont {Chubykalo-Fesenko},
  \citenamefont {Ruta}, \citenamefont {Chantrell},\ and\ \citenamefont
  {Hovorka}}]{serantes2020local}%
  \BibitemOpen
  \bibfield  {author} {\bibinfo {author} {\bibfnamefont {D.}~\bibnamefont
  {Serantes}}, \bibinfo {author} {\bibfnamefont {P.}~\bibnamefont {Nieves}},
  \bibinfo {author} {\bibfnamefont {O.}~\bibnamefont {Chubykalo-Fesenko}},
  \bibinfo {author} {\bibfnamefont {S.}~\bibnamefont {Ruta}}, \bibinfo {author}
  {\bibfnamefont {R.}~\bibnamefont {Chantrell}},\ and\ \bibinfo {author}
  {\bibfnamefont {O.}~\bibnamefont {Hovorka}},\ }\bibfield  {title} {\bibinfo
  {title} {Local heat release in magnetic nanoparticle hyperthermia: When
  individual particle hysteresis loops do not represent the local heating},\
  }\href {https://doi.org/10.1103/PhysRevB.102.214412} {\bibfield  {journal}
  {\bibinfo  {journal} {Physical Review B}\ }\textbf {\bibinfo {volume}
  {102}},\ \bibinfo {pages} {214412} (\bibinfo {year} {2020})}\BibitemShut
  {NoStop}%
\bibitem [{\citenamefont {Weaver}\ \emph {et~al.}(2009)\citenamefont {Weaver},
  \citenamefont {Rauwerdink},\ and\ \citenamefont {Hansen}}]{Weaveretal2009}%
  \BibitemOpen
  \bibfield  {author} {\bibinfo {author} {\bibfnamefont {J.~B.}\ \bibnamefont
  {Weaver}}, \bibinfo {author} {\bibfnamefont {A.~M.}\ \bibnamefont
  {Rauwerdink}},\ and\ \bibinfo {author} {\bibfnamefont {E.~W.}\ \bibnamefont
  {Hansen}},\ }\bibfield  {title} {\bibinfo {title} {Magnetic nanoparticle
  temperature estimation},\ }\href {https://doi.org/10.1118/1.3106342}
  {\bibfield  {journal} {\bibinfo  {journal} {Medical Physics}\ }\textbf
  {\bibinfo {volume} {36}},\ \bibinfo {pages} {1822} (\bibinfo {year}
  {2009})}\BibitemShut {NoStop}%
\bibitem [{\citenamefont {Tan}\ \emph {et~al.}(2014)\citenamefont {Tan},
  \citenamefont {Carrey},\ and\ \citenamefont {Respaud}}]{tan14prb}%
  \BibitemOpen
  \bibfield  {author} {\bibinfo {author} {\bibfnamefont {R.~P.}\ \bibnamefont
  {Tan}}, \bibinfo {author} {\bibfnamefont {J.}~\bibnamefont {Carrey}},\ and\
  \bibinfo {author} {\bibfnamefont {M.}~\bibnamefont {Respaud}},\ }\bibfield
  {title} {\bibinfo {title} {Magnetic hyperthermia properties of nanoparticles
  inside lysosomes using kinetic monte carlo simulations: Influence of key
  parameters and dipolar interactions, and evidence for strong spatial
  variation of heating power},\ }\bibfield  {journal} {\bibinfo  {journal}
  {Physical Review B}\ }\textbf {\bibinfo {volume} {90}},\ \href
  {https://doi.org/10.1103/physrevb.90.214421} {10.1103/physrevb.90.214421}
  (\bibinfo {year} {2014})\BibitemShut {NoStop}%
\bibitem [{\citenamefont {{I. Astefanoaei and A.
  Stancu}}(2019)}]{AstAle_jap19}%
  \BibitemOpen
  \bibfield  {author} {\bibinfo {author} {\bibnamefont {{I. Astefanoaei and A.
  Stancu}}},\ }\bibfield  {title} {\bibinfo {title} {{A computational study of
  the bioheat transfer in magnetic hyperthermia cancer therapy}},\ }\href@noop
  {} {\bibfield  {journal} {\bibinfo  {journal} {J. Appl. Phys.}\ }\textbf
  {\bibinfo {volume} {125}},\ \bibinfo {pages} {194701} (\bibinfo {year}
  {2019})}\BibitemShut {NoStop}%
\bibitem [{\citenamefont {Gu}\ \emph {et~al.}(2023)\citenamefont {Gu},
  \citenamefont {Pi{\~n}ol}, \citenamefont {Moreno-Loshuertos}, \citenamefont
  {Brites}, \citenamefont {Zeler}, \citenamefont {Mart{\'i}nez}, \citenamefont
  {Maurin-Pasturel}, \citenamefont {Fern{\'a}ndez-Silva}, \citenamefont
  {Marco-Brualla}, \citenamefont {T{\'e}llez}, \citenamefont {Cases},
  \citenamefont {Belsu{\'e}}, \citenamefont {Bonvin}, \citenamefont {Carlos},\
  and\ \citenamefont {Mill{\'a}n}}]{Guetal2023}%
  \BibitemOpen
  \bibfield  {author} {\bibinfo {author} {\bibfnamefont {Y.}~\bibnamefont
  {Gu}}, \bibinfo {author} {\bibfnamefont {R.}~\bibnamefont {Pi{\~n}ol}},
  \bibinfo {author} {\bibfnamefont {R.}~\bibnamefont {Moreno-Loshuertos}},
  \bibinfo {author} {\bibfnamefont {C.~D.~S.}\ \bibnamefont {Brites}}, \bibinfo
  {author} {\bibfnamefont {J.}~\bibnamefont {Zeler}}, \bibinfo {author}
  {\bibfnamefont {A.}~\bibnamefont {Mart{\'i}nez}}, \bibinfo {author}
  {\bibfnamefont {G.}~\bibnamefont {Maurin-Pasturel}}, \bibinfo {author}
  {\bibfnamefont {P.}~\bibnamefont {Fern{\'a}ndez-Silva}}, \bibinfo {author}
  {\bibfnamefont {J.}~\bibnamefont {Marco-Brualla}}, \bibinfo {author}
  {\bibfnamefont {P.}~\bibnamefont {T{\'e}llez}}, \bibinfo {author}
  {\bibfnamefont {R.}~\bibnamefont {Cases}}, \bibinfo {author} {\bibfnamefont
  {R.~N.}\ \bibnamefont {Belsu{\'e}}}, \bibinfo {author} {\bibfnamefont
  {D.}~\bibnamefont {Bonvin}}, \bibinfo {author} {\bibfnamefont {L.~D.}\
  \bibnamefont {Carlos}},\ and\ \bibinfo {author} {\bibfnamefont
  {A.}~\bibnamefont {Mill{\'a}n}},\ }\bibfield  {title} {\bibinfo {title}
  {Local temperature increments and induced cell death in intracellular
  magnetic hyperthermia},\ }\href {https://doi.org/10.1021/acsnano.3c00388}
  {\bibfield  {journal} {\bibinfo  {journal} {ACS Nano}\ }\textbf {\bibinfo
  {volume} {17}},\ \bibinfo {pages} {6822} (\bibinfo {year}
  {2023})}\BibitemShut {NoStop}%
\bibitem [{\citenamefont {Mai}\ \emph {et~al.}(2019)\citenamefont {Mai},
  \citenamefont {Balakrishnan}, \citenamefont {Barthel}, \citenamefont
  {Piccardi}, \citenamefont {Niculaes}, \citenamefont {Marinaro}, \citenamefont
  {Fernandes}, \citenamefont {Curcio}, \citenamefont {Kakwere}, \citenamefont
  {Autret}, \citenamefont {Cingolani}, \citenamefont {Gazeau},\ and\
  \citenamefont {Pellegrino}}]{mai19acs}%
  \BibitemOpen
  \bibfield  {author} {\bibinfo {author} {\bibfnamefont {B.~T.}\ \bibnamefont
  {Mai}}, \bibinfo {author} {\bibfnamefont {P.~B.}\ \bibnamefont
  {Balakrishnan}}, \bibinfo {author} {\bibfnamefont {M.~J.}\ \bibnamefont
  {Barthel}}, \bibinfo {author} {\bibfnamefont {F.}~\bibnamefont {Piccardi}},
  \bibinfo {author} {\bibfnamefont {D.}~\bibnamefont {Niculaes}}, \bibinfo
  {author} {\bibfnamefont {F.}~\bibnamefont {Marinaro}}, \bibinfo {author}
  {\bibfnamefont {S.}~\bibnamefont {Fernandes}}, \bibinfo {author}
  {\bibfnamefont {A.}~\bibnamefont {Curcio}}, \bibinfo {author} {\bibfnamefont
  {H.}~\bibnamefont {Kakwere}}, \bibinfo {author} {\bibfnamefont
  {G.}~\bibnamefont {Autret}}, \bibinfo {author} {\bibfnamefont
  {R.}~\bibnamefont {Cingolani}}, \bibinfo {author} {\bibfnamefont
  {F.}~\bibnamefont {Gazeau}},\ and\ \bibinfo {author} {\bibfnamefont
  {T.}~\bibnamefont {Pellegrino}},\ }\bibfield  {title} {\bibinfo {title}
  {Thermoresponsive iron oxide nanocubes for an effective clinical translation
  of magnetic hyperthermia and heat-mediated chemotherapy},\ }\href
  {https://doi.org/10.1021/acsami.8b16226} {\bibfield  {journal} {\bibinfo
  {journal} {ACS Applied Materials \& Interfaces}\ }\textbf {\bibinfo {volume}
  {11}},\ \bibinfo {pages} {5727–5739} (\bibinfo {year} {2019})}\BibitemShut
  {NoStop}%
\bibitem [{\citenamefont {Mehdaoui}\ \emph {et~al.}(2013)\citenamefont
  {Mehdaoui}, \citenamefont {Tan}, \citenamefont {Meffre}, \citenamefont
  {Carrey}, \citenamefont {Lachaize}, \citenamefont {Chaudret},\ and\
  \citenamefont {Respaud}}]{Mehdaoui_prb2013}%
  \BibitemOpen
  \bibfield  {author} {\bibinfo {author} {\bibfnamefont {B.}~\bibnamefont
  {Mehdaoui}}, \bibinfo {author} {\bibfnamefont {R.~P.}\ \bibnamefont {Tan}},
  \bibinfo {author} {\bibfnamefont {A.}~\bibnamefont {Meffre}}, \bibinfo
  {author} {\bibfnamefont {J.}~\bibnamefont {Carrey}}, \bibinfo {author}
  {\bibfnamefont {S.}~\bibnamefont {Lachaize}}, \bibinfo {author}
  {\bibfnamefont {B.}~\bibnamefont {Chaudret}},\ and\ \bibinfo {author}
  {\bibfnamefont {M.}~\bibnamefont {Respaud}},\ }\bibfield  {title} {\bibinfo
  {title} {Increase of magnetic hyperthermia efficiency due to dipolar
  interactions in low-anisotropy magnetic nanoparticles: Theoretical and
  experimental results},\ }\href {https://doi.org/10.1103/PhysRevB.87.174419}
  {\bibfield  {journal} {\bibinfo  {journal} {Phys. Rev. B}\ }\textbf {\bibinfo
  {volume} {87}},\ \bibinfo {pages} {174419} (\bibinfo {year}
  {2013})}\BibitemShut {NoStop}%
\bibitem [{\citenamefont {Déjardin}\ \emph {et~al.}(2017)\citenamefont
  {Déjardin}, \citenamefont {Vernay}, \citenamefont {Respaud},\ and\
  \citenamefont {Kachkachi}}]{dejardin17jap}%
  \BibitemOpen
  \bibfield  {author} {\bibinfo {author} {\bibfnamefont {J.-L.}\ \bibnamefont
  {Déjardin}}, \bibinfo {author} {\bibfnamefont {F.}~\bibnamefont {Vernay}},
  \bibinfo {author} {\bibfnamefont {M.}~\bibnamefont {Respaud}},\ and\ \bibinfo
  {author} {\bibfnamefont {H.}~\bibnamefont {Kachkachi}},\ }\bibfield  {title}
  {\bibinfo {title} {Effect of dipolar interactions and dc magnetic field on
  the specific absorption rate of an array of magnetic nanoparticles},\
  }\bibfield  {journal} {\bibinfo  {journal} {Journal of Applied Physics}\
  }\textbf {\bibinfo {volume} {121}},\ \href
  {https://doi.org/10.1063/1.4984013} {10.1063/1.4984013} (\bibinfo {year}
  {2017})\BibitemShut {NoStop}%
\bibitem [{\citenamefont {Boucher}\ \emph {et~al.}(2011)\citenamefont
  {Boucher}, \citenamefont {Lacroix}, \citenamefont {Carignan}, \citenamefont
  {Yelon},\ and\ \citenamefont {M{\'e}nard}}]{boucheretal11apl}%
  \BibitemOpen
  \bibfield  {author} {\bibinfo {author} {\bibfnamefont {V.}~\bibnamefont
  {Boucher}}, \bibinfo {author} {\bibfnamefont {C.}~\bibnamefont {Lacroix}},
  \bibinfo {author} {\bibfnamefont {L.~P.}\ \bibnamefont {Carignan}}, \bibinfo
  {author} {\bibfnamefont {A.}~\bibnamefont {Yelon}},\ and\ \bibinfo {author}
  {\bibfnamefont {D.}~\bibnamefont {M{\'e}nard}},\ }\href@noop {} {\bibfield
  {journal} {\bibinfo  {journal} {Appl. Phys. Lett}\ }\textbf {\bibinfo
  {volume} {98}},\ \bibinfo {pages} {112502} (\bibinfo {year}
  {2011})}\BibitemShut {NoStop}%
\bibitem [{\citenamefont {Ilg}\ and\ \citenamefont
  {Kröger}(2020)}]{ilg20phys}%
  \BibitemOpen
  \bibfield  {author} {\bibinfo {author} {\bibfnamefont {P.}~\bibnamefont
  {Ilg}}\ and\ \bibinfo {author} {\bibfnamefont {M.}~\bibnamefont {Kröger}},\
  }\bibfield  {title} {\bibinfo {title} {Dynamics of interacting magnetic
  nanoparticles: effective behavior from competition between brownian and néel
  relaxation},\ }\href {https://doi.org/10.1039/d0cp04377j} {\bibfield
  {journal} {\bibinfo  {journal} {Physical Chemistry Chemical Physics}\
  }\textbf {\bibinfo {volume} {22}},\ \bibinfo {pages} {22244–22259}
  (\bibinfo {year} {2020})}\BibitemShut {NoStop}%
\bibitem [{\citenamefont {Figueiredo}\ and\ \citenamefont
  {Schwarzacher}(2007)}]{figueiredo07jpcm}%
  \BibitemOpen
  \bibfield  {author} {\bibinfo {author} {\bibfnamefont {W.}~\bibnamefont
  {Figueiredo}}\ and\ \bibinfo {author} {\bibfnamefont {W.}~\bibnamefont
  {Schwarzacher}},\ }\bibfield  {title} {\bibinfo {title} {Magnetic relaxation
  and thermal properties of a two-dimensional array of dipolar-coupled
  nanoparticles},\ }\href {https://doi.org/10.1088/0953-8984/19/27/276203}
  {\bibfield  {journal} {\bibinfo  {journal} {Journal of Physics: Condensed
  Matter}\ }\textbf {\bibinfo {volume} {19}},\ \bibinfo {pages} {276203}
  (\bibinfo {year} {2007})}\BibitemShut {NoStop}%
\bibitem [{\citenamefont {Sánchez}\ \emph {et~al.}(2022)\citenamefont
  {Sánchez}, \citenamefont {Vasilakaki}, \citenamefont {Lee}, \citenamefont
  {Normile}, \citenamefont {Andersson}, \citenamefont {Mathieu}, \citenamefont
  {López‐Ortega}, \citenamefont {Pichon}, \citenamefont {Peddis},
  \citenamefont {Binns}, \citenamefont {Nordblad}, \citenamefont {Trohidou},
  \citenamefont {Nogués},\ and\ \citenamefont {De~Toro}}]{sanchez22small}%
  \BibitemOpen
  \bibfield  {author} {\bibinfo {author} {\bibfnamefont {E.~H.}\ \bibnamefont
  {Sánchez}}, \bibinfo {author} {\bibfnamefont {M.}~\bibnamefont
  {Vasilakaki}}, \bibinfo {author} {\bibfnamefont {S.~S.}\ \bibnamefont {Lee}},
  \bibinfo {author} {\bibfnamefont {P.~S.}\ \bibnamefont {Normile}}, \bibinfo
  {author} {\bibfnamefont {M.~S.}\ \bibnamefont {Andersson}}, \bibinfo {author}
  {\bibfnamefont {R.}~\bibnamefont {Mathieu}}, \bibinfo {author} {\bibfnamefont
  {A.}~\bibnamefont {López‐Ortega}}, \bibinfo {author} {\bibfnamefont
  {B.~P.}\ \bibnamefont {Pichon}}, \bibinfo {author} {\bibfnamefont
  {D.}~\bibnamefont {Peddis}}, \bibinfo {author} {\bibfnamefont
  {C.}~\bibnamefont {Binns}}, \bibinfo {author} {\bibfnamefont
  {P.}~\bibnamefont {Nordblad}}, \bibinfo {author} {\bibfnamefont
  {K.}~\bibnamefont {Trohidou}}, \bibinfo {author} {\bibfnamefont
  {J.}~\bibnamefont {Nogués}},\ and\ \bibinfo {author} {\bibfnamefont {J.~A.}\
  \bibnamefont {De~Toro}},\ }\bibfield  {title} {\bibinfo {title} {Crossover
  from individual to collective magnetism in dense nanoparticle systems: Local
  anisotropy versus dipolar interactions},\ }\bibfield  {journal} {\bibinfo
  {journal} {Small}\ }\textbf {\bibinfo {volume} {18}},\ \href
  {https://doi.org/10.1002/smll.202106762} {10.1002/smll.202106762} (\bibinfo
  {year} {2022})\BibitemShut {NoStop}%
\bibitem [{\citenamefont {{F. Vernay, Z. Sabsabi, H.
  Kachkachi}}(2014)}]{vernayetal14prb}%
  \BibitemOpen
  \bibfield  {author} {\bibinfo {author} {\bibnamefont {{F. Vernay, Z. Sabsabi,
  H. Kachkachi}}},\ }\href@noop {} {\bibfield  {journal} {\bibinfo  {journal}
  {{Phys. Rev. B}}\ }\textbf {\bibinfo {volume} {90}},\ \bibinfo {pages}
  {094416} (\bibinfo {year} {2014})}\BibitemShut {NoStop}%
\bibitem [{\citenamefont {Davis}\ \emph {et~al.}(2020)\citenamefont {Davis},
  \citenamefont {Kang}, \citenamefont {Lee}, \citenamefont {Shin},
  \citenamefont {Putterman}, \citenamefont {Cheon},\ and\ \citenamefont
  {Shapiro}}]{davis20biop}%
  \BibitemOpen
  \bibfield  {author} {\bibinfo {author} {\bibfnamefont {H.~C.}\ \bibnamefont
  {Davis}}, \bibinfo {author} {\bibfnamefont {S.}~\bibnamefont {Kang}},
  \bibinfo {author} {\bibfnamefont {J.-H.}\ \bibnamefont {Lee}}, \bibinfo
  {author} {\bibfnamefont {T.-H.}\ \bibnamefont {Shin}}, \bibinfo {author}
  {\bibfnamefont {H.}~\bibnamefont {Putterman}}, \bibinfo {author}
  {\bibfnamefont {J.}~\bibnamefont {Cheon}},\ and\ \bibinfo {author}
  {\bibfnamefont {M.~G.}\ \bibnamefont {Shapiro}},\ }\bibfield  {title}
  {\bibinfo {title} {Nanoscale heat transfer from magnetic nanoparticles and
  ferritin in an alternating magnetic field},\ }\href
  {https://doi.org/10.1016/j.bpj.2020.01.028} {\bibfield  {journal} {\bibinfo
  {journal} {Biophysical Journal}\ }\textbf {\bibinfo {volume} {118}},\
  \bibinfo {pages} {1502–1510} (\bibinfo {year} {2020})}\BibitemShut
  {NoStop}%
\bibitem [{\citenamefont {Hergt}\ \emph {et~al.}(2004)\citenamefont {Hergt},
  \citenamefont {Hiergeist}, \citenamefont {Hilger}, \citenamefont {Kaiser},
  \citenamefont {Lapatnikov}, \citenamefont {Margel},\ and\ \citenamefont
  {Richter}}]{hergt04jmmm}%
  \BibitemOpen
  \bibfield  {author} {\bibinfo {author} {\bibfnamefont {R.}~\bibnamefont
  {Hergt}}, \bibinfo {author} {\bibfnamefont {R.}~\bibnamefont {Hiergeist}},
  \bibinfo {author} {\bibfnamefont {I.}~\bibnamefont {Hilger}}, \bibinfo
  {author} {\bibfnamefont {W.}~\bibnamefont {Kaiser}}, \bibinfo {author}
  {\bibfnamefont {Y.}~\bibnamefont {Lapatnikov}}, \bibinfo {author}
  {\bibfnamefont {S.}~\bibnamefont {Margel}},\ and\ \bibinfo {author}
  {\bibfnamefont {U.}~\bibnamefont {Richter}},\ }\bibfield  {title} {\bibinfo
  {title} {Maghemite nanoparticles with very high ac-losses for application in
  rf-magnetic hyperthermia},\ }\href
  {https://doi.org/10.1016/j.jmmm.2003.09.001} {\bibfield  {journal} {\bibinfo
  {journal} {Journal of Magnetism and Magnetic Materials}\ }\textbf {\bibinfo
  {volume} {270}},\ \bibinfo {pages} {345–357} (\bibinfo {year}
  {2004})}\BibitemShut {NoStop}%
\bibitem [{\citenamefont {Déjardin}\ and\ \citenamefont
  {Kachkachi}(2022)}]{DejKac2022}%
  \BibitemOpen
  \bibfield  {author} {\bibinfo {author} {\bibfnamefont {J.-L.}\ \bibnamefont
  {Déjardin}}\ and\ \bibinfo {author} {\bibfnamefont {H.}~\bibnamefont
  {Kachkachi}},\ }\bibfield  {title} {\bibinfo {title} {{Time profile of
  temperature rise in assemblies of nanomagnets}},\ }\href@noop {} {\bibfield
  {journal} {\bibinfo  {journal} {Journal of Magnetism and Magnetic Materials}\
  }\textbf {\bibinfo {volume} {556}},\ \bibinfo {pages} {169354} (\bibinfo
  {year} {2022})}\BibitemShut {NoStop}%
\bibitem [{\citenamefont {Lyeo}\ and\ \citenamefont
  {Cahill}(2006)}]{lyeo06prb}%
  \BibitemOpen
  \bibfield  {author} {\bibinfo {author} {\bibfnamefont {H.-K.}\ \bibnamefont
  {Lyeo}}\ and\ \bibinfo {author} {\bibfnamefont {D.~G.}\ \bibnamefont
  {Cahill}},\ }\bibfield  {title} {\bibinfo {title} {Thermal conductance of
  interfaces between highly dissimilar materials},\ }\bibfield  {journal}
  {\bibinfo  {journal} {Physical Review B}\ }\textbf {\bibinfo {volume} {73}},\
  \href {https://doi.org/10.1103/physrevb.73.144301}
  {10.1103/physrevb.73.144301} (\bibinfo {year} {2006})\BibitemShut {NoStop}%
\bibitem [{\citenamefont {{D. G. Cahill, P. V. Braun, G. Chen, D. R. Clarke, S.
  Fan, K. E. Goodson, P. Keblinski, W. P. King, G. D. Mahan, A. Majumdar, H. J.
  Maris, S. R. Phillpot, E. Pop,and L. Shi}}(2014)}]{Cahill_et_al_APR2014}%
  \BibitemOpen
  \bibfield  {author} {\bibinfo {author} {\bibnamefont {{D. G. Cahill, P. V.
  Braun, G. Chen, D. R. Clarke, S. Fan, K. E. Goodson, P. Keblinski, W. P.
  King, G. D. Mahan, A. Majumdar, H. J. Maris, S. R. Phillpot, E. Pop,and L.
  Shi}}},\ }\href@noop {} {\bibfield  {journal} {\bibinfo  {journal} {Appl.
  Phys. Reviews}\ }\textbf {\bibinfo {volume} {1}},\ \bibinfo {pages} {011305}
  (\bibinfo {year} {2014})}\BibitemShut {NoStop}%
\bibitem [{\citenamefont {{J.-L. D\'ejardin, H.
  Kachkachi}}(2024)}]{dejkac_as24}%
  \BibitemOpen
  \bibfield  {author} {\bibinfo {author} {\bibnamefont {{J.-L. D\'ejardin, H.
  Kachkachi}}},\ }\bibfield  {title} {\bibinfo {title} {Heat generation and
  diffusion in an assembly of magnetic nanoparticles: Application to magnetic
  hyperthermia.},\ }\href@noop {} {\bibfield  {journal} {\bibinfo  {journal}
  {Applied Sciences}\ }\textbf {\bibinfo {volume} {14}},\ \bibinfo {pages}
  {5757} (\bibinfo {year} {2024})}\BibitemShut {NoStop}%
\bibitem [{\citenamefont {{R. A. Rytov, V.A Bautin and N.A.
  Usov}}(2022)}]{RytovEtAl_sr22}%
  \BibitemOpen
  \bibfield  {author} {\bibinfo {author} {\bibnamefont {{R. A. Rytov, V.A
  Bautin and N.A. Usov}}},\ }\bibfield  {title} {\bibinfo {title} {{Towards
  optimal thermal distribution in magnetic hyperthermia}},\ }\href@noop {}
  {\bibfield  {journal} {\bibinfo  {journal} {Scientific. Rep.}\ }\textbf
  {\bibinfo {volume} {12}},\ \bibinfo {pages} {1} (\bibinfo {year}
  {2022})}\BibitemShut {NoStop}%
\bibitem [{\citenamefont {Talapatra}\ and\ \citenamefont
  {Adeyeye}(2020)}]{talapatra20nanoscale}%
  \BibitemOpen
  \bibfield  {author} {\bibinfo {author} {\bibfnamefont {A.}~\bibnamefont
  {Talapatra}}\ and\ \bibinfo {author} {\bibfnamefont {A.~O.}\ \bibnamefont
  {Adeyeye}},\ }\bibfield  {title} {\bibinfo {title} {Linear chains of
  nanomagnets: engineering the effective magnetic anisotropy},\ }\href
  {https://doi.org/10.1039/d0nr06026g} {\bibfield  {journal} {\bibinfo
  {journal} {Nanoscale}\ }\textbf {\bibinfo {volume} {12}},\ \bibinfo {pages}
  {20933–20944} (\bibinfo {year} {2020})}\BibitemShut {NoStop}%
\bibitem [{\citenamefont {Anand}\ \emph {et~al.}(2019)\citenamefont {Anand},
  \citenamefont {Banerjee},\ and\ \citenamefont {Carrey}}]{anand19prb}%
  \BibitemOpen
  \bibfield  {author} {\bibinfo {author} {\bibfnamefont {M.}~\bibnamefont
  {Anand}}, \bibinfo {author} {\bibfnamefont {V.}~\bibnamefont {Banerjee}},\
  and\ \bibinfo {author} {\bibfnamefont {J.}~\bibnamefont {Carrey}},\
  }\bibfield  {title} {\bibinfo {title} {Relaxation in one-dimensional chains
  of interacting magnetic nanoparticles: Analytical formula and kinetic monte
  carlo simulations},\ }\bibfield  {journal} {\bibinfo  {journal} {Physical
  Review B}\ }\textbf {\bibinfo {volume} {99}},\ \href
  {https://doi.org/10.1103/physrevb.99.024402} {10.1103/physrevb.99.024402}
  (\bibinfo {year} {2019})\BibitemShut {NoStop}%
\bibitem [{\citenamefont {Huízar-Félix}\ \emph {et~al.}(2016)\citenamefont
  {Huízar-Félix}, \citenamefont {Muñoz}, \citenamefont {Orue}, \citenamefont
  {Magén}, \citenamefont {Ibarra}, \citenamefont {Barandiarán}, \citenamefont
  {Muela},\ and\ \citenamefont {Fdez-Gubieda}}]{huizarfelix16apl}%
  \BibitemOpen
  \bibfield  {author} {\bibinfo {author} {\bibfnamefont {A.~M.}\ \bibnamefont
  {Huízar-Félix}}, \bibinfo {author} {\bibfnamefont {D.}~\bibnamefont
  {Muñoz}}, \bibinfo {author} {\bibfnamefont {I.}~\bibnamefont {Orue}},
  \bibinfo {author} {\bibfnamefont {C.}~\bibnamefont {Magén}}, \bibinfo
  {author} {\bibfnamefont {A.}~\bibnamefont {Ibarra}}, \bibinfo {author}
  {\bibfnamefont {J.~M.}\ \bibnamefont {Barandiarán}}, \bibinfo {author}
  {\bibfnamefont {A.}~\bibnamefont {Muela}},\ and\ \bibinfo {author}
  {\bibfnamefont {M.~L.}\ \bibnamefont {Fdez-Gubieda}},\ }\bibfield  {title}
  {\bibinfo {title} {Assemblies of magnetite nanoparticles extracted from
  magnetotactic bacteria: A magnetic study},\ }\bibfield  {journal} {\bibinfo
  {journal} {Applied Physics Letters}\ }\textbf {\bibinfo {volume} {108}},\
  \href {https://doi.org/10.1063/1.4941835} {10.1063/1.4941835} (\bibinfo
  {year} {2016})\BibitemShut {NoStop}%
\bibitem [{\citenamefont {Grmela}\ \emph {et~al.}(2005)\citenamefont {Grmela},
  \citenamefont {Lebon}, \citenamefont {Dauby},\ and\ \citenamefont
  {Bousmina}}]{Grmela2005Ballisticdiffusive}%
  \BibitemOpen
  \bibfield  {author} {\bibinfo {author} {\bibfnamefont {M.}~\bibnamefont
  {Grmela}}, \bibinfo {author} {\bibfnamefont {G.}~\bibnamefont {Lebon}},
  \bibinfo {author} {\bibfnamefont {P.~C.}\ \bibnamefont {Dauby}},\ and\
  \bibinfo {author} {\bibfnamefont {M.}~\bibnamefont {Bousmina}},\ }\bibfield
  {title} {\bibinfo {title} {Ballistic-diffusive heat conduction at nanoscale:
  Generic approach},\ }\href {https://doi.org/10.1016/j.physleta.2005.03.048}
  {\bibfield  {journal} {\bibinfo  {journal} {Physics Letters A}\ }\textbf
  {\bibinfo {volume} {339}},\ \bibinfo {pages} {237} (\bibinfo {year}
  {2005})}\BibitemShut {NoStop}%
\bibitem [{\citenamefont {Yang}\ \emph {et~al.}(2010)\citenamefont {Yang},
  \citenamefont {Zhang},\ and\ \citenamefont {Li}}]{Yang2010}%
  \BibitemOpen
  \bibfield  {author} {\bibinfo {author} {\bibfnamefont {N.}~\bibnamefont
  {Yang}}, \bibinfo {author} {\bibfnamefont {G.}~\bibnamefont {Zhang}},\ and\
  \bibinfo {author} {\bibfnamefont {B.}~\bibnamefont {Li}},\ }\bibfield
  {title} {\bibinfo {title} {Violation of {F}ourier's law and anomalous heat
  diffusion in silicon},\ }\href {https://doi.org/10.1016/j.nantod.2010.02.002}
  {\bibfield  {journal} {\bibinfo  {journal} {Nano Today}\ }\textbf {\bibinfo
  {volume} {5}},\ \bibinfo {pages} {85} (\bibinfo {year} {2010})},\ \bibinfo
  {note} {also available as arXiv:1002.3419}\BibitemShut {NoStop}%
\bibitem [{\citenamefont {Jou}\ \emph {et~al.}(2011)\citenamefont {Jou},
  \citenamefont {Sellitto},\ and\ \citenamefont {Alvarez}}]{Jou2011}%
  \BibitemOpen
  \bibfield  {author} {\bibinfo {author} {\bibfnamefont {D.}~\bibnamefont
  {Jou}}, \bibinfo {author} {\bibfnamefont {A.}~\bibnamefont {Sellitto}},\ and\
  \bibinfo {author} {\bibfnamefont {F.~X.}\ \bibnamefont {Alvarez}},\
  }\bibfield  {title} {\bibinfo {title} {Heat waves and phonon--wall collisions
  in nanowires},\ }\href {https://doi.org/10.1098/rspa.2010.0645} {\bibfield
  {journal} {\bibinfo  {journal} {Proceedings of the Royal Society A:
  Mathematical, Physical and Engineering Sciences}\ }\textbf {\bibinfo {volume}
  {467}},\ \bibinfo {pages} {2520} (\bibinfo {year} {2011})}\BibitemShut
  {NoStop}%
\bibitem [{\citenamefont {{G. Cahill, W. K. Ford, K. E. Goodson, G. D. Mahan,
  A. Majumdar, H. J. Maris, R. Merlin, S. R.
  Phillpot}}(2003)}]{Cahill_et_al_JAP2003}%
  \BibitemOpen
  \bibfield  {author} {\bibinfo {author} {\bibnamefont {{G. Cahill, W. K. Ford,
  K. E. Goodson, G. D. Mahan, A. Majumdar, H. J. Maris, R. Merlin, S. R.
  Phillpot}}},\ }\href@noop {} {\bibfield  {journal} {\bibinfo  {journal}
  {Appl. Physics}\ }\textbf {\bibinfo {volume} {93}},\ \bibinfo {pages} {793}
  (\bibinfo {year} {2003})}\BibitemShut {NoStop}%
\bibitem [{\citenamefont {Ziman}(1960)}]{Ziman1960}%
  \BibitemOpen
  \bibfield  {author} {\bibinfo {author} {\bibfnamefont {J.~M.}\ \bibnamefont
  {Ziman}},\ }\href@noop {} {\emph {\bibinfo {title} {Electrons and Phonons:
  The Theory of Transport Phenomena in Solids}}}\ (\bibinfo  {publisher}
  {Oxford University Press},\ \bibinfo {address} {Oxford},\ \bibinfo {year}
  {1960})\BibitemShut {NoStop}%
\bibitem [{\citenamefont {Rezgui}(2023)}]{Rezgui2023}%
  \BibitemOpen
  \bibfield  {author} {\bibinfo {author} {\bibfnamefont {H.}~\bibnamefont
  {Rezgui}},\ }\bibfield  {title} {\bibinfo {title} {Phonon hydrodynamic
  transport: Observation of thermal wave-like flow and second sound propagation
  in graphene at 100 k},\ }\href {https://doi.org/10.1021/acsomega.3c02558}
  {\bibfield  {journal} {\bibinfo  {journal} {ACS Omega}\ }\textbf {\bibinfo
  {volume} {8}},\ \bibinfo {pages} {23964} (\bibinfo {year}
  {2023})}\BibitemShut {NoStop}%
\bibitem [{\citenamefont {Ziambaras}\ and\ \citenamefont
  {Hyldgaard}(2005)}]{Ziambarasetal2005}%
  \BibitemOpen
  \bibfield  {author} {\bibinfo {author} {\bibfnamefont {E.}~\bibnamefont
  {Ziambaras}}\ and\ \bibinfo {author} {\bibfnamefont {P.}~\bibnamefont
  {Hyldgaard}},\ }\bibfield  {title} {\bibinfo {title} {{Phonon Knudsen flow in
  nanostructured semiconductor systems}},\ }\href
  {https://doi.org/10.1063/1.2175474} {\bibfield  {journal} {\bibinfo
  {journal} {Journal of Applied Physics}\ }\textbf {\bibinfo {volume} {99}},\
  \bibinfo {pages} {054303} (\bibinfo {year} {2005})}\BibitemShut {NoStop}%
\bibitem [{\citenamefont {Vernotte}(1958)}]{Vernotte1958}%
  \BibitemOpen
  \bibfield  {author} {\bibinfo {author} {\bibfnamefont {P.}~\bibnamefont
  {Vernotte}},\ }\bibfield  {title} {\bibinfo {title} {Les paradoxes de la
  th{\'e}orie continue de l’{\'e}quation de la chaleur},\ }\href@noop {}
  {\bibfield  {journal} {\bibinfo  {journal} {Comptes Rendus de l'Acad{\'e}mie
  des Sciences}\ }\textbf {\bibinfo {volume} {246}},\ \bibinfo {pages} {3154}
  (\bibinfo {year} {1958})}\BibitemShut {NoStop}%
\bibitem [{\citenamefont {Cattaneo}(1958)}]{Cattaneo1958}%
  \BibitemOpen
  \bibfield  {author} {\bibinfo {author} {\bibfnamefont {C.}~\bibnamefont
  {Cattaneo}},\ }\bibfield  {title} {\bibinfo {title} {Sur une forme de
  l’{\'e}quation de la chaleur {\'e}liminant le paradoxe d'une propagation
  instantan{\'e}e},\ }\href@noop {} {\bibfield  {journal} {\bibinfo  {journal}
  {Comptes Rendus de l'Acad{\'e}mie des Sciences}\ }\textbf {\bibinfo {volume}
  {247}},\ \bibinfo {pages} {431} (\bibinfo {year} {1958})}\BibitemShut
  {NoStop}%
\bibitem [{\citenamefont {Majumdar}(1993)}]{Majumdar1993}%
  \BibitemOpen
  \bibfield  {author} {\bibinfo {author} {\bibfnamefont {A.}~\bibnamefont
  {Majumdar}},\ }\bibfield  {title} {\bibinfo {title} {Microscale heat
  conduction in dielectric thin films},\ }\href
  {https://doi.org/10.1115/1.2910673} {\bibfield  {journal} {\bibinfo
  {journal} {Journal of Heat Transfer}\ }\textbf {\bibinfo {volume} {115}},\
  \bibinfo {pages} {7} (\bibinfo {year} {1993})}\BibitemShut {NoStop}%
\bibitem [{\citenamefont {Chen}(2000)}]{Chen2000}%
  \BibitemOpen
  \bibfield  {author} {\bibinfo {author} {\bibfnamefont {G.}~\bibnamefont
  {Chen}},\ }\bibfield  {title} {\bibinfo {title} {Particularities of heat
  conduction in nanostructures},\ }\href
  {https://doi.org/10.1023/A:1010005910368} {\bibfield  {journal} {\bibinfo
  {journal} {Journal of Nanoparticle Research}\ }\textbf {\bibinfo {volume}
  {2}},\ \bibinfo {pages} {199} (\bibinfo {year} {2000})}\BibitemShut {NoStop}%
\bibitem [{\citenamefont {Déjardin}\ and\ \citenamefont
  {Kachkachi}(2024)}]{DejKac2024}%
  \BibitemOpen
  \bibfield  {author} {\bibinfo {author} {\bibfnamefont {J.-L.}\ \bibnamefont
  {Déjardin}}\ and\ \bibinfo {author} {\bibfnamefont {H.}~\bibnamefont
  {Kachkachi}},\ }\bibfield  {title} {\bibinfo {title} {Heat generation and
  diffusion in an assembly of magnetic nanoparticles: Application to magnetic
  hyperthermia},\ }\href@noop {} {\bibfield  {journal} {\bibinfo  {journal}
  {Applied Sciences}\ }\textbf {\bibinfo {volume} {14}},\ \bibinfo {pages}
  {5757} (\bibinfo {year} {2024})}\BibitemShut {NoStop}%
\bibitem [{\citenamefont {{H.L. Rodr{\'\i}guez-Luccioni, M.M. Latorre-Esteves,
  J.J. M{\'e}ndez-Vega, O.O. Soto, A. R. Rodr{\'\i}guez, C.C. Rinaldi, M.M.
  Torres-Lugo}}(2011)}]{RodriguezEtAl_ijn11}%
  \BibitemOpen
  \bibfield  {author} {\bibinfo {author} {\bibnamefont {{H.L.
  Rodr{\'\i}guez-Luccioni, M.M. Latorre-Esteves, J.J. M{\'e}ndez-Vega, O.O.
  Soto, A. R. Rodr{\'\i}guez, C.C. Rinaldi, M.M. Torres-Lugo}}},\ }\bibfield
  {title} {\bibinfo {title} {{Enhanced reduction in cell viability by
  hyperthermia induced by magnetic nanoparticles}},\ }\href@noop {} {\bibfield
  {journal} {\bibinfo  {journal} {Int. J. Nanomed.}\ }\textbf {\bibinfo
  {volume} {6}},\ \bibinfo {pages} {373} (\bibinfo {year} {2011})}\BibitemShut
  {NoStop}%
\bibitem [{\citenamefont {Mehdaoui}\ \emph {et~al.}(2011)\citenamefont
  {Mehdaoui}, \citenamefont {Meffre}, \citenamefont {Carrey}, \citenamefont
  {Lachaize}, \citenamefont {Lacroix}, \citenamefont {Gougeon}, \citenamefont
  {Chaudret},\ and\ \citenamefont {Respaud}}]{Mehdaoui_AFM2011}%
  \BibitemOpen
  \bibfield  {author} {\bibinfo {author} {\bibfnamefont {B.}~\bibnamefont
  {Mehdaoui}}, \bibinfo {author} {\bibfnamefont {A.}~\bibnamefont {Meffre}},
  \bibinfo {author} {\bibfnamefont {J.}~\bibnamefont {Carrey}}, \bibinfo
  {author} {\bibfnamefont {S.}~\bibnamefont {Lachaize}}, \bibinfo {author}
  {\bibfnamefont {L.-M.}\ \bibnamefont {Lacroix}}, \bibinfo {author}
  {\bibfnamefont {M.}~\bibnamefont {Gougeon}}, \bibinfo {author} {\bibfnamefont
  {B.}~\bibnamefont {Chaudret}},\ and\ \bibinfo {author} {\bibfnamefont
  {M.}~\bibnamefont {Respaud}},\ }\bibfield  {title} {\bibinfo {title}
  {{Optimal Size of Nanoparticles for Magnetic Hyperthermia: A Combined
  Theoretical and Experimental Study}},\ }\href@noop {} {\bibfield  {journal}
  {\bibinfo  {journal} {Adv. Functional Mater.}\ }\textbf {\bibinfo {volume}
  {21}},\ \bibinfo {pages} {4573} (\bibinfo {year} {2011})}\BibitemShut
  {NoStop}%
\bibitem [{\citenamefont {Mehdaoui}\ \emph {et~al.}(2012)\citenamefont
  {Mehdaoui}, \citenamefont {Carrey}, \citenamefont {Stadler}, \citenamefont
  {Cornejo}, \citenamefont {Nayral}, \citenamefont {Delpech}, \citenamefont
  {Chaudret},\ and\ \citenamefont {Respaud}}]{mehdaouietal12apl}%
  \BibitemOpen
  \bibfield  {author} {\bibinfo {author} {\bibfnamefont {B.}~\bibnamefont
  {Mehdaoui}}, \bibinfo {author} {\bibfnamefont {J.}~\bibnamefont {Carrey}},
  \bibinfo {author} {\bibfnamefont {M.}~\bibnamefont {Stadler}}, \bibinfo
  {author} {\bibfnamefont {A.}~\bibnamefont {Cornejo}}, \bibinfo {author}
  {\bibfnamefont {C.}~\bibnamefont {Nayral}}, \bibinfo {author} {\bibfnamefont
  {F.}~\bibnamefont {Delpech}}, \bibinfo {author} {\bibfnamefont
  {B.}~\bibnamefont {Chaudret}},\ and\ \bibinfo {author} {\bibfnamefont
  {M.}~\bibnamefont {Respaud}},\ }\bibfield  {title} {\bibinfo {title}
  {{Influence of a transverse static magnetic field on the magnetic
  hyperthermia...}},\ }\href@noop {} {\bibfield  {journal} {\bibinfo  {journal}
  {Appl. Phys. Lett.}\ }\textbf {\bibinfo {volume} {100}},\ \bibinfo {pages}
  {052403} (\bibinfo {year} {2012})}\BibitemShut {NoStop}%
\bibitem [{\citenamefont {Haase}\ and\ \citenamefont
  {Nowak}(2012)}]{Haase_Nowak_PRB85_2012}%
  \BibitemOpen
  \bibfield  {author} {\bibinfo {author} {\bibfnamefont {C.}~\bibnamefont
  {Haase}}\ and\ \bibinfo {author} {\bibfnamefont {U.}~\bibnamefont {Nowak}},\
  }\bibfield  {title} {\bibinfo {title} {{Role of dipole-dipole interactions
  for hyperthermia heating of magnetic nanoparticle ensembles}},\ }\href
  {https://doi.org/10.1103/PhysRevB.85.045435} {\bibfield  {journal} {\bibinfo
  {journal} {Phys. Rev. B}\ }\textbf {\bibinfo {volume} {85}},\ \bibinfo
  {pages} {045435} (\bibinfo {year} {2012})}\BibitemShut {NoStop}%
\bibitem [{\citenamefont {Vallejo-Fernandez}\ and\ \citenamefont
  {O'Grady}(2013)}]{OGRADY_APL2013}%
  \BibitemOpen
  \bibfield  {author} {\bibinfo {author} {\bibfnamefont {G.}~\bibnamefont
  {Vallejo-Fernandez}}\ and\ \bibinfo {author} {\bibfnamefont {K.}~\bibnamefont
  {O'Grady}},\ }\bibfield  {title} {\bibinfo {title} {{Effect of the
  distribution of anisotropy constants on hysteresis losses for magnetic
  hyperthermia applications}},\ }\href {https://doi.org/10.1063/1.4824649}
  {\bibfield  {journal} {\bibinfo  {journal} {Applied Physics Letters}\
  }\textbf {\bibinfo {volume} {103}},\ \bibinfo {pages} {142417} (\bibinfo
  {year} {2013})},\ \Eprint
  {https://arxiv.org/abs/https://doi.org/10.1063/1.4824649}
  {https://doi.org/10.1063/1.4824649} \BibitemShut {NoStop}%
\bibitem [{\citenamefont {Conde-Leboran}\ \emph {et~al.}(2015)\citenamefont
  {Conde-Leboran}, \citenamefont {Baldomir}, \citenamefont {Martinez-Boubeta},
  \citenamefont {Chubykalo-Fesenko}, \citenamefont {del Morales}, \citenamefont
  {Salas}, \citenamefont {Cabrera}, \citenamefont {Camarero}, \citenamefont
  {Teran},\ and\ \citenamefont {Serantes}}]{condeetal15jpcc}%
  \BibitemOpen
  \bibfield  {author} {\bibinfo {author} {\bibfnamefont {I.}~\bibnamefont
  {Conde-Leboran}}, \bibinfo {author} {\bibfnamefont {D.}~\bibnamefont
  {Baldomir}}, \bibinfo {author} {\bibfnamefont {C.}~\bibnamefont
  {Martinez-Boubeta}}, \bibinfo {author} {\bibfnamefont {O.}~\bibnamefont
  {Chubykalo-Fesenko}}, \bibinfo {author} {\bibfnamefont {M.~P.}\ \bibnamefont
  {del Morales}}, \bibinfo {author} {\bibfnamefont {G.}~\bibnamefont {Salas}},
  \bibinfo {author} {\bibfnamefont {D.}~\bibnamefont {Cabrera}}, \bibinfo
  {author} {\bibfnamefont {J.}~\bibnamefont {Camarero}}, \bibinfo {author}
  {\bibfnamefont {F.~J.}\ \bibnamefont {Teran}},\ and\ \bibinfo {author}
  {\bibfnamefont {D.}~\bibnamefont {Serantes}},\ }\bibfield  {title} {\bibinfo
  {title} {{A Single Picture Explains Diversity of Hyperthermia Response of
  Magnetic Nanoparticles}},\ }\href {https://doi.org/10.1021/acs.jpcc.5b02555}
  {\bibfield  {journal} {\bibinfo  {journal} {The Journal of Physical Chemistry
  C}\ }\textbf {\bibinfo {volume} {119}},\ \bibinfo {pages} {15698} (\bibinfo
  {year} {2015})}\BibitemShut {NoStop}%
\bibitem [{\citenamefont {Ruta}\ \emph {et~al.}(2015)\citenamefont {Ruta},
  \citenamefont {Chantrell},\ and\ \citenamefont
  {Hovorka}}]{Ruta_ScientificReports_2015}%
  \BibitemOpen
  \bibfield  {author} {\bibinfo {author} {\bibfnamefont {S.}~\bibnamefont
  {Ruta}}, \bibinfo {author} {\bibfnamefont {R.}~\bibnamefont {Chantrell}},\
  and\ \bibinfo {author} {\bibfnamefont {O.}~\bibnamefont {Hovorka}},\
  }\bibfield  {title} {\bibinfo {title} {{Unified model of hyperthermia via
  hysteresis heating in systems of interacting magnetic nanoparticles}},\
  }\href {http://dx.doi.org/10.1038/srep09090} {\bibfield  {journal} {\bibinfo
  {journal} {Sci. Rep.}\ }\textbf {\bibinfo {volume} {5}},\ \bibinfo {pages}
  {9090} (\bibinfo {year} {2015})}\BibitemShut {NoStop}%
\bibitem [{\citenamefont {Morse}\ and\ \citenamefont
  {Feshbach}(1953)}]{MorseFeschbach_mgh53}%
  \BibitemOpen
  \bibfield  {author} {\bibinfo {author} {\bibfnamefont {P.~M.}\ \bibnamefont
  {Morse}}\ and\ \bibinfo {author} {\bibfnamefont {H.}~\bibnamefont
  {Feshbach}},\ }\href@noop {} {\emph {\bibinfo {title} {Methods of theoretical
  physics}}}\ (\bibinfo  {publisher} {McGraw-Hill},\ \bibinfo {address} {New
  York},\ \bibinfo {year} {1953})\BibitemShut {NoStop}%
\bibitem [{\citenamefont {Duffy}(2015)}]{duffy2015green}%
  \BibitemOpen
  \bibfield  {author} {\bibinfo {author} {\bibfnamefont {D.~G.}\ \bibnamefont
  {Duffy}},\ }\href@noop {} {\emph {\bibinfo {title} {Green's functions with
  applications}}}\ (\bibinfo  {publisher} {Chapman and Hall/CRC},\ \bibinfo
  {year} {2015})\BibitemShut {NoStop}%
\bibitem [{\citenamefont {{J. L. Garcia-Palacios}}(2007)}]{garpal00acp}%
  \BibitemOpen
  \bibfield  {author} {\bibinfo {author} {\bibnamefont {{J. L.
  Garcia-Palacios}}},\ }\bibinfo {title} {{On the Statics and Dynamics of
  Magnetoanisotropic Nanoparticles}},\ in\ \href
  {https://doi.org/10.1002/9780470141717.ch1} {\emph {\bibinfo {booktitle}
  {{Advances in Chemical Physics}}}},\ Vol.\ \bibinfo {volume} {112}\ (\bibinfo
   {publisher} {John Wiley \& Sons, Inc.},\ \bibinfo {year} {2007})\ pp.\
  \bibinfo {pages} {1--210}\BibitemShut {NoStop}%
\bibitem [{\citenamefont {{Z. Sabsabi, F. Vernay, O. Iglesias, H.
  Kachkachi}}(2013)}]{sabsabietal13prb}%
  \BibitemOpen
  \bibfield  {author} {\bibinfo {author} {\bibnamefont {{Z. Sabsabi, F. Vernay,
  O. Iglesias, H. Kachkachi}}},\ }\bibfield  {title} {\bibinfo {title}
  {{Interplay between surface anisotropy and dipolar interactions in an
  assembly of nanomagnets}},\ }\href@noop {} {\bibfield  {journal} {\bibinfo
  {journal} {Phys. Rev. B}\ }\textbf {\bibinfo {volume} {88}},\ \bibinfo
  {pages} {104424} (\bibinfo {year} {2013})}\BibitemShut {NoStop}%
\bibitem [{\citenamefont {D.~Ledue}\ and\ \citenamefont
  {Kachkachi}()}]{LedueEtal_2026}%
  \BibitemOpen
  \bibfield  {author} {\bibinfo {author} {\bibfnamefont {F.~V.}\ \bibnamefont
  {D.~Ledue}}\ and\ \bibinfo {author} {\bibfnamefont {H.}~\bibnamefont
  {Kachkachi}},\ }\bibfield  {title} {\bibinfo {title} {Magnetization
  relaxation of interacting chains of nanomagnets},\ }\href@noop {} {\bibinfo
  {journal} {In preparation}\ }\BibitemShut {NoStop}%
\bibitem [{\citenamefont {Timko}\ \emph {et~al.}(2009)\citenamefont {Timko},
  \citenamefont {Dzarova}, \citenamefont {Kovac}, \citenamefont {Skumiel},
  \citenamefont {Jozefczak}, \citenamefont {Hornowski}, \citenamefont
  {Gojzewski}, \citenamefont {Zavisova}, \citenamefont {Koneracka},
  \citenamefont {Sprincova}, \citenamefont {Strbak}, \citenamefont
  {Kopcansky},\ and\ \citenamefont {Tomasovcova}}]{timko2009magnetic}%
  \BibitemOpen
\bibfield  {journal} {  }\bibfield  {author} {\bibinfo {author} {\bibfnamefont
  {M.}~\bibnamefont {Timko}}, \bibinfo {author} {\bibfnamefont
  {A.}~\bibnamefont {Dzarova}}, \bibinfo {author} {\bibfnamefont
  {J.}~\bibnamefont {Kovac}}, \bibinfo {author} {\bibfnamefont
  {A.}~\bibnamefont {Skumiel}}, \bibinfo {author} {\bibfnamefont
  {A.}~\bibnamefont {Jozefczak}}, \bibinfo {author} {\bibfnamefont
  {T.}~\bibnamefont {Hornowski}}, \bibinfo {author} {\bibfnamefont
  {H.}~\bibnamefont {Gojzewski}}, \bibinfo {author} {\bibfnamefont
  {V.}~\bibnamefont {Zavisova}}, \bibinfo {author} {\bibfnamefont
  {M.}~\bibnamefont {Koneracka}}, \bibinfo {author} {\bibfnamefont
  {A.}~\bibnamefont {Sprincova}}, \bibinfo {author} {\bibfnamefont
  {O.}~\bibnamefont {Strbak}}, \bibinfo {author} {\bibfnamefont
  {P.}~\bibnamefont {Kopcansky}},\ and\ \bibinfo {author} {\bibfnamefont
  {N.}~\bibnamefont {Tomasovcova}},\ }\bibfield  {title} {\bibinfo {title}
  {Magnetic properties and heating effect in bacterial magnetic
  nanoparticles},\ }\href {https://doi.org/10.1016/j.jmmm.2009.02.120}
  {\bibfield  {journal} {\bibinfo  {journal} {Journal of Magnetism and Magnetic
  Materials}\ }\textbf {\bibinfo {volume} {321}},\ \bibinfo {pages} {1521}
  (\bibinfo {year} {2009})}\BibitemShut {NoStop}%
\bibitem [{\citenamefont {{Fdez-Gubieda}}\ \emph {et~al.}(2020)\citenamefont
  {{Fdez-Gubieda}}, \citenamefont {{Alonso}}, \citenamefont
  {{Garc{\'\i}a-Prieto}}, \citenamefont {{Garc{\'\i}a-Arribas}}, \citenamefont
  {{Fern{\'a}ndez Barqu{\'\i}n}},\ and\ \citenamefont
  {{Muela}}}]{Fdez_GubiedaEtAl_JAP2020}%
  \BibitemOpen
  \bibfield  {author} {\bibinfo {author} {\bibfnamefont {M.~L.}\ \bibnamefont
  {{Fdez-Gubieda}}}, \bibinfo {author} {\bibfnamefont {J.}~\bibnamefont
  {{Alonso}}}, \bibinfo {author} {\bibfnamefont {A.}~\bibnamefont
  {{Garc{\'\i}a-Prieto}}}, \bibinfo {author} {\bibfnamefont {A.}~\bibnamefont
  {{Garc{\'\i}a-Arribas}}}, \bibinfo {author} {\bibfnamefont {L.}~\bibnamefont
  {{Fern{\'a}ndez Barqu{\'\i}n}}},\ and\ \bibinfo {author} {\bibfnamefont
  {A.}~\bibnamefont {{Muela}}},\ }\bibfield  {title} {\bibinfo {title}
  {{Magnetotactic bacteria for cancer therapy}},\ }\href@noop {} {\bibfield
  {journal} {\bibinfo  {journal} {J. Appl. Phys.}\ }\textbf {\bibinfo {volume}
  {128}},\ \bibinfo {pages} {070902} (\bibinfo {year} {2020})}\BibitemShut
  {NoStop}%
\end{thebibliography}
%

\cleardoublepage{}

\appendix
\section{Linearized SLP Coefficients from Magnetic Response}
\label{app:sar_coeffs}

In the main text, the local heating power at site $\xi_{n}$ was linearized
as a temperature-dependent source, see Eq. (\ref{eq:SLP_linear}). Here we provide the explicit expressions
of the coefficients $a_p, b_p$ in terms of nanomagnet and field parameters, including dipolar-interaction (DI) effects.

\subsection*{Linearized SLP: linear response theory}

In linear-response theory, the SLP is given by Eq.~\eqref{eq:SAR_LRT}, where $\chi_{\mathrm{eq}}$ is the equilibrium susceptibility and $\Gamma\left(T\right)$ the relaxation rate, expressed in terms of the energy barrier $\sigma=KV/k_{\mathrm{B}}T$ and the DI coefficient $\lambda$. Approximate analytical expressions for the latter were developed in Refs. \onlinecite{garpal00acp, sabsabietal13prb, vernayetal14prb}.
In zero DC magnetic field, the following approximate expression of $\Gamma\left(T\right)$ was obtained in Ref.~\onlinecite{LedueEtal_2026},
\begin{equation}
 \Gamma_{\parallel}\left(0,\sigma,\xi\right) =\frac{2\sqrt{\sigma}}{\sqrt{\pi}\tau_{0}}e^{-\sigma\left(1+4\tilde{\lambda}\right)}.
 \label{eq:GammaParal_h0}
\end{equation}

Then, expanding in the relative temperature elevation $\theta=(T-T_0)/T_0$, we obtain the linearized SLP in Eq.~\eqref{eq:SLP_linear} with the following baseline and linear temperature-feedback coefficients
\begin{align}
a_p & =A_{0}\left[\left(1-\frac{1}{\sigma_{0}}\right)+\left(1-\frac{2}{\sigma_{0}}\right)\tilde{\lambda}\right]\frac{\eta_{0}}{1+\eta_{0}^{2}},\label{eq:a_sar_app}\\[0.5em]
b_p & =A_{0}\frac{\eta_{0}}{1+\eta_{0}^{2}}\Bigg\{-1+\frac{1}{2}\left(1-\frac{1}{\sigma_{0}}\right)(2\sigma_{0}-1)\left(\frac{\eta_{0}^{2}-1}{\eta_{0}^{2}+1}\right)\nonumber \\
 & \hspace{3.2em}+\left[-1+\frac{1}{2}\left(10\sigma_{0}-13+\frac{2}{\sigma_{0}}\right)\left(\frac{\eta_{0}^{2}-1}{\eta_{0}^{2}+1}\right)\right]\tilde{\lambda}\Bigg\},\label{eq:b_sar_app}
\end{align}
where we have introduced the standard dimensionless parameters
\begin{align}
\sigma_{0} & =\frac{KV}{k_{B}T_{0}},\quad V=\frac{\pi}{6}D^{3},\\
\varpi_{0} & =\omega\tau_{0},\\
\lambda & =\frac{\mu_{0}}{4\pi}\frac{M_{s}^{2}V^{2}}{d^{3}}\,\frac{1}{2KV}.\label{eq:dim_params_app}
\end{align}
and the effective relaxation parameter (including DI)
reads~\cite{LedueEtal_2026}
\begin{equation}
\eta_{0}=\frac{\sqrt{\pi}}{2}\,\varpi_{0}\,\frac{\exp\!\left[\sigma_{0}\left(1+4\tilde{\lambda}\right)\right]}{\sqrt{\sigma_{0}}}.\label{eq:eta0_app}
\end{equation}

Dipolar interactions renormalize $\lambda$ via the Riemann zeta factor:
\begin{equation}
\tilde{\lambda}=\lambda\,\zetaR(3).\label{eq:lambda_tilde_app}
\end{equation}

Here $\zetaR$ denotes the Riemann zeta function and should not be confused with the thermal coupling coefficient $\zeta$ introduced in the main text; for a chain $\zetaR(3)\simeq 1.202$.

The SLP prefactor
\begin{equation}
A_{0}=\frac{\mu_{0}M_{s}^{2}V}{k_{B}T_{0}}\cdot\frac{\mu_{0}h_{0}^{2}\omega}{2}\label{eq:A0_app}
\end{equation}
gives the scale of magnetic energy dissipation, with dimensions of power per unit volume (W/m$^3$).

Equations \eqref{eq:a_sar_app}--\eqref{eq:b_sar_app} provide the
explicit mapping from magnetic and geometrical parameters $(K,M_{s},V,d)$
and AC-field parameters $(h_{0},\omega)$ to the effective temperature-dependent
heating law used throughout this work.

\subsection*{Relationship to dimensionless coefficients in the main text}

The SLP coefficients $a_p$ and $b_p$ enter the thermal problem at
three successive description levels, each involving a different
rescaling:
\begin{enumerate}
\item \textbf{Nanomagnet-scale coefficients} $a = \Upsilon_0 a_p$
  and $b = \Upsilon_0 b_p$, with the scaling prefactor~$\Upsilon_0$
  given in Eq.~\eqref{eq:Upsilon}.
\item \textbf{Renormalized coefficients} $\tilde{a}$ and
  $\tilde{b}$ [Eq.~\eqref{eq:renorm_coeffs}], which account for
  finite interfacial thermal resistance through the dimensionless
  coupling strength~$\gamma$ and appear in the nanomagnet-scale heat
  equation~\eqref{eq:HE_nm_final}.
\item \textbf{Coarse-grained coefficients}
  $a_{\mathrm{cg}} = a/\beta_N$ and
  $b_{\mathrm{cg}} = b/\beta_N$
  [Eq.~\eqref{eq:acg_bcg}], which govern the assembly-scale
  source term~\eqref{eq:SAR_coarse}, with
  $\beta_N = L_N t_d/(\rho_m c_{v,m})$ the assembly-scale loss
  parameter.
\end{enumerate}
This hierarchy clarifies that the same underlying magnetic physics
(encoded in $a_p$ and $b_p$) gives rise
to different effective coefficients at different description levels,
due to interfacial effects ($\gamma$), loss scaling ($L_{m}$ vs
$L_{N}$), and spatial averaging.

\subsection*{Parameter study for magnetite nanomagnets}

For magnetite nanomagnets, we use the physical parameters in Table \ref{tab:MagnetiteParams} to study the dependence of SLP coefficients $a_p$ and $b_p$ on key parameters. The reference values correspond to those used in Sec.~\ref{sec:Results}.

\begin{table}[h]
\centering \caption{\label{tab:MagnetiteParams}Physical parameters used in the SAR coefficient
study.}
\begin{tabular}{@{}lcc@{}}
\toprule
\textbf{Parameter}  & \textbf{Symbol}  & \textbf{Default Value} \tabularnewline
\midrule
Nanoparticle diameter  & $D$  & 12.0 nm \tabularnewline
Interfacial coefficient  & $h_{s}$  & 0.330 W/(m$^{2}$$\cdot$K) \tabularnewline
AC field amplitude  & $h_{0}$  & 38.2 kA/m \tabularnewline
AC field frequency  & $f_{0}$  & 194.0 kHz \tabularnewline
Material  & ---  & Magnetite \tabularnewline
\bottomrule
\end{tabular}
\end{table}

\begin{figure}[H]
\centering \includegraphics[width=\columnwidth]{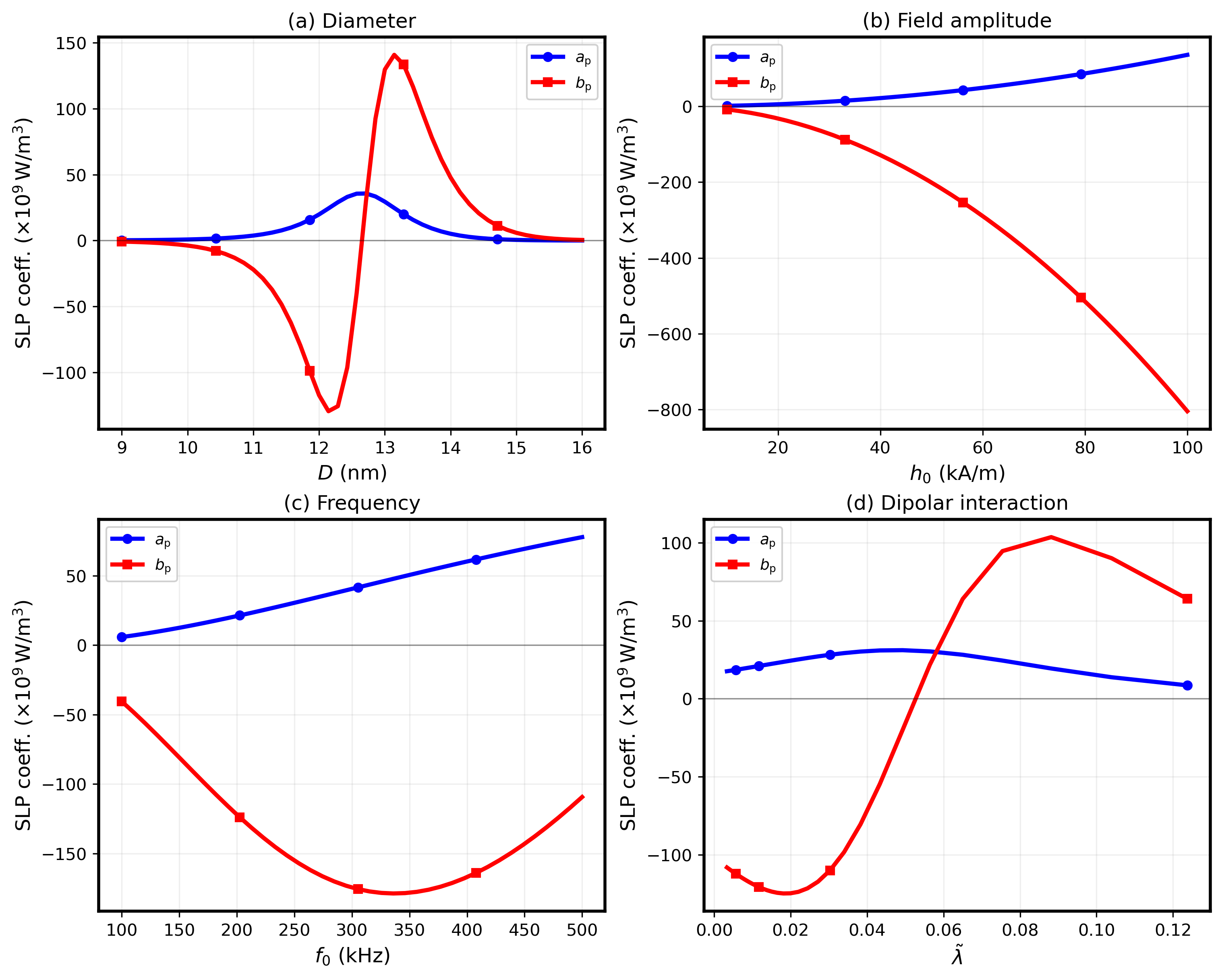}
\caption{ Dependence of SLP coefficients on (a) nanoparticle diameter $D$
and (b) AC field amplitude $h_{0}$. Solid lines: $a_p$;
dashed lines: $b_p$.}
\label{fig:SLP_coefficients}
\end{figure}

Figure~\ref{fig:SLP_coefficients} shows how $a_p$
and $b_p$ vary with nanomagnet diameter and field
amplitude. The baseline heating strength $a_p$ generally
increases with both $D$ and $h_{0}$, while the feedback coefficient
$b_p$ remains negative (indicating self-limiting heating)
and becomes more negative with increasing parameters, promoting thermal
localization.

\begin{figure}[H]
\centering \includegraphics[width=\columnwidth]{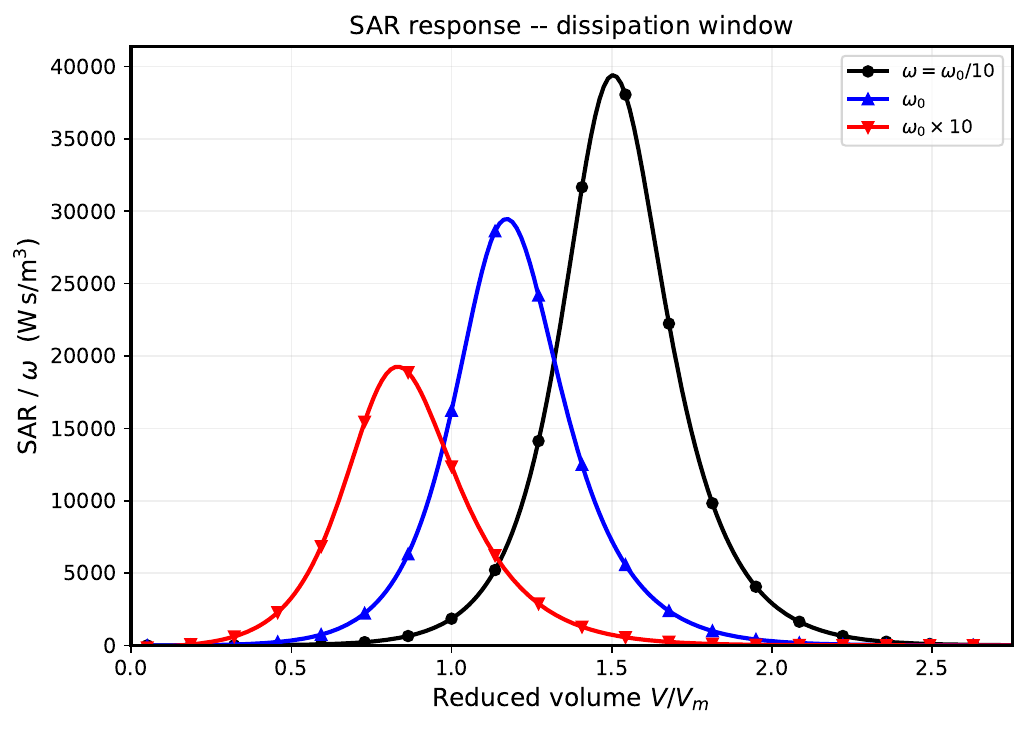}
\caption{ Specific loss power (SLP) normalized by frequency $\omega$
as a function of reduced volume $V/V_{m}$ for three different driving
frequencies: $\omega=\omega_{0}/10$, $\omega_{0}$, and $\omega_{0}\times10$.
The peak position shifts systematically with frequency, demonstrating
the frequency-tunable dissipation window. }
\label{fig:sar_response}
\end{figure}

The existence of the dissipation window, revealed by the SLP coefficients
$a_p$ and $b_p$ in Fig.~\ref{fig:SLP_coefficients},
directly manifests in the SLP response shown in Fig.~\ref{fig:sar_response}.
These plots demonstrate how the SLP($V/V_{m}$) curves exhibit distinct
peaks that shift systematically with driving frequency, confirming
the theoretical prediction of a frequency-tunable dissipation window.

At the reference frequency $\omega_{0}$, a well-defined peak emerges
where the driving frequency matches the natural timescales of the
nanomagnet system, enabling optimal energy absorption. When the frequency
decreases to $\omega_{0}/10$, the peak shifts toward larger volumes
($V/V_{m}$), indicating that lower frequencies couple efficiently
to larger magnetic elements. Conversely, at $\omega_{0}\times10$,
the peak moves to smaller volumes, showing that higher frequencies
effectively drive smaller nanomagnets.

This frequency-dependent tunability of the dissipation window underscores
the critical importance of matching the driving frequency to the nanomagnet
volume distribution for optimal SLP performance in applications such
as magnetic hyperthermia. Additionally, the observed variations in
peak broadening and amplitude across frequencies reflect changes in
the resonance quality factor and energy transfer efficiency within
the dissipation window.

\section{Eigenfunctions and modal solutions for Dirichlet and Neumann boundary conditions}
\label{app:BC_comparison}

The dimensionless heat equation~\eqref{eq:HE_nm_final},
\begin{equation}
\frac{\partial\theta}{\partial\tau}
  -\frac{\partial^{2}\theta}{\partial\xi^{2}}
  = \sum_{n=0}^{\mathcal{N}-1}
      \bigl[\tilde{a}+\tilde{b}\,\theta(\xi_{n},\tau)\bigr]
      \delta(\xi-\xi_{n})
    -\beta\,\theta(\xi,\tau),
\tag{\ref{eq:HE_nm_final}}
\end{equation}
is defined on the domain $0\le\xi\le\Lambda$ with
$\xi_{n}=n$ ($n=0,\dots,\mathcal{N}-1$) and $\Lambda=\mathcal{N}-1$.
Its solution depends on the boundary conditions (BC) imposed at
$\xi=0$ and $\xi=\Lambda$.
In this appendix we solve Eq.~\eqref{eq:HE_nm_final} for two limiting
choices --- Dirichlet (ideal heat bath) and Neumann (perfect insulation)
--- and investigate the role of the BC in the formation and persistence
of localized thermal hotspots at the nanomagnet positions.

\subsection*{Motivation}

The Dirichlet condition $\theta=0$ at both ends models an assembly
whose boundary is in perfect thermal contact with a reservoir at the
ambient temperature $T_{0}$.
This is appropriate, for example, for nanomagnets deposited on a
well-conducting substrate.
The Neumann (no-flux) condition $\partial_{\xi}\theta=0$ at both ends
models a thermally insulated boundary, as encountered in biological
systems (e.g.\ magnetotactic bacteria in a low-conductivity aqueous
environment \cite{timko2009magnetic, Fdez_GubiedaEtAl_JAP2020}) or in free-standing nanocomposite films.
Because the boundary conditions control whether heat can escape through
the system edges, they directly affect the amplitude and contrast of
the temperature field at the nanomagnet positions and therefore influence
the very existence of hotspots.

\subsection*{Dirichlet boundary conditions}

We impose
\begin{equation}
\theta(0,\tau)=\theta(\Lambda,\tau)=0,
\label{eq:BC_Dirichlet}
\end{equation}
corresponding to ideal thermal baths at the chain ends.
The orthonormal eigenfunctions of the operator
$-\partial_{\xi}^{2}$ satisfying these conditions are
\begin{equation}
\phi_{r}^{(\mathrm{D})}(\xi)
  =\sqrt{\frac{2}{\Lambda}}\,
   \sin\!\Bigl(\frac{r\pi\xi}{\Lambda}\Bigr),
\qquad r=1,2,\dots,
\label{eq:eigD}
\end{equation}
with eigenvalues
\begin{equation}
\lambda_{r}^{(\mathrm{D})}
  =\Bigl(\frac{r\pi}{\Lambda}\Bigr)^{2},
\qquad r=1,2,\dots
\label{eq:evalD}
\end{equation}
The lowest eigenvalue is $\lambda_{1}^{(\mathrm{D})}=(\pi/\Lambda)^{2}>0$; no zero mode exists.
Expanding $\theta(\xi,\tau)=\sum_{r\ge1}c_{r}(\tau)\,\phi_{r}^{(\mathrm{D})}(\xi)$, one obtains the modal system
\begin{equation}
\frac{d\mathbf{c}}{d\tau}
  =\mathbf{B}^{(\mathrm{D})}\,\mathbf{c}+\mathbf{d}^{(\mathrm{D})},
\label{eq:modal_D}
\end{equation}
where, after truncation to $R$ modes, $\mathbf{c}=(c_{1},\dots,c_{R})^{\!\top}$,
\begin{align}
B_{rs}^{(\mathrm{D})}
  &= -\bigl(\lambda_{r}^{(\mathrm{D})}+\beta\bigr)\delta_{rs}
     +\tilde{b}\,\mathcal{M}_{rs}^{(\mathrm{D})},
\label{eq:BD}\\[4pt]
d_{r}^{(\mathrm{D})}
  &= \tilde{a}\,\mathcal{S}_{r}^{(\mathrm{D})},
\label{eq:dD}
\end{align}
and the modal coupling matrix and source vector are
\begin{align}
\mathcal{M}_{rs}^{(\mathrm{D})}
  &= \sum_{n=0}^{\mathcal{N}-1}
     \phi_{r}^{(\mathrm{D})}(\xi_{n})\,
     \phi_{s}^{(\mathrm{D})}(\xi_{n}),
\label{eq:MrsD}\\[4pt]
\mathcal{S}_{r}^{(\mathrm{D})}
  &= \sum_{n=0}^{\mathcal{N}-1}
     \phi_{r}^{(\mathrm{D})}(\xi_{n}).
\label{eq:SrD}
\end{align}
A crucial observation is that
$\phi_{r}^{(\mathrm{D})}(0)=\phi_{r}^{(\mathrm{D})}(\Lambda)=0$
for every $r$.
Consequently, the boundary nanomagnets $n=0$ and $n=\mathcal{N}-1$
contribute nothing to either $\mathcal{M}_{rs}^{(\mathrm{D})}$ or
$\mathcal{S}_{r}^{(\mathrm{D})}$: they inject heat into the medium,
but the Dirichlet condition immediately drains it, rendering them
\emph{thermally invisible} in the modal decomposition.

For vanishing initial conditions, $\mathbf{c}(0)=\mathbf{0}$,
the exact solution reads
\begin{equation}
\mathbf{c}(\tau)
  =\bigl(\mathbf{B}^{(\mathrm{D})}\bigr)^{-1}
   \bigl(e^{\mathbf{B}^{(\mathrm{D})}\tau}-\mathbf{I}\bigr)\,
   \mathbf{d}^{(\mathrm{D})},
\label{eq:solD}
\end{equation}
provided $\mathbf{B}^{(\mathrm{D})}$ is invertible (i.e.\ the system
admits a steady state).

\subsection*{Neumann boundary conditions}

We now impose homogeneous Neumann (no-flux) conditions,
\begin{equation}
\left.\frac{\partial\theta}{\partial\xi}\right|_{\xi=0}
  =\left.\frac{\partial\theta}{\partial\xi}\right|_{\xi=\Lambda}
  =0,
\label{eq:BC_Neumann}
\end{equation}
corresponding to perfectly insulated chain ends.

The orthonormal eigenfunctions of $-\partial_{\xi}^{2}$ with these
conditions are
\begin{equation}
\phi_{r}^{(\mathrm{N})}(\xi)=
\begin{cases}
\displaystyle\frac{1}{\sqrt{\Lambda}},
  & r=0,\\[8pt]
\displaystyle\sqrt{\frac{2}{\Lambda}}\,
  \cos\!\Bigl(\frac{r\pi\xi}{\Lambda}\Bigr),
  & r=1,2,\dots,
\end{cases}
\label{eq:eigN}
\end{equation}
with eigenvalues
\begin{equation}
\lambda_{r}^{(\mathrm{N})}
  =\Bigl(\frac{r\pi}{\Lambda}\Bigr)^{2},
\qquad r=0,1,2,\dots
\label{eq:evalN}
\end{equation}
In particular, $\lambda_{0}^{(\mathrm{N})}=0$: the Neumann spectrum
admits a \emph{zero mode} --- a spatially uniform temperature
perturbation --- that is absent from the Dirichlet spectrum.
For $r\ge1$ the eigenvalues coincide with the Dirichlet ones.

Expanding in the Neumann basis,
$\theta(\xi,\tau)=\sum_{r\ge0}c_{r}(\tau)\,\phi_{r}^{(\mathrm{N})}(\xi)$,
one obtains an analogous modal system,
\begin{equation}
\frac{d\mathbf{c}}{d\tau}
  =\mathbf{B}^{(\mathrm{N})}\,\mathbf{c}+\mathbf{d}^{(\mathrm{N})},
\label{eq:modal_N}
\end{equation}
where now $\mathbf{c}=(c_{0},c_{1},\dots,c_{R})^{\!\top}$, and
\begin{align}
B_{rs}^{(\mathrm{N})}
  &= -\bigl(\lambda_{r}^{(\mathrm{N})}+\beta\bigr)\delta_{rs}
     +\tilde{b}\,\mathcal{M}_{rs}^{(\mathrm{N})},
\label{eq:BN}\\[4pt]
d_{r}^{(\mathrm{N})}
  &= \tilde{a}\,\mathcal{S}_{r}^{(\mathrm{N})},
\end{align}
with
\begin{align}
\mathcal{M}_{rs}^{(\mathrm{N})}
  &=\sum_{n=0}^{\mathcal{N}-1}
    \phi_{r}^{(\mathrm{N})}(\xi_{n})\,
    \phi_{s}^{(\mathrm{N})}(\xi_{n}),
\label{eq:MrsN}\\[4pt]
\mathcal{S}_{r}^{(\mathrm{N})}
  &=\sum_{n=0}^{\mathcal{N}-1}
    \phi_{r}^{(\mathrm{N})}(\xi_{n}).
\label{eq:SrN}
\end{align}

Unlike the Dirichlet case,
$\phi_{r}^{(\mathrm{N})}(0)=1/\sqrt{\Lambda}\ne 0$ (for $r=0$)
and $\phi_{r}^{(\mathrm{N})}(0)=\sqrt{2/\Lambda}\ne 0$ (for $r\ge1$).
Therefore, the boundary nanomagnets at $n=0$ and $n=\mathcal{N}-1$
contribute fully to the source vector and coupling matrix.
All $\mathcal{N}$ particles are \emph{thermally active} in the Neumann
formulation.

The formal solution is again
\begin{equation}
\mathbf{c}(\tau)
  =\bigl(\mathbf{B}^{(\mathrm{N})}\bigr)^{-1}
   \bigl(e^{\mathbf{B}^{(\mathrm{N})}\tau}-\mathbf{I}\bigr)\,
   \mathbf{d}^{(\mathrm{N})},
\label{eq:solN}
\end{equation}
provided $\mathbf{B}^{(\mathrm{N})}$ is invertible.

\subsection*{The zero mode and global heat accumulation}

The $r=0$ component of the Neumann modal system deserves special
attention.  Projecting onto $\phi_{0}^{(\mathrm{N})}=1/\sqrt{\Lambda}$
gives
\begin{equation}
\begin{split}
\frac{dc_{0}}{d\tau}
  ={}& -\beta\,c_{0}
    +\tilde{b}\,\frac{\mathcal{N}}{\Lambda}\,c_{0}\\
    &+\tilde{b}\sum_{s\ge1}\mathcal{M}_{0s}^{(\mathrm{N})}\,c_{s}
    +\tilde{a}\,\frac{\mathcal{N}}{\sqrt{\Lambda}},
\end{split}
\label{eq:c0_eqn}
\end{equation}
where we have used
$\mathcal{M}_{00}^{(\mathrm{N})}
  =\sum_{n}\bigl[\phi_{0}^{(\mathrm{N})}(\xi_{n})\bigr]^{2}
  =\mathcal{N}/\Lambda$.

In the absence of mode coupling (diagonal approximation), the zero-mode
amplitude grows or decays at rate
\begin{equation}
\sigma_{0}
  = -\beta+\tilde{b}\,\frac{\mathcal{N}}{\Lambda}.
\label{eq:sigma0}
\end{equation}
This rate is zero when
\begin{equation}
\tilde{b}
  = \tilde{b}_{c}^{(0)}
  \equiv \frac{\beta\,\Lambda}{\mathcal{N}}.
\label{eq:bc0_Neumann}
\end{equation}
For $\beta\to0$ (negligible nanoscale losses),
$\tilde{b}_{c}^{(0)}\to0$:
any nonzero positive feedback destabilises the uniform channel.
Physically, this reflects the fact that, with insulated boundaries
and no volumetric losses, there is no mechanism to remove heat from
the system; the temperature drifts upward without bound.

Contrast this with the Dirichlet case, where the fundamental-mode
stability criterion reads (cf.\ Eq.~\eqref{eq:bc_analytical})
\begin{equation}
\tilde{b}_{c}^{(\mathrm{D})}
  = \frac{\lambda_{1}^{(\mathrm{D})}+\beta}
         {\displaystyle\frac{2}{\Lambda}
          \sum_{n=0}^{\mathcal{N}-1}
          \sin^{2}\!\Bigl(\frac{\pi\xi_{n}}{\Lambda}\Bigr)},
\label{eq:bc_Dirichlet_recall}
\end{equation}
which remains finite even for $\beta=0$ because
$\lambda_{1}^{(\mathrm{D})}>0$.
Diffusion toward the boundary heat sinks alone is sufficient to stabilise
the system, independently of the nanoscale loss coefficient.

\emph{Summary}: the existence of the zero mode makes the Neumann system
generically less stable than its Dirichlet counterpart.  In the
physically relevant regime $\tilde{b}<0$ (self-limiting feedback) and
$\beta>0$, both systems admit steady states, but their temperature
levels and spatial structures differ qualitatively, as discussed next.

\subsection*{Steady-state temperature fields}

In steady state ($d\mathbf{c}/d\tau=0$), the temperature field
($\alpha=\mathrm{D}$ or $\mathrm{N}$) is
\begin{equation}
\theta_{\mathrm{ss}}^{(\alpha)}(\xi)
  = -\sum_{r}\frac{d_{r}^{(\alpha)}}{B_{rr}^{(\alpha)}}\,
     \phi_{r}^{(\alpha)}(\xi)
  + \text{(off-diag.\ corrections)},
\label{eq:ss_general}
\end{equation}
where the diagonal approximation (neglecting mode coupling through
$\tilde{b}\,\mathcal{M}_{rs}$, $r\ne s$) captures the leading
structure.  The full solution is obtained from
$\mathbf{c}_{\mathrm{ss}}=-(\mathbf{B}^{(\alpha)})^{-1}\mathbf{d}^{(\alpha)}$.

Two structural differences are noteworthy:

\begin{enumerate}
\item \textbf{Uniform temperature offset (Neumann only).}
The zero-mode contribution to the Neumann steady state is (using Eq.~\eqref{eq:c0_eqn})
\begin{equation}
\theta_{\mathrm{ss}}^{(\mathrm{N})}\Big|_{r=0}
  = \frac{\tilde{a}\,\mathcal{N}/\Lambda}
         {\beta-\tilde{b}\,\mathcal{N}/\Lambda}\,,
\label{eq:ss_uniform}
\end{equation}
which represents a spatially uniform temperature elevation of the
entire chain.  No analogous contribution exists in the Dirichlet
solution, where $\theta$ vanishes at the boundaries by construction.
For small $\beta$ and $\tilde{b}<0$,
\begin{equation}
\theta_{\mathrm{ss}}^{(\mathrm{N})}\Big|_{r=0}
  \simeq \frac{\tilde{a}}{|\tilde{b}|}\,,
\label{eq:ss_uniform_approx}
\end{equation}
which depends only on the ratio of baseline (or background) heating to feedback
strength and is independent of $\beta$ and the system size.

\item \textbf{Boundary nanomagnet temperatures.}
For the Dirichlet case,
$\theta_{\mathrm{ss}}^{(\mathrm{D})}(\xi_{0})
 =\theta_{\mathrm{ss}}^{(\mathrm{D})}(\xi_{\mathcal{N}-1})=0$:
the boundary nanomagnets always sit at the ambient temperature,
regardless of the heating strength.
For the Neumann case,
$\theta_{\mathrm{ss}}^{(\mathrm{N})}(\xi_{0})$ and
$\theta_{\mathrm{ss}}^{(\mathrm{N})}(\xi_{\mathcal{N}-1})$ are finite
and, in many configurations, comparable to the temperature of interior
particles.
\end{enumerate}

\subsection*{Hotspot contrast and the role of boundary conditions}

A natural measure of the spatial localization of heating is the
\emph{hotspot contrast}, defined as the temperature difference between
a nanomagnet position and the midpoint between two adjacent
nanomagnets.  For interior nanomagnet $n$ ($\alpha=\mathrm{D}$ or $\mathrm{N}$),
\begin{equation}
\begin{split}
\Delta\theta_{n}^{(\alpha)}
  ={}& \theta_{\mathrm{ss}}^{(\alpha)}(\xi_{n})\\
  &-\tfrac{1}{2}\bigl[\theta_{\mathrm{ss}}^{(\alpha)}(\xi_{n}-\tfrac{1}{2})
   +\theta_{\mathrm{ss}}^{(\alpha)}(\xi_{n}+\tfrac{1}{2})\bigr].
\end{split}
\label{eq:hotspot_contrast}
\end{equation}
Because the zero mode ($r=0$) is spatially uniform, it cancels
exactly in $\Delta\theta_{n}^{(\mathrm{N})}$:
\begin{equation}
\begin{split}
\Delta\theta_{n}^{(\mathrm{N})}
  ={}& \sum_{r\ge1}c_{r,\mathrm{ss}}^{(\mathrm{N})}
    \bigl[\phi_{r}^{(\mathrm{N})}(\xi_{n})\\
   &\quad -\tfrac{1}{2}\phi_{r}^{(\mathrm{N})}(\xi_{n}-\tfrac{1}{2})
          -\tfrac{1}{2}\phi_{r}^{(\mathrm{N})}(\xi_{n}+\tfrac{1}{2})\bigr].
\end{split}
\label{eq:Dtheta_Neumann}
\end{equation}
Therefore, the hotspot contrast is controlled entirely by the $r\ge1$
modes in both the Dirichlet and Neumann cases.  Since these modes
share the same eigenvalues
$\lambda_{r}=(r\pi/\Lambda)^{2}$, the contrast depends on two
factors that differ between the boundary conditions:
(i)~the \emph{source projections}
$\mathcal{S}_{r}^{(\alpha)}$, which determine how efficiently the
nanomagnet array drives each mode, and
(ii)~the \emph{modal coupling matrices}
$\mathcal{M}_{rs}^{(\alpha)}$, which control mode mixing through
the feedback term $\tilde{b}$.

A key difference arises from the behaviour at
the chain boundaries.  With Dirichlet BC, the sine functions vanish at
$\xi=0$ and $\xi=\Lambda$, so the two boundary nanomagnets
($n=0$, $n=\mathcal{N}-1$) do not contribute to any mode.
With Neumann BC, the cosine functions attain their extremal values
at the boundaries
($\cos 0=1$, $\cos r\pi=(-1)^{r}$), so the boundary nanomagnets
couple \emph{maximally} to the modal decomposition.
As a result, the effective number of active sources is
\begin{equation}
\mathcal{N}_{\mathrm{eff}}^{(\mathrm{D})}=\mathcal{N}-2,
\qquad
\mathcal{N}_{\mathrm{eff}}^{(\mathrm{N})}=\mathcal{N}.
\label{eq:Neff}
\end{equation}
For long chains ($\mathcal{N}\gg1$) this difference is negligible,
but for short chains or near the edges it becomes significant.

\subsection*{Single-source Green's function and local temperature peaks}

A complementary perspective is obtained from the diagonal Green's
function
$G^{(\alpha)}(\xi,\xi)=\sum_{r}[\phi_{r}^{(\alpha)}(\xi)]^{2}/(\lambda_{r}^{(\alpha)}+\beta)$,
which measures the steady-state temperature at a point $\xi$ due to a
unit point source at the same location.  Explicitly,
\begin{align}
G^{(\mathrm{D})}(\xi,\xi)
  &= \frac{2}{\Lambda}\sum_{r=1}^{\infty}
     \frac{\sin^{2}(r\pi\xi/\Lambda)}{(r\pi/\Lambda)^{2}+\beta},
\label{eq:GD}\\[6pt]
G^{(\mathrm{N})}(\xi,\xi)
  &= \frac{1}{\Lambda\,\beta}
    +\frac{2}{\Lambda}
  \sum_{r=1}^{\infty}
     \frac{\cos^{2}(r\pi\xi/\Lambda)}{(r\pi/\Lambda)^{2}+\beta}.
\label{eq:GN}
\end{align}
Using $\sin^{2}\!x+\cos^{2}\!x=1$, the difference is
\begin{equation}
\begin{split}
&G^{(\mathrm{N})}(\xi,\xi)-G^{(\mathrm{D})}(\xi,\xi)\\
  &\quad= \frac{1}{\Lambda\,\beta}
    +\frac{2}{\Lambda}\sum_{r=1}^{\infty}
     \frac{\cos(2r\pi\xi/\Lambda)}{(r\pi/\Lambda)^{2}+\beta}
  > 0
\end{split}
\label{eq:GN_minus_GD}
\end{equation}
for all $\xi\in(0,\Lambda)$.
Therefore,
\begin{equation}
G^{(\mathrm{N})}(\xi,\xi) > G^{(\mathrm{D})}(\xi,\xi)
\label{eq:GN_gt_GD}
\end{equation}
for all interior points.

\emph{The local temperature peak produced by each nanomagnet is always
higher under Neumann BC than under Dirichlet BC.}
The dominant contribution to the excess comes from the zero-mode
term $1/(\Lambda\beta)$, which diverges as $\beta\to0$.
This reflects the global heat accumulation that occurs when no heat
can escape through the boundaries.

\subsection*{Comparative stability criteria}

The stability of the steady state against perturbations is governed
by the eigenvalues of $\mathbf{B}^{(\alpha)}$.
In the diagonal approximation, the growth rate of mode $r$ is
\begin{equation}
\sigma_{r}^{(\alpha)}
  = -\bigl(\lambda_{r}^{(\alpha)}+\beta\bigr)
    +\tilde{b}\,\mathcal{M}_{rr}^{(\alpha)}.
\label{eq:growth_rate}
\end{equation}
The most damaging mode (highest $\sigma_{r}$, assuming $\tilde{b}>0$)
is the one with the smallest bare decay rate
$\lambda_{r}^{(\alpha)}+\beta$.

\textbf{Dirichlet:}
The slowest mode is $r=1$ with bare decay rate
$\lambda_{1}^{(\mathrm{D})}+\beta=(\pi/\Lambda)^{2}+\beta>0$.
The system is stable for
\begin{equation}
\tilde{b}<\tilde{b}_{c}^{(\mathrm{D})}
  =\frac{(\pi/\Lambda)^{2}+\beta}
        {\mathcal{M}_{11}^{(\mathrm{D})}}\,.
\label{eq:stab_D}
\end{equation}
Even for $\beta=0$, the boundary-induced diffusive decay
$(\pi/\Lambda)^{2}$ ensures a finite stability margin.

\textbf{Neumann:}
The slowest mode is $r=0$ with bare decay rate
$\lambda_{0}^{(\mathrm{N})}+\beta=\beta$.
The system is stable for
\begin{equation}
\tilde{b}<\tilde{b}_{c}^{(0)}
  =\frac{\beta\,\Lambda}{\mathcal{N}}\,.
\label{eq:stab_N}
\end{equation}
For $\beta\to0$, $\tilde{b}_{c}^{(0)}\to0$:
any positive feedback, however small, drives a global thermal
runaway through the uniform channel.

In the physically prevalent regime of self-limiting feedback
($\tilde{b}<0$, as is the case for the reference magnetite--PMMA
parameters of Table~\ref{table:reference_params}; see also
Appendix~\ref{app:sar_coeffs}), both systems are unconditionally
stable.  However, the Neumann system attains a higher steady-state
temperature level, as shown above.

For the higher modes ($r\ge1$), the stability criteria involve
the same eigenvalues $\lambda_{r}=(r\pi/\Lambda)^{2}$ in both
cases, with only the coupling constants $\mathcal{M}_{rr}^{(\alpha)}$
differing.  These differences are significant primarily for
low-$r$ modes and near the boundaries.

\subsection*{Physical discussion: which BC favours hotspots?}

The comparison reveals that the answer depends on whether ``hotspot''
is defined in absolute or relative terms.

\textbf{1.\ Absolute temperature at the nanomagnet positions.}
The Neumann BC yields strictly higher temperatures at every source
position [Eq.~\eqref{eq:GN_gt_GD}], for two reasons:
(a)~the zero mode accumulates heat globally when no boundary drain
is available, and
(b)~all $\mathcal{N}$ nanomagnets contribute as active sources,
including the two boundary nanomagnets that are suppressed under
Dirichlet BC.

\textbf{2.\ Hotspot contrast (temperature peak relative to local baseline).}
The zero mode cancels in the hotspot contrast~\eqref{eq:hotspot_contrast},
which depends only on $r\ge1$ modes.
For interior particles far from the boundaries, the difference between
sine and cosine projections diminishes and the contrasts are comparable.
Near the boundaries, however, the Neumann BC provides substantially
higher contrast because boundary nanomagnets are active sources
rather than thermally dead sinks.

\textbf{3.\ Relative temperature variance.}
The relative spatial variance
\begin{equation}
\mathcal{V}_{\mathrm{rel}}^{(\alpha)}
  = \frac{\operatorname{Var}(\theta_{\mathrm{ss}}^{(\alpha)})}
         {\langle\theta_{\mathrm{ss}}^{(\alpha)}\rangle^{2}}\,,
\label{eq:Vrel}
\end{equation}
where
$\langle\cdot\rangle=\mathcal{N}^{-1}\sum_{n}(\cdot)$ denotes the average over nanomagnet positions.
In the Dirichlet case, the mean
$\langle\theta_{\mathrm{ss}}^{(\mathrm{D})}\rangle$ is moderate
because boundary nanomagnets contribute zero; the variance is
controlled by the dome-shaped profile that vanishes at the edges.
In the Neumann case, the mean is raised by the zero mode
[Eq.~\eqref{eq:ss_uniform}], while the variance is set by
higher modes.  For small $\beta$, the zero-mode pedestal dominates
the mean, suppressing the relative variance:
\begin{equation}
\mathcal{V}_{\mathrm{rel}}^{(\mathrm{N})}
  \ll\mathcal{V}_{\mathrm{rel}}^{(\mathrm{D})}
\quad\text{when }
  \beta\ll|\tilde{b}|\,\frac{\mathcal{N}}{\Lambda}.
\label{eq:Vrel_comparison}
\end{equation}
\emph{In relative terms, the Dirichlet BC preserves sharper spatial
heterogeneity}, because the zero-temperature boundary condition forces
the profile to vary between zero and its maximum, whereas the
Neumann profile fluctuates about a high pedestal.

\subsection*{Summary}

Table~\ref{tab:BC_summary} collects the main differences.

\begin{table}[t]
\centering
\footnotesize
\caption{Comparison of Dirichlet and Neumann boundary conditions
for the nanomagnet--scale heat equation.}
\label{tab:BC_summary}
\begin{tabular}{@{}lp{1.1in}p{1.1in}@{}}
\toprule
\textbf{Property}
  & \textbf{Dirichlet}
  & \textbf{Neumann}
\tabularnewline
\midrule
Zero mode
  & absent
  & present ($\lambda_0=0$)
\tabularnewline
Boundary NM activity
  & suppressed
  & full
\tabularnewline
Active sources
  & $\mathcal{N}-2$
  & $\mathcal{N}$
\tabularnewline
SS ($\beta=0$, $\tilde{b}\le0$)
  & exists
  & only if $\tilde{b}<0$
\tabularnewline
SS ($\beta=0$, $\tilde{b}>0$)
  & if $\tilde{b}<\tilde{b}_c^{(\mathrm{D})}$
  & does not exist
\tabularnewline
Absolute NM temp.\
  & lower
  & higher
\tabularnewline
Hotspot contrast (int.)
  & comparable
  & comparable
\tabularnewline
Hotspot contrast (bdy.)
  & zero
  & finite
\tabularnewline
Relative spatial var.\
  & higher
  & lower (pedestal)
\tabularnewline
\bottomrule
\end{tabular}
\end{table}

The choice of boundary condition is therefore not merely a mathematical
convenience but has direct physical consequences for the thermal
landscape in nanomagnet assemblies.
Dirichlet conditions model well-thermalized boundaries and preserve
strong spatial heterogeneity, but at the cost of suppressing edge
effects and underestimating absolute temperatures.
Neumann conditions model thermal confinement and yield higher absolute
temperatures at all positions, but the resulting pedestal reduces the
relative contrast of hotspots.
In practice, real systems are likely intermediate (Robin BC), with the
effective Biot number $\mathrm{Bi}_{b}=h_{b}d/\kappa_{m}$ interpolating
between the two limits:
$\mathrm{Bi}_{b}\to\infty$ recovers Dirichlet, while
$\mathrm{Bi}_{b}\to0$ recovers Neumann.

For applications requiring \emph{maximum local temperature} at the
nanomagnet sites (e.g.\ bond-breaking, catalysis), Neumann-like
confinement is favourable.
For applications requiring \emph{maximum spatial contrast} between hot
and cold regions (e.g.\ targeted drug release, site-selective
activation), Dirichlet-like boundary thermalization is preferable.

\section{Green's function formulation of the nanomagnet-scale heat equation}
\label{app:GF_nanoscale}

For completeness, we present an alternative derivation of the temperature
field at the nanomagnet scale based on the Green's function of the
one-dimensional diffusion operator with Dirichlet boundary conditions \cite{duffy2015green}.
This formulation provides a real-space interpretation of the modal
solution derived in Sec.~\ref{sec:nanoscale} and is useful for benchmarking,
physical interpretation, and future extensions.

Throughout this appendix, we work in dimensional variables and consider the nanomagnet-resolved heat
equation, valid on time scales comparable to or slightly larger than
an AC-field period and on spatial scales comparable to the nanomagnet
size and interparticle spacing.

\subsection*{Green's function of the nanoscale diffusion operator with losses}

At the nanomagnet scale, the temperature elevation $\Delta T(x,t)=T(x,t)-T_{0}$
in the embedding matrix obeys the heat equation with localized sources
and nanoscale environmental losses [see Eq.~\eqref{eq:HE_nm_dim}]:
\begin{equation}
\rho_{m}c_{v,m}\,\frac{\partial\Delta T}{\partial t}=\kappa_{m}\,\partial_{x}^{2}\Delta T-L_{m}\Delta T(x,t)+\sum_{n=0}^{\mathcal{N}-1}P_{n}\,\delta(x-x_{n}),\label{eq:HE_nano_app}
\end{equation}
where $\kappa_{m}$ and $\rho_{m}c_{v,m}$ are the thermal conductivity
and volumetric heat capacity of the matrix, $L_{m}$ is the nanoscale
volumetric Newton cooling coefficient, and $P_{n}$ is the time-averaged
power dissipated by nanomagnet $n$. Note that interfacial thermal
exchange is incorporated through the self-consistent relation between
$P_{n}$ and the local temperature difference, as given by Eq.~\eqref{eq:self_consistent_T}
in the main text.

We define the linear diffusion-with-loss operator
\begin{equation}
\mathcal{L}=-\kappa_{m}\,\partial_{x}^{2}+L_{m},
\end{equation}
acting on the interval $0\le x\le L$ with Dirichlet boundary conditions
\begin{equation}
\Delta T(0,t)=\Delta T(L,t)=0,
\end{equation}
corresponding to a thermally regulated environment.

The Green's function $G(x,x';t)$ is defined as the solution of
\begin{equation}
\rho_{m}c_{v,m}\,\frac{\partial G}{\partial t}+\mathcal{L}G=\delta(t)\,\delta(x-x'),\label{eq:G_def_nano}
\end{equation}
with $G(x,x';t<0)=0$ and homogeneous Dirichlet boundary conditions\cite{MorseFeschbach_mgh53, duffy2015green}.

\subsection*{Modal representation of the Green's function}

Expanding $G$ in the Dirichlet eigenfunctions
\begin{equation}
\phi_{r}(x)=\sqrt{\frac{2}{L}}\sin\!\left(\frac{r\pi x}{L}\right),\qquad r=1,2,\dots,
\end{equation}
with eigenvalues
\begin{equation}
k_{r}^2=\left(\frac{r\pi}{L}\right)^{2},
\end{equation}
and noting that these are eigenfunctions of both $-\partial_{x}^{2}$
and the identity operator, we obtain\cite{MorseFeschbach_mgh53, duffy2015green}
\begin{equation}
G(x,x';t)=\sum_{r=1}^{\infty}\phi_{r}(x)\,\phi_{r}(x')\,\exp\!\left[-\left(\frac{\kappa_{m}k_{r}^2+L_{m}}{\rho_{m}c_{v,m}}\right)t\right].\label{eq:G_modal_nano}
\end{equation}

Each mode relaxes with a characteristic time
\begin{equation}
\tau_{r}=\frac{\rho_{m}c_{v,m}}{\kappa_{m}k_{r}^2+L_{m}},
\end{equation}
which includes contributions from both spatial diffusion and nanoscale
environmental losses.

\subsection*{Convolution representation of the temperature field}

Using Duhamel's principle, the solution of Eq.~\eqref{eq:HE_nano_app}
with zero initial condition can be written as
\begin{equation}
\Delta T(x,t)=\sum_{n=0}^{\mathcal{N}-1}\int_{0}^{t}P_{n}\,G(x,x_{n};t-t')\,dt'.\label{eq:T_G_conv}
\end{equation}

Evaluating this expression at the nanomagnet positions $x=x_{m}$
gives
\begin{equation}
\Delta T_{m}(t)=\sum_{n=0}^{\mathcal{N}-1}P_{n}\int_{0}^{t}G(x_{m},x_{n};t-t')\,dt',
\end{equation}
which describes the mutual thermal coupling between nanomagnets mediated
by the embedding matrix, including the effects of nanoscale losses.

In steady state, the convolution reduces to
\begin{equation}
\Delta T_{m}^{(\mathrm{ss})}=\sum_{n=0}^{\mathcal{N}-1}P_{n}\,\mathcal{G}(x_{m},x_{n}),
\end{equation}
where the static Green's function is
\begin{equation}
\mathcal{G}(x,x')=\sum_{r=1}^{\infty}\frac{\phi_{r}(x)\,\phi_{r}(x')}{\kappa_{m}k_{r}^2+L_{m}}.\label{eq:G_static_nano}
\end{equation}

\subsection*{Relation to the modal solution in Sec.~\ref{sec:nanoscale}}

The connection to the dimensionless modal system of Sec.~\ref{sec:nanoscale}
is established as follows.  Substituting the linearized
source~\eqref{eq:Pn_linear}, after eliminating the nanomagnet
internal temperature via the self-consistent
relation~\eqref{eq:self_consistent_T}, into
Eq.~\eqref{eq:HE_nano_app} and introducing the dimensionless
variables $\xi = x/d$, $\tau = t/t_d$, and
$\theta = \Delta T/T_0$~[Eq.~\eqref{eq:xi-tau-td}], one recovers
exactly the nanomagnet-scale heat
equation~\eqref{eq:HE_nm_final} with renormalized
coefficients $\tilde{a}$ and
$\tilde{b}$~[Eq.~\eqref{eq:renorm_coeffs}].  Projecting onto the
dimensionless eigenfunctions~\eqref{eq:eigenfunctions} then yields
the coupled ODE system~\eqref{eq:modal_nm_full} with decay rates
$\lambda_r + \beta = (r\pi/\Lambda)^2 + \beta$, where the first
term arises from spatial diffusion (the dimensional eigenvalue
$\kappa_m k_{r}^2$ rescaled by $t_d$) and the second from nanoscale
losses ($\beta = L_m t_d/(\rho_m c_{v,m})$).

The Green's-function formulation thus provides a complementary real-space
view of the same physics captured by the modal expansion: localized
heating by discrete nanomagnets, diffusive spreading in the matrix,
thermal coupling through the temperature field, and nanoscale environmental
losses.

\subsection*{Scope and limitations}

It is important to emphasize that the present formulation applies
to the \emph{nanomagnet-resolved} temperature field, prior to any
spatial or temporal coarse-graining. The interfacial coefficient $h_{s}$
enters through the self-consistent relation between $P_{n}$ and $\Delta T(x_{n},t)$
as given in Eq.~\eqref{eq:self_consistent_T} of the main text, and
is incorporated into the renormalized coefficients $\tilde{a}$ and
$\tilde{b}$. The nanoscale loss coefficient $L_{m}$ describes direct
environmental coupling at the nanomagnet level and must not be confused
with the effective volumetric leakage coefficient $L_{N}$ used in
coarse-grained, assembly-scale models. The mathematical connection
between these two descriptions and the hierarchical relationship $L_{N}=L_{m}+L_{\text{emergent}}$
is addressed in Sec.~\ref{subsec:scale_link}.

\end{document}